%% file: KrahEngelsSchneiderReiss_wPOD2020.tex
\documentclass{svjour3}                     

\smartqed  

\makeatletter
\def\cl@chapter{\@elt {theorem}}
\makeatother

\usepackage{graphicx}
\input{packages}
\input{math_shortcuts}
\graphicspath{{figures/}{svg-inkscape/}}  

\renewcommand{\includesvg}[2][]{%
\includegraphics[#1]{#2_svg-tex.pdf}}
\journalname{Noname}
\begin{document}

\title{Wavelet Adaptive Proper Orthogonal Decomposition for Large Scale Flow Data
}

\titlerunning{Wavelet Adaptive Proper Orthogonal Decomposition}        

\author{Philipp Krah \and
        Thomas Engels \and
        {Kai~~Schneider} \and
        Julius Reiss
}


\institute{
Philipp Krah,
Technische Universität Berlin,
Institute of Mathematics,
Straße des 17. Juni 136, 10623 Berlin, Germany\\
\email{krah@math.tu-berlin.de}
\vspace{0.2cm}
\\
Thomas Engels,
Department of Animal Physiology, Institute of Biological Sciences, University of Rostock, Albert-Einstein-Str. 3, Rostock 18059, Germany\\
\email{thomas.engels@uni-rostock.de}
\vspace{0.2cm}
\\
Kai Schneider,
Aix-Marseille Université, CNRS, Centrale Marseille, Institut de Math\'ematiques de Marseille (I2M), 39 rue Joliot-Curie, 13453 Marseille cedex 13, France\\
\email{kai.schneider@univ-amu.fr}
\vspace{0.2cm}
\\
Julius Reiss,
Technische Universität Berlin,
Institute of Fluid Mechanics and Engineering Acoustics,
Müller-Breslau-Str. 15,
10623 Berlin, Germany\\
\email{julius.reiss@tnt.tu-berlin.de}
\vspace{0.2cm}
\\
}

\date{Received: date / Accepted: date}

\maketitle

\input{abstract.tex}
\input{introduction.tex}
\input{numerics.tex}
\newpage
\input{algorithm.tex}
\input{errors.tex}
\input{results.tex}

\input{conclusion.tex}

\begin{acknowledgements}
The authors gratefully acknowledge the support of the Deutsche Forsch-ungs- gemeinschaft (DFG) as part of GRK2433 DAEDALUS.
The authors were granted access to the HPC resources of IDRIS under the Allocation No. 2018-91664 attributed by GENCI (Grand \'Equipement National de Calcul Intensif).
Centre de Calcul Intensif d’Aix-Marseille Universit\'e is acknowledged for granting access to its high performance computing resources.
\end{acknowledgements}

\section*{Conflict of Interest}
The authors declare that they have no conflict of interest.

\section*{Code and Data Availability}
Software can be found under \cite{WABBIT_github,WABBIT_pythongithub}. All scripts to reproduce the results are available at \cite{WABBIT_convergence}.

\bibliographystyle{spmpsci}      
\bibliography{references.bib}   

\appendix
\input{appendix.tex}

\end{document}

%% file: packages.tex
\usepackage{amsmath,amssymb,mathtools,wasysym}    
\usepackage{hyperref}
\usepackage{cleveref}
\usepackage{regexpatch}
\usepackage{algpseudocode}
\usepackage{algorithm}
\usepackage{ulem} 
\usepackage{tikz}
\usepackage{todonotes}
\makeatletter
\xpatchcmd{\@todo}{\setkeys{todonotes}{#1}}{\setkeys{todonotes}{inline,#1}}{}{}
\makeatother
\usepackage{parskip}
\usepackage{booktabs}
\usepackage{calc,intcalc}
\usepackage[list=true, font=large, labelfont=bf,labelformat=brace, position=top]{subcaption}
\usepackage[numbers]{natbib}
\usepackage{ifthen}
\usepackage{pdflscape}
\usepackage{siunitx}      
\usepackage{blkarray,bigstrut} 
\usepackage{wrapfig}
\usepackage{grffile}        
\usepackage{svg}            

\DeclareRobustCommand\dashed{\tikz[baseline=-0.6ex]\draw[thick,dashed] (0,0)--(0.54,0);}

%% file: abstract.tex
\begin{abstract}
The proper orthogonal decomposition (POD) is a powerful classical tool in fluid mechanics used, for instance, for model reduction and extraction of coherent flow features.
However, its applicability to high-resolution data, as produced by three-dimensional direct numerical simulations, is limited owing to its computational complexity.
Here, we propose a wavelet-based adaptive version of the POD (the wPOD), in order to overcome this limitation.
The amount of data to be analyzed is reduced by compressing them using biorthogonal wavelets, yielding a sparse representation while conveniently providing control of the compression error. 
Numerical analysis shows how the distinct error contributions of wavelet compression and POD truncation can be balanced under certain assumptions, allowing us to efficiently process 
high-resolution data from  three-dimensional simulations of flow problems. 
Using a synthetic academic test case, we compare our algorithm with the randomized singular value decomposition. Furthermore, we demonstrate the ability of our method analyzing data of a 2D wake flow and a 3D flow generated by a flapping insect computed with direct numerical simulation.


\keywords{
  Proper Orthogonal Decomposition \and
  Biorthogonal Wavelets \and
  Wavelet Adaptive Block-Based Grids \and
  Fluid Dynamics \and
  Reduced Order Models}
\end{abstract}

%% file: introduction.tex

\section{Introduction}

The Proper Orthogonal Decomposition (POD) \cite{SiL1987} is one of the most important methods in modern data analysis of fluid flows. For large-scale data, as typically produced by high resolution direct numerical simulation or high resolution imaging, the POD finds low dimensional descriptions by approximating the snapshot data $\mathcal{U}=\{\u_1,\dots,\u_{\NSnapshots}\}$ in terms of a few orthogonal basis functions $\vec{\Mode}_k$, called modes:
  \begin{equation}
    \label{eq:basicPODansatz}
    \tilde{\u}_i=\sum_{k=1}^r a_{ki}\vec{\Mode}_k \quad \text{ with }\quad a_{ki} = \sprod{\u_i}{ \vec{\Mode}_k}, \qquad r\ll \NSnapshots\,.
  \end{equation}
This effectively reduces the high dimensional data $\mathcal{U}$ to a set of $N_s\times r$ coefficients $a_{ki}$. It is known that this choice of basis is optimal in minimizing the approximation error $\sum_i\Vert \u_i - \tilde{\u}_i\Vert$ {within a linear subspace} (see, for example \cite{Volkwein2011} or, in the context of matrices, the Eckart Young theorem \cite{EcY1936}). However, it is not optimal with respect to memory efficiency and applicability, as for reconstructing any approximation on $\mathcal{U}$, the high dimensional snapshots $\u$ and basis vectors $\Mode_k$ need to be stored and processed. This often makes any non-linear observable of  $\mathcal{U}$ difficult to process on a desktop computer. Therefore, in this article, we aim at reducing the amount of data that is needed to calculate $\Mode_k$, exploiting sparsity enabled by wavelet adaptation techniques.

POD has been applied successfully in Model Order Reduction (MOR) (for a review, see \cite{BennerOhlbergerPateraRozzaUrban2017,BennerCohenOhlbergerWillcox2017} or the lecture notes of  \cite{Volkwein2011} for POD-MOR). In combination with Galerkin projection methods, POD is used to obtain reduced models of discretized partial differential equations (PDEs).
It is the basis of many modern decompositions in fluid dynamics, like the shifted POD \cite{ReS2018} used for analysis of transport phenomena or the spectral \cite{SieberPaschereitOberleithner2016} and multiscale POD \cite{MendezBalabaneBuchlin2018}, which identify coherent structures with specific energies.
However, most of the applications mentioned above have been only applied in one or two spatial dimensions,
since the tremendous amount of data in fluid dynamics {in three dimensions} makes the computation of POD modes extremely expensive, if not unfeasible. Recent attempts to improve this are going in two directions: adaptivity and randomization.

Randomized methods are inspired by randomized numerical linear algebra (see the surveys of \cite{HalkoMartinssonTropp2011,Mahoney2011}) using random projection matrices to approximate the column space of a possible tall and skinny snapshot matrix build from $\mathcal{U}$ onto a smaller surrogate matrix to solve the classical POD problem using snapshot POD \cite{YuChakravorty2015} or singular value decomposition (SVD) \cite{AllaKutz2019}.
After the POD modes are identified for the small system, they are projected back onto the original high dimensional space. Although these methods allow for rapidly solving the POD problem for large scale systems, they have two main drawbacks: the resulting modes are not sparse and the algorithm does not converge if the singular values do not decay rapidly \cite{HalkoMartinssonTropp2011}.

Adaptive methods benefit from the sparse representation of the data already in the stage of production. For example, when generating the data numerically using finite element solvers like \Softwarename{FEniCS} \cite{Fenics15_2015,FEniCS_page} or wavelet adaptive solvers such as \Softwarename{WABBIT} \cite{WABBIT2018,EngelsSchneiderReissFarge2019,WABBIT_github}. In contrast to randomized methods, the representation of the data is seen from an infinite dimensional perspective, where each snapshot corresponds to a function over an infinite dimensional Hilbert space. This fact has both advantages and drawbacks:
On the one hand, adaptation techniques allow direct relations to the {}"truth", i.e. the exact solution of the PDE or the underlying physical system. For example, \cite{AliSteihUrban2017,AliUrban2016} obtain approximations of the exact solution of the PDE within a given tolerance by using dual wavelet expansions of the PDE systems' continuous residual together with greedy reduced basis methods.
Furthermore, adaptation is advantageous over equidistant grids because it distributes computational efforts to places where relevant information is located.
On the other hand, adaptive methods require specifically tailored techniques for storing and processing the data. When ignoring gains in precision, adaptive methods are thus more complex and computationally demanding compared to non-adaptive equivalents for the same amount of processed data. Transferred to the POD problem this implies that the POD basis cannot be formed by means of a simple SVD or the \textit{direct method} (as coined in \cite{SiL1987}), since no snapshot matrix or correlations between (a theoretically infinite number of) space points can be computed. Alongside \cite{UllmannRotkvicLang2016,GrassleHinzeLangUllmann2019,GraessleHinze2018,FangPainNavonPiggotGormanAllisonGoddard2009}, we therefore use the so-called \textit{method of snapshots or strobes} \cite{SiL1987}, which only relies on correlations between a finite number of snapshots, computed with a simple inner product.
Nevertheless, snapshot and direct methods are in principle  equivalent in solving the POD problem \cite{Volkwein2001,HolmesLumleyBerkoozRowley2012}.

Although most authors use the method of snapshots to obtain POD modes, their precise implementations differ.
In early works, Ullmann et al. \cite{UllmannRotkvicLang2016} used two different approaches to compute the POD basis. In the first approach, they represent all snapshots and the resulting POD basis on a common finite element (FE) grid/space, which is somehow similar to the approach in \cite{FangPainNavonPiggotGormanAllisonGoddard2009}. This has the advantage that the snapshots and modes can be interpreted as Euclidean vectors of the same size with a single weighted inner product. In the second approach, the authors build common FE-spaces for pairs of snapshots to compute the correlation matrix and define the POD modes implicitly as a linear combination of the snapshots on their original FE-space. A variation of the first approach has been used by \cite{GrassleHinzeLangUllmann2019} to build reduced order models for {the} incompressible Navier-Stokes equations. Most recent advances \cite{GraessleHinze2018} go one step further to avoid any common finite element space. Therefore, {the authors of} \cite{GraessleHinze2018} reformulate the inner product between two snapshots in terms of the underlying finite element basis. Based on the work of \cite{MassingLarsonLogg2013}, the authors in \cite{GraessleHinze2018} are able to compute inner products between arbitrary FE- grids, when including cut finite elements. Instead of lifting the grids to a common reference grid, the overlap between different finite elements has to be calculated.
In \cite{FBdSBF2011} the POD is compared with orthogonal wavelets for extracting coherent vortices out of turbulent flows considering direct numerical simulation data of 2D drift-wave turbulence. Issues of computational complexity and memory requirements for storing the POD modes were discussed.

The present paper approaches  the method of snapshots from a wavelet point of view as it entails various desirable conceptual features that are not shared by the other adaptive methods outlined above.
It shares {some basic ideas} with \cite{CastrillonAmaratunga2002,UytterhoevenRoose1997}, although the results presented {there} are only in one space dimension. The approach presented here is integrated as a post-processing routine into an open source software called \Softwarename{WABBIT} ((W)avelet (A)daptive (B)lock (B)ased Solver for (I)nteractions with (T)urbulence), which is freely available at \cite{WABBIT_github}. The basic adaptation technique grounds on the idea of \cite{DominguesGomesDiaz2003} and the recent works \cite{WABBIT2018,EngelsSchneiderReissFarge2019}, in which a block-structured grid is refined or coarsened using biorthogonal wavelets. The framework is implemented using the MPI Library to exploit parallel computing architectures. The novelty of the present approach lies in the combination of POD and wavelet adaptation and the ability to balance {wavelet compression and POD truncation} errors.
Here, the hierarchical nature of wavelets enables efficient refinement and coarsening due to the scaling relations of wavelets.
Moreover, the underlying multiresolution analysis (MRA) \cite{Mallat1989} allows to approximate any finite energy function with a given precision. Nonlinear approximation \cite{DeVore1998} using thresholding of the wavelet coefficients yields then sparse and efficient representations to reduce memory and CPU time requirements.
Furthermore, the locally uniform Cartesian grid structure of each block enables us to apply the method to images or other 2D/3D equidistant fields in space. This allows us to directly compare our results with randomized methods, which need several passes over the data in case the singular values of the snapshot matrix decay slowly \cite{HalkoMartinssonTropp2011}. Our results show indeed  that we can avoid this by sparse adaptation of the data. This serves as motivation to propose a strategy of how to balance a priori rank truncation and wavelet compression error.


 The present paper is organized as follows. In section \ref{sec:numerics}, we introduce the wavelet adaptive framework with a detailed description of our implementation  (\cref{sec:GridAndImplementation}) and a brief summary of the applied wavelet adaptation scheme (\cref{sec:bbwadapt}). The wavelet adaptive POD is outlined in section \ref{sec:wPODAlgo}, followed by an error estimation in \cref{sec:ErrorAnalysis}, where we provide a strategy for balancing adaptation and truncation errors. Next, we examine and discuss the behavior of our algorithm with the help of three different examples in \cref{sec:numerical-results}. In the first example a 2D synthetic test case is provided, where
 the influence of various parameters is studied and the results are compared to the randomized SVD.
Thereafter, \cref{subsec:directNumericalandAdaptive} presents two data sets in the context of computational fluid dynamics:  data from a direct numerical simulation of a 2D wake flow past a cylinder (\cref{sec:vortex_street}) and a 3D block-based adaptive simulation of a bumblebee in forward flight (see section \cref{sec:3DInsectsFlight}). Finally, we summarize our results and provide a short outlook for future research (\cref{sec:conclusion_and_outlook}).

%% file: numerics.tex
\newcommand{\prefactor}{a_j}
\section{Numerical Methods and Implementation}
\label{sec:numerics}
In the following, we describe the numerical methods used in the wPOD algorithm and
give detailed insight into its implementation, when handling multiple block-based adaptive grids. The basic wavelet adaptation technique used for our algorithm has been already discussed in \cite{WABBIT2018,EngelsSchneiderReissFarge2019}. We hence limit the presentation here to changes specific to our algorithm. In the interest of readability, we will assume two-dimensional data, thus all quantities with a subscript $\alpha$ are indexed over $\alpha=1,2$.

\subsection{Block Structured Grid and Implementation}
\label{sec:GridAndImplementation}

Multiresolution representations require a dedicated data structure. Here, spatial data is divided into a set of nested blocks, which are organized in a tree. We use a collection of trees, which we call forest, in order to store multiple snapshots together with their designated tree simultaneously.

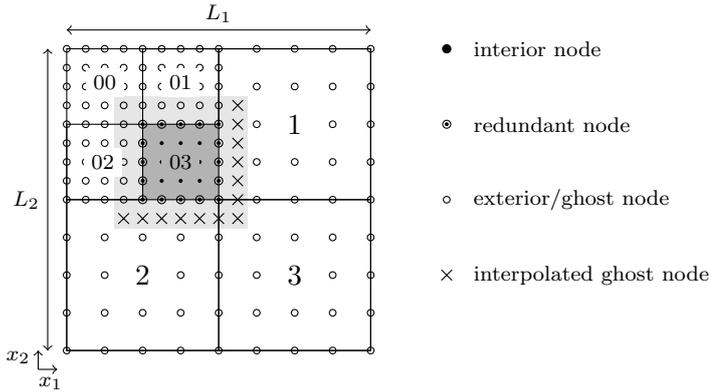
\begin{figure}[htbp]
  \centering
  \input{lattice}
  \caption{ A 2D grid composed of seven blocks. A single block (dark grey)
            consists of an odd number of interior grid nodes $\Bs_\alpha=5$
            in direction $x_\alpha$, where $\alpha=1,2$. The block includes a ghost node layer (light gray) which is synchronized with neighboring blocks. The size of this layer depends on the {support of} the chosen wavelet. Ghost nodes are interpolated, if the levels of the neighboring blocks differ.
         }
  \label{fig:blockstructure}
\end{figure}

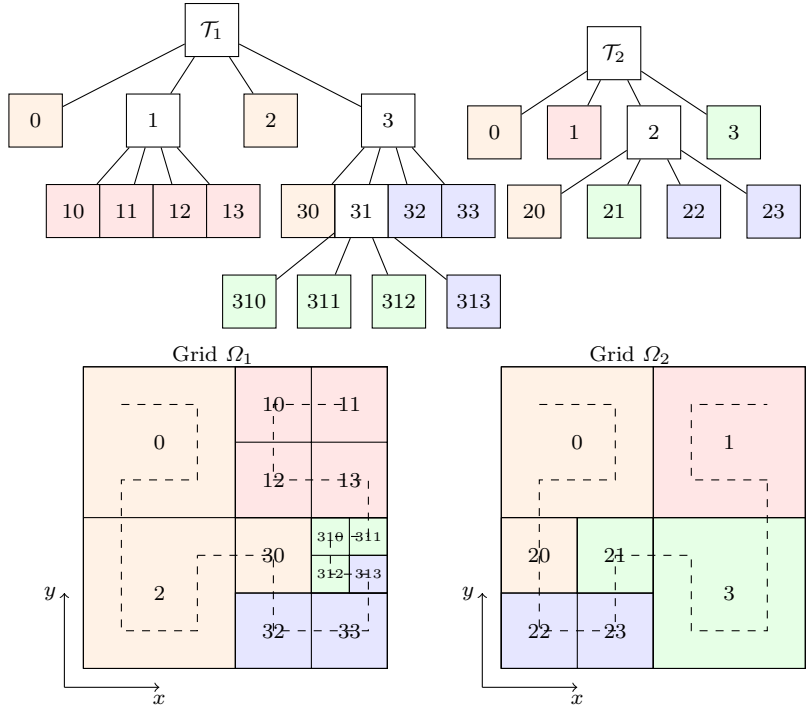
\begin{figure}[h]
  \centering
  \input{treestructure}
  \caption{ A forest of two {trees} and their associated grids $\Omega_i$. Top part visualizes the tree structure for tree $i=1,2$. Colored leafs at the end of a branch correspond to blocks on the grid as shown in bottom part. The block color encodes the processor that holds the block. The blocks are distributed among processors using space filling curves (dashed lines).}
  \label{fig:treestructure}
\end{figure}

\paragraph{Computational Grid:}
With each tree in the forest $\Forest$, we associate a multiresolution
grid $\Omega_i$ on a rectangular domain $\Domain\defeq [0,L_1]\times [0,L_2]\subset \mathbb{R}_+^2$.
As {illustrated} in \cref{fig:treestructure} for the 2D case,
the grid $\Omega_i$ is composed of blocks
\begin{equation}
  \Block_p^j=\{x= x_p + (k_1 \Delta x_1^j,k_2 \Delta x_2^j) \mid k_1=0,\dots,\Bs_1\,, k_2 = 0,\dots,\Bs_2\}
\end{equation}
of equal size ${\Bs}_1 \times {\Bs}_2$.
The subdivision of the grid is controlled by the tree level $j=\Jmin,\Jmin +1,\dots,\Jmax$.
With increasing tree level $j\to j+1$ the lattice spacing of the block is divided by
two, i.e. {$\Delta x^j_\alpha = 2^{-j} L_\alpha/(\Bs_\alpha-1)$}.
Here, $L_\alpha>0$ is the size of the compuational domain.
Gradedness of the resulting grid is enforced by allowing adjacent blocks to differ only by one tree-level.
The level and location of a block is encoded in a unique tree code as indicated in fig.~\ref{fig:blockstructure}. The computational grid $\Omega_i$ is the union of all blocks in the tree,
\begin{equation}
  \Grid_i = \bigcup_{j=\Jmin}^{\Jmax}\bigcup_{p\in\Lambda^j_i} \Block_p^j,
  \label{eq-def:MultiresolutionMesh}
\end{equation}
where $\Lambda^j_i$ is the set of all block IDs at a given tree-level $j$ and tree $i$.
The block IDs $p\in\Lambda^j_i$ are called \textit{tree code}.
 The synchronization between blocks is done by using an overlapping layer covering $g$ points (light gray area in  \cref{fig:blockstructure}). These additional points are called \textit{ghost nodes} and they form the \textit{ghost node layer}.
The size of the ghost node layer is adjusted according to the spatial support of the chosen wavelet, here $g=6$ for CDF 4/4 wavelets.
If neighboring blocks are not on the same tree level, we interpolate or decimate the corresponding nodes when synchronizing the ghost nodes. Redundant nodes on the boundary of adjacent blocks
always belong to the block with lower tree level, i.e. coarser lattice spacing.  The definition of the grid as well as the reasons for choosing this particular design are detailed in appendix B of \cite{EngelsSchneiderReissFarge2019}.

\paragraph{Forest Composed of Multiple Trees:}

For the administration of all blocks in the forest, we use a \textit{multi-tree structure}
illustrated in \cref{fig:treestructure}. Here each tree $\Tree_i$ holds a collection
of blocks $\Block_p$ and block values $ u^{(p)}$
\begin{equation}
  \label{eq-def:tree}
  \Tree_i = \{(\Block_p, u^{(p)}(\vec{x})) \mid \vec{x} \in \Block_p, p\in\Lambda_i^j, j=\Jmin,\dots,\Jmax\}\,,
\end{equation}
where a block is a leaf at
the end of a branch, which can be uniquely identified by a tree code $p\in \Lambda_i^j$
and a tree ID $i$. The tree ID identifies the corresponding grid $\Omega_i$ and the tree code
determines the topology, such as block position $\vec{x}_p$ and lattice
spacing $\Delta x_i$.
In the following, the collection of trees is called \textit{forest}
\begin{equation}
  \Forest = \{\Tree_i \mid i=1,\dots,\NTree\}\,.
  \label{eq-def:forest}
\end{equation}
In \cref{sec:wPODAlgo} we will use the forest to hold multiple spatial fields as time or parameter samples of the wPOD algorithm.

\paragraph{Light and Heavy Data Storage:}
To distinguish between administrative and physical information, we separate our data
structures into light and heavy data.

The \textit{light data} (\texttt{lgt\_n}, \texttt{lgt\_active}, \texttt{lgt\_block})
is the minimal information necessary to organize the topology of our grids.
From the light data we can determine neighbor relations between the blocks and keep
track of the processors holding the block.
It is therefore synchronized among all processing units.
\Cref{fig:lgtstructure} in \cref{appxC} illustrates an example of light data in the case of the tree structure
shown in \cref{fig:treestructure}.
All light data are stored in the \texttt{lgt\_block} array.
Each row holds the necessary information of one block, such as tree ID, grid refinement level, tree code and \textit{refinement status/coarsening indicator} (see appendix \cref{appx:block-based_wavelet_adaptation}).
The row index is called light-ID (\texttt{lgt\_id}).
It is ordered lexicographically in the process-ID:
$\texttt{lgt\_id}=(i_\mathrm{proc}-1) N_\mathrm{blocks} + j$, with $j=1,\dots,N_\mathrm{blocks}$. In this way, we relate
the position of the block with the process-ID $i_\mathrm{proc}$.
During the execution of the algorithm, blocks can be created or deleted. To avoid expensive memory allocation, we set the block inactive by marking the rows in \texttt{lgt\_block} with -1.
From \texttt{lgt\_block} we compute active block lists (\texttt{lgt\_active}) for each tree.
In this way, we are able to manipulate blocks on different grids efficiently, by only
looping over the active lists of a given tree specified by its \texttt{tree\_id}
(compare with \cref{fig:lgtstructure}).

Besides the light administrative structure, we have to store large data fields
with the physical information of the state vector quantity $ u(\vec{x})$ and all neighbor
relations of the blocks.
These data are called \textit{heavy data} and they are equally distributed among the processors
using the index of the space filling curve (Hilbert/Z-curve) \cite{Zumbusch2012}. The index can be easily calculated from the tree code.
In \cref{fig:treestructure} the \textit{Hilbert-curve} is visualized with the dashed line passing through
the lattice and the processors-IDs are encoded with the color of the block.

Using \textit{space filling curves} has two major advantages for our algorithm.
Firstly, due to locality properties of the space filling curve the communication
between adjacent blocks which do not share the same processors is kept at a minimum.
Secondly, the uniqueness of the curve ensures that trees with the same tree structure
(i.e. same grid $\Omega_i$)
have identical processor distribution. This is advantageous when performing pointwise operations between trees.

\subsection{Block-Based Wavelet Adaptation}
\label{sec:bbwadapt}

\begin{figure}[htp!]
  \centering
  \includegraphics[width=1\linewidth]{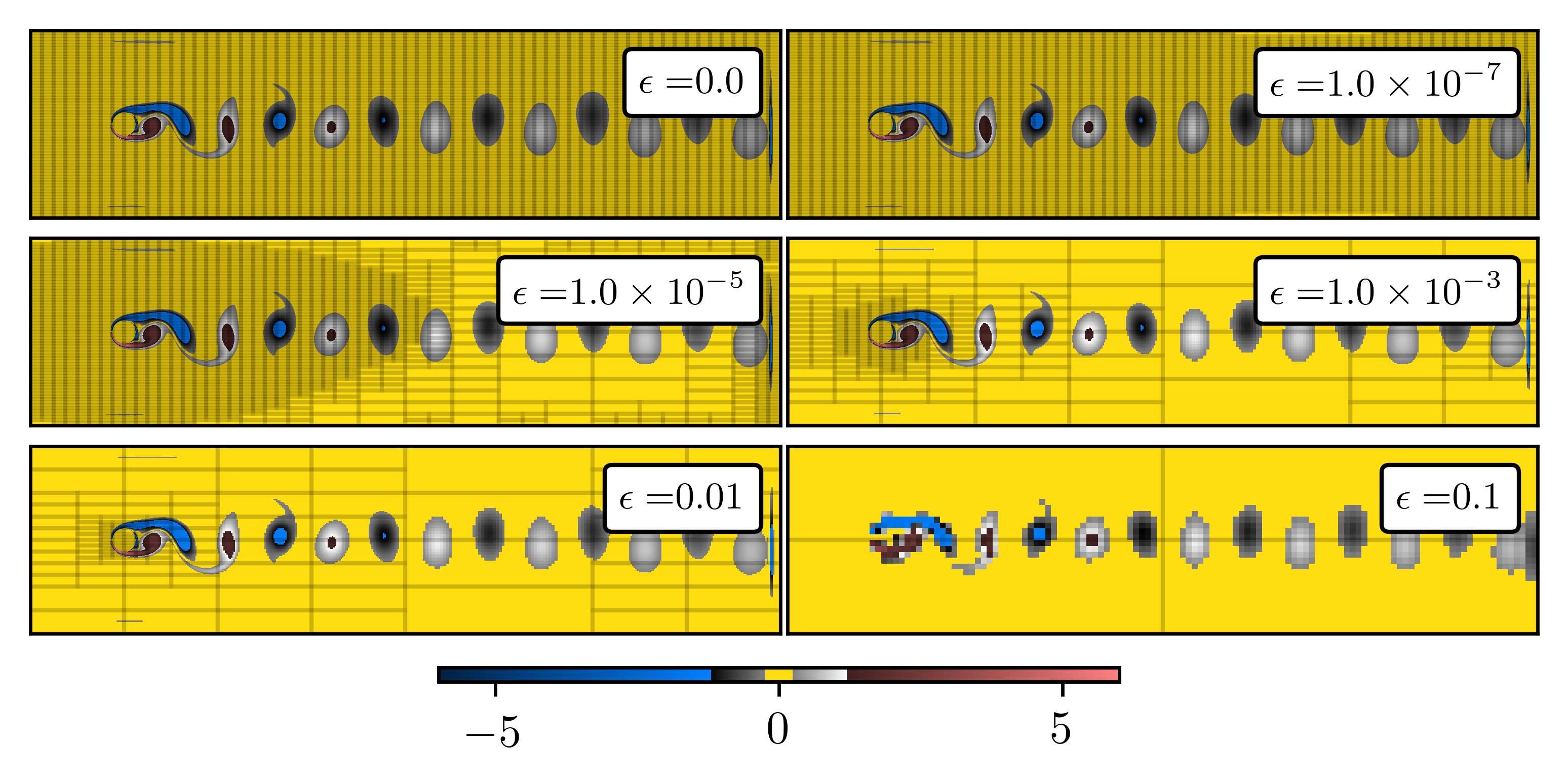}
  \caption{Block-based adaptation of a flow past a cylinder for different thresholds $\epsilon$. Shown is the vorticity field $\omega^\epsilon=\partial_x v_y^\epsilon - \partial_y v_x^\epsilon$, which  is computed after having applied the wavelet adaptation to the full state vector $\u^\epsilon=(v_x^\epsilon,v_y^\epsilon,p^\epsilon)$.
  Each block represents $\Bs_1\times\Bs_2=65\times17$ points.
  More details can be found in \cref{sec:vortex_street}.
  }
  \label{fig:threshold-vorx}
\end{figure}

Our adaptation algorithm is block-based, meaning that blocks
are coarsened or refined depending on the local regularity of the sampled function.
Blocks can be coarsened by leaving out every second point and merged with their four
or eight neighboring blocks in two and three space dimensions, respectively. Blocks can be refined via interpolation
at all dyadic points and
subdividing them into four (2D) or respectively eight (3D) new blocks.
The procedure of recursive dyadic refinement is known as interpolatory subdivision
and was first introduced by Deslauriers and Dubuc in \cite{DeslauriersDubuc1987,DeslauriersDubuc1989}.
It was later shown by Donoho in \cite{Donoho1992} that the resulting interpolation or scaling
function can be used for constructing a multiresolution analysis. Our approach is based on the discrete point-value multiresolution framework of Harten \cite{Harten1993,Harten1996,Harten1997}, which uses interpolating scaling functions of Deslauriers and Dubuc for discrete data representation.
However, to obtain a better scale and frequency separation between the different scales we use lifted wavelets. Hence a low pass filter is applied before downsampling the block. The filter coefficients of the coarsening and refinement procedure (see \cref{tabl:dd24}) define the underlying wavelet scheme.
The used lifted Deslauriers-Dubuc wavelets correspond to biorthogonal Cohen-Daubechies-Feauveau (CDF) wavelets \cite{Cohen1992}.
We refer the reader to \cref{appx:block-based_wavelet_adaptation} for a detailed explanation of the adaptation algorithm. In the following the
adaptation algorithm will be denoted by $u^\epsilon=$\texttt{adapt}($u,\epsilon,\Jmin,\Jmax)$.
Here, the input and output functions $u,u^\epsilon\in L^2(\Domain)$ are continuous and given in the block-based format outlined above. The adaptation algorithm is limited to a minimal and maximal refinement level $\Jmin,\Jmax$.
Depending on the wavelet threshold $\epsilon$ blocks are coarsened if the absolute difference between a block sample and its interpolated values from a coarser level is smaller than $\epsilon$. The difference is referred to as a \textit{detail coefficient} of the wavelet basis.
Therefore, the error between the input and output, i.e. the wavelet compression, can be bounded by the wavelet threshold:
\begin{equation}
  \label{eq:errWavelet}
  \frac{\norm{u-u^\epsilon}}{\norm{u}}\le \epsilon.
\end{equation}
Here, the normalization of the wavelet basis determines the norm used in \cref{eq:errWavelet}. For this study we use biorthogonal wavelets, which are normalized in $L^2$.
In the following we will denote all fields with an upper index $\epsilon$, where details
have been filtered with the adaptation algorithm and can be thus expressed in a wavelet series with filtered coefficients. \Cref{fig:threshold-vorx} shows
an example of such a field. Here one snapshot of a direct numerical simulation is block-decomposed and filtered for different threshold values $\epsilon$.
In fig. ~\ref{fig:threshold-vorx} the block boundaries are visualized by a rectangular box.
All blocks have the same resolution $\Bs_1\times\Bs_2=65\times17$. Smaller blocks correspond to higher tree levels, i.e. smaller lattice spacing. With increasing $\epsilon$
the number of blocks decreases, since less details are above the threshold.

%% file: lattice.tex
\colorlet{colP1}{gray!60}
\colorlet{colP2}{gray!20}
\colorlet{colG}{blue!40}        
\colorlet{colredundant}{red!50} 
\usetikzlibrary{shapes.misc}

\tikzset{cross/.style={cross out, draw=black,minimum size=4pt, inner sep=0pt, outer sep=0pt}}
\begin{tikzpicture}[scale=0.25,font=\footnotesize]

  \coordinate (origin) at (1,1);
  \coordinate (upperleft) at (1,17);
  \coordinate (upperright) at (17,17);
  \coordinate (lowerright) at (17,1);

	\coordinate (A) at (3.5,2.5);
	\coordinate (B) at (10.5,8.5);
	\fill[colP2,thick] (3.5,7.5) rectangle (10.5,14.5); 
	\fill[dashed,colP1,thick] (5,9) rectangle (9,13); 	

	\coordinate (P1a) at (5,7);
	\coordinate (P1b) at (9,4);
	\coordinate (P2a) at (P1b);
	\coordinate (P2b) at (17,1);
	\coordinate (P3a) at (P1b);
	\coordinate (P3b) at (17,9);
	\coordinate (P4a) at (3,3);
	\coordinate (P4b) at (17,11);

	\coordinate (P5a) at (3,3);
	\coordinate (P5b) at (1,1);
	\coordinate (P6a) at (P1a);
	\coordinate (P6b) at (1,11);
	\coordinate (P4a) at (P3b);
	\coordinate (P4b) at (15,11);

   \draw (origin) rectangle (upperright);
   \draw (1,9) rectangle (5,13);
   \draw (5,13) rectangle (9,17);
   \draw (origin) rectangle (9,9);
   \draw (9,1) rectangle (17,9);
   \draw (9,9) rectangle (17,17);


	\draw[->] (-0.5,0 ) -- (0.5,-0) node [pos=0.66,below] {$x_{1}$};
	\draw[->] (- 0.5, 0 ) -- (-0.5, 1) node [pos=0.66,left] {$x_{2}$};

	\foreach \x in {1,...,17}
	{
		\foreach \y in {1,...,17}
		{
                                \ifthenelse{\x=10\AND \y>7 \AND \y<15 \OR \y=8 \AND \x>3 \AND \x<11}
                                {
																	\ifthenelse{\x=11 \AND \y>10 \AND \intcalcMod{\y-1}{2}=0}
																	{

																	}{
																		\node[cross] at (\x,\y) {};
																	}
                                }{}
																\ifthenelse{\y=9 \AND \x>4 \AND \x<10}
																{
																			\draw (\x,\y) circle (6pt);
																}{}

																\ifthenelse{\x=5 \AND \y>9 \AND \y<14}
																{
																			\draw (\x,\y) circle (6pt);
																}{}

																\ifthenelse{\y=13 \AND \x>5 \AND \x<10
																 						 }
																{
																			\draw (\x,\y) circle (6pt);
																}{}

																\ifthenelse{ \x=9 \AND \y>9 \AND \y<13}
																{
																			\draw (\x,\y) circle (6pt);
																}{}

                                \ifthenelse{\x>4\AND \y>8 \AND \x<10 \AND \y<14}
                                {
                                        \fill (\x,\y)  circle[radius=3pt];
                                }{

																	\ifthenelse{\x<10 \AND \y>8}
                                	{
																		\ifthenelse{\x=11 \AND \y<15 \AND \intcalcMod{\y}{2}=0}
																		{
																		}{
                                        \draw (\x,\y) circle [radius =5pt] ;
																		}
                                	}{
                                		\ifthenelse{\intcalcMod{\x-1}{2}=0 \AND \intcalcMod{\y-1}{2}=0}
                                  	{
                                        \draw (\x,\y) circle [radius =5pt] ;
                                  	}
                                  	}{}
                                }
		}
	}

	\node [fill=white] at (5,5) {\large2};
	\node [fill=white] at (3,15.2) {00};
	\node [fill=white] at (7,15.2) {01};
	\node [fill=white] at (2.9,11) {02};
	\node [fill=colP1] at (7,11) {03};
	\node [fill=white] at (13,13) {\large1};
	\node [fill=white] at (13,5) {\large3};
			  \coordinate (C) at (upperright);
        \coordinate [right = of C, label={[label distance=0.25cm] right:interior node}] (PBlock);
        \fill (PBlock) circle (7pt);

        \coordinate [below = of PBlock, label={[label distance=0.25cm] right:redundant node}] (PRed);
        \draw (PRed) circle [radius =6pt] ;
				\fill (PRed) circle (3pt);

        \coordinate [below = of PRed, label={[label distance=0.25cm] right:exterior/ghost node}] (PGhost);
        \draw (PGhost) circle [radius =5pt] ;
        \coordinate [below = of PGhost, label={[label distance=0.25cm] right:interpolated ghost node}] (INTP);
        \node [cross] at (INTP) {};

       \draw [<->] (0,1) -- (0,17) node[left, pos=0.5] {$L_2$};
       \draw [<->] (1,18) -- (17,18) node[above, pos=0.5] {$L_1$};

\end{tikzpicture}

%% file: treestructure.tex

\colorlet{colp1}{orange!10!white}
\colorlet{colp2}{blue!10!white}
\colorlet{colp3}{green!10!white}
\colorlet{colp4}{red!10!white}

\begin{minipage}[t]{0.45\textwidth}
\centering
\begin{tikzpicture}[sibling distance=30pt,level 1/.style={sibling distance=55pt},
  level 2/.style={sibling distance=25pt},level 3/.style={sibling distance=35pt},scale=0.8]
  \tikzstyle{every node}=[draw, rectangle, minimum width=20pt, minimum height = 20pt]
  \node {$\Tree_1$}
  child {node [fill=colp1]{0}}
  child {node {1}
    child {node [fill=colp4]{10}}
    child {node [fill=colp4]{11}}
    child {node [fill=colp4]{12}}
    child {node [fill=colp4]{13}}
  }
  child {node [fill=colp1]{2}}
  child {node {3}
    child {node [fill=colp1]{30}}
    child {node {31}
          child {node [fill=colp3]{310}}
          child {node [fill=colp3]{311}}
          child {node [fill=colp3]{312}}
          child {node [fill=colp2]{313}}
          }
    child {node [fill=colp2]{32}}
    child {node [fill=colp2]{33}}
  };
\end{tikzpicture}
\end{minipage}%
\begin{minipage}[t]{0.45\textwidth}
\centering
  \begin{tikzpicture}[scale=0.7]
  \tikzstyle{every node}=[draw, node distance = 1.2cm,rectangle, minimum width=20pt, minimum height = 20pt]

  \node {$\Tree_2$}
  child {node [fill=colp1]{0}}
  child {node [fill=colp4]{1}}
  child {node {2}
    child {node [fill=colp1]{20}}
    child {node [fill=colp3]{21}}
    child {node [fill=colp2]{22}}
    child {node (D)[fill=colp2]{23}
    }
  }
  child {node [fill=colp3]{3}};
  \node [draw=none,below of=D] {};
\end{tikzpicture}
\end{minipage}\\

\vspace{0.1cm}

\begin{minipage}[t]{0.45\textwidth}
\centering
    Grid $\Omega_1$\\
  \begin{tikzpicture}[scale=0.5]
   \draw [fill=colp1](0,0) rectangle (8,8);
   \draw [fill=colp3](6,0) rectangle (8,4);
   \draw [fill=colp2](4,0) rectangle (8,2);
   \draw [fill=colp2](7,2) rectangle (8,3);
   \draw [fill=colp4](4,4) rectangle (8,8);
   \draw (0,0) grid [step=4] (8,8);
  \draw (4,0) grid [step=2] (8,4);
  \draw (4,4) grid [step=2] (8,8);
  \draw (6,2) grid [step=1] (8,4);
  \draw[->] (-0.5,-0.50) --(2,-0.50) node[below] {$x$};
  \draw[->] (-0.5,-0.50) --(-0.50,2) node[left] {$y$};
  \draw[dashed] (1,7) -- (3,7) -- (3,5) -- (1,5)--(1,3) --(1,1) -- (3,1) --(3,3);
  \draw[dashed] (3,3) -- (5,3) -- (5,1) -- (7.5,1)--(7.5,2.5)--(6.5,2.5)--(6.5,3.5)--(7.5,3.5);
  \draw[dashed] (7.5,3.5)--(7.5,5)--(5,5)--(5,7)--(7,7);
  \node at (2,2) {2};
  \node at (7,1) {33};
  \node at (5,1) {32};
  \node at (5,3) {30};
  \node at (2,6) {0};
  \node at (6.5,3.5) {\tiny310};
  \node at (7.5,3.5) {\tiny311};
  \node at (7.5,2.5) {\tiny313};
  \node at (6.5,2.5) {\tiny312};
  \node at (5,7) {10};
  \node at (7,7) {11};
  \node at (5,5) {12};
  \node at (7,5) {13};
\end{tikzpicture}
\end{minipage}%
\begin{minipage}[t]{0.45\textwidth}
\centering
Grid $\Omega_2$\\
  \begin{tikzpicture}[scale=0.5]
   \draw [fill=colp1](0,0) rectangle (8,8);
   \draw [fill=colp1](2,2) rectangle (4,4);
   \draw [fill=colp2](0,0) rectangle (4,2);
   \draw [fill=colp3](2,2) rectangle (4,4);
   \draw [fill=colp3](4,0) rectangle (8,4);
   \draw [fill=colp4](4,4) rectangle (8,8);
  \draw (0,0) grid [step=4] (8,8);
  \draw (0,0) grid [step=2] (4,4);
  \draw[->] (-0.5,-0.50) --(2,-0.50) node[below] {$x$};
  \draw[->] (-0.5,-0.50) --(-0.50,2) node[left] {$y$};
  \draw[dashed] (1,7) -- (3,7) -- (3,5) -- (1,5)--(1,3) --(1,1) -- (3,1) --(3,3);
  \draw[dashed] (3,3) -- (5,3) -- (5,1) -- (7,1)--(7,2)--(7,5);
  \draw[dashed] (7,5)--(5,5)--(5,7)--(7,7);
\node at (1,1) {22};
\node at (3,1) {23};
\node at (3,3) {21};
\node at (1,3) {20};
\node at (2,6) {0};
\node at (6,2) {3};
\node at (6,6) {1};

\end{tikzpicture}
\end{minipage}

%% file: algorithm.tex
\renewcommand{\ucoef}{\underline{\boldsymbol{\mathsf{u}}}}
\section{Algorithm}
In the following, we introduce the proper orthogonal decomposition (POD) as the basis of our algorithm in \cref{sec:snapshot_POD}. This section aims at providing a broad overview on the state of the art techniques to compute PODs for large data sets in fluid dynamics, namely the snapshot POD and the randomized singular value decomposition (rSVD). For a more detailed explanation  we refer the reader to \cite{Volkwein2001} for the POD and \cite{HalkoMartinssonTropp2011}
for the rSVD.
In \cref{sec:wPODAlgo} we generalize the snapshot POD using wavelet adapted snapshot fields.
We also propose a strategy for balancing wavelet thresholding and POD truncation error in \cref{sec:ErrorAnalysis}.

\subsection{The Snapshot POD and randomized SVD}
\label{sec:snapshot_POD}
Given are samples of a continuous vector-valued $L^2$-function $\u(x,\mu)\in \mathbb{R}^K,\, K>0$ sampled in some parameter interval $\{\mu_i\}_{i=1}^{\NSnapshots}$. The samples are called snapshots and are in the following indexed by $\u_i=\u(x,\mu_i)$.
Our algorithm solves the following POD optimization problem

\begin{equation}
  \label{eq:minPOD}
  \min_{\{\Mode_k\}}\sum_{i=1}^{\NSnapshots}\norm{\u_i -  \sum_{k=1}^r\sprod{\u_i}{ \vec{\Mode}_k}\Mode_k}^2,\quad \text{such that} \quad \sprod{\Mode_k}{ \vec{\Mode}_l}=\delta_{kl}\,,
\end{equation}
with the $L^2$ inner product $\sprod{\cdot}{\cdot}$ and associated norm $\norm{\cdot}=\sqrt{\sprod{\cdot}{\cdot}}$.
We solve it with the \textit{method of snapshots or strobes} \cite{SiL1987} since for data where the spatial resolution is much larger then the number of snapshots $\NSnapshots$, \cref{eq:minPOD} is reduced to a {small eigenvalue problem of size} $\NSnapshots\times\NSnapshots\sim\ord{100}$
\begin{equation}
	\label{eq:smallEVproblem}
	\mathbf{C} \mathbf{v}_k = \lambda_k \mathbf{v}_k \quad\text{for}\quad k = 1,\dots,r\,,
\end{equation}
for the correlation matrix
\begin{equation}
	\label{eq:eucliedean}
	\mathbf{C}_{ij}=\frac{1}{\NSnapshots V}
          \sprod{\u_i}{\u_j}\,,
\end{equation}
together with the relation:
\begin{equation}
	\label{eq:modes}
	\Mode_k = \frac{1}{\sqrt{\lambda_k \NSnapshots }}\sum_{i=1}^{\NSnapshots} (\vec{v}_k)_{i} \u_i\quad k = 1,\dots,r\,.
\end{equation}

The method of snapshots is strongly connected to the \textit{singular value decomposition} (SVD)~\cite{Volkwein2001}, because left singular vectors correspond to the solution of \cref{eq:minPOD} and right singular vectors to $\vec{v}_k$, when assuming Euclidean space \cite{Volkwein2001}. Respectively, the eigenvalues $\lambda_k=\sigma^2_k\ge0$ are squares of the singular values. Furthermore, it is known from the Eckart-Young-Mirsky theorem \cite{EcY1936} that the resulting approximation error in the Frobenius norm, when truncating after the $r$-th mode, is given by the sum $\sum_{k=r+1}^{\NSnapshots}\sigma_k^2$ of the remaining singular values.

However, caution must be taken when using this method instead of the SVD, because the condition number $\kappa(\Smat)\defeq\sigma_{\max}(\Smat)/\sigma_{\min}(\Smat)$ of the associated snapshot matrix $\Smat$ is squared: $\kappa(\Smat^T \Smat)=\kappa(\Smat)^2$. This can lead to inaccuracy of POD modes with small singular values (see the famous example of L\"achli \cite{Lauchli1961}). Nevertheless, one is often willing to accept this potential error in favor of a smaller problem size.

Another way of reducing the problem size, without squaring the condition number, is using a \textit{randomized SVD} algorithm,
which is outlined in \cite{HalkoMartinssonTropp2011}.
Here, an orthogonal matrix $\mathbf{Q}\in\mathbb{R}^{M\times q}$, $q\ll\NSnapshots$
is formed which approximates the column space of $\Smat\approx \mathbf{QB}$.
With its help, a singular value decomposition of the small $q\times\NSnapshots$ matrix $\mathbf{B}=\mathbf{Q}^T\Smat$ is computed and the left singular vectors $\Mode_k^\mathbf{B}$ are projected back onto the full space via: $\Mode_k = \mathbf{Q}\Mode_k^\mathbf{B}$.
Usually the orthogonal matrix is formed by a QR decomposition taking $q$ random samples of the column space of $\Smat$. For a target number of $r$ modes one usually oversamples  $q=r+n$ by taking $n=5$ or $n=10$ additional random samples \cite{HalkoMartinssonTropp2011}. However, if the singular values decay slowly, $\mathbf{Q}$ may not represent $\mathbf{U}$ well enough and costly tricks, like \textit{Power Iterations} using additional passes over the data, have to be applied \cite{HalkoMartinssonTropp2011}.
Moreover, as pointed out in \cite{HalkoMartinssonTropp2011} for very large matrices $\Smat$ the data cannot be loaded into fast memory and therefore the transfer from the slow memory typically dominates the arithmetic.
In contrast, the wPOD algorithm presented in the next section seeks to avoid these problems by reducing the relevant information of each single snapshot to fit it into the fast memory.

\subsection{The Weighted Inner Product and Pointwise Tree Operations}
\label{sec:weightedInnerProduct}
\begin{figure}[t]
  \centering
  \begin{subfigure}[c]{0.35\textwidth}
    \centering
	  \includegraphics[width=0.45\linewidth]{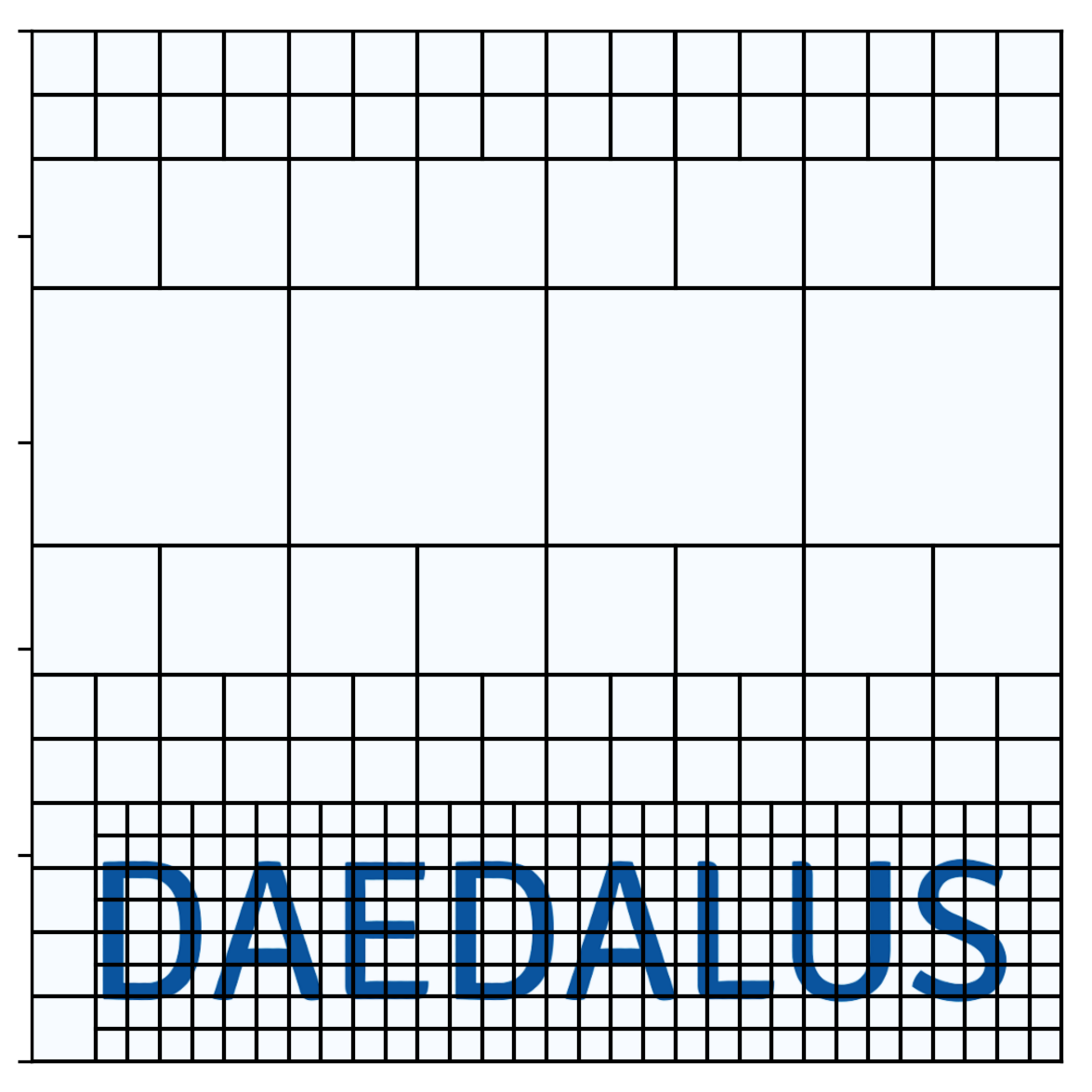}%
          \includegraphics[width=0.45\linewidth]{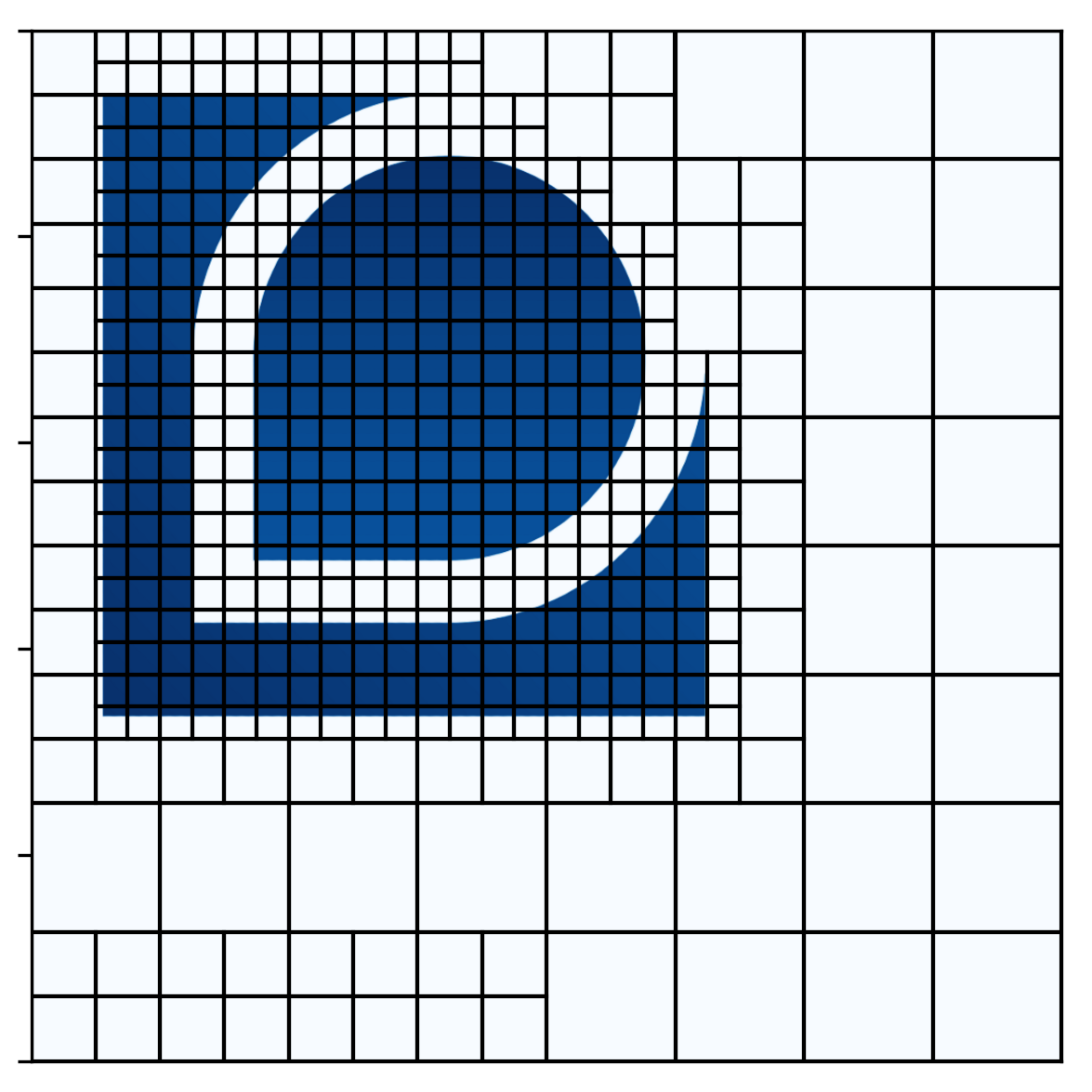}\\
	  \includegraphics[width=0.45\linewidth]{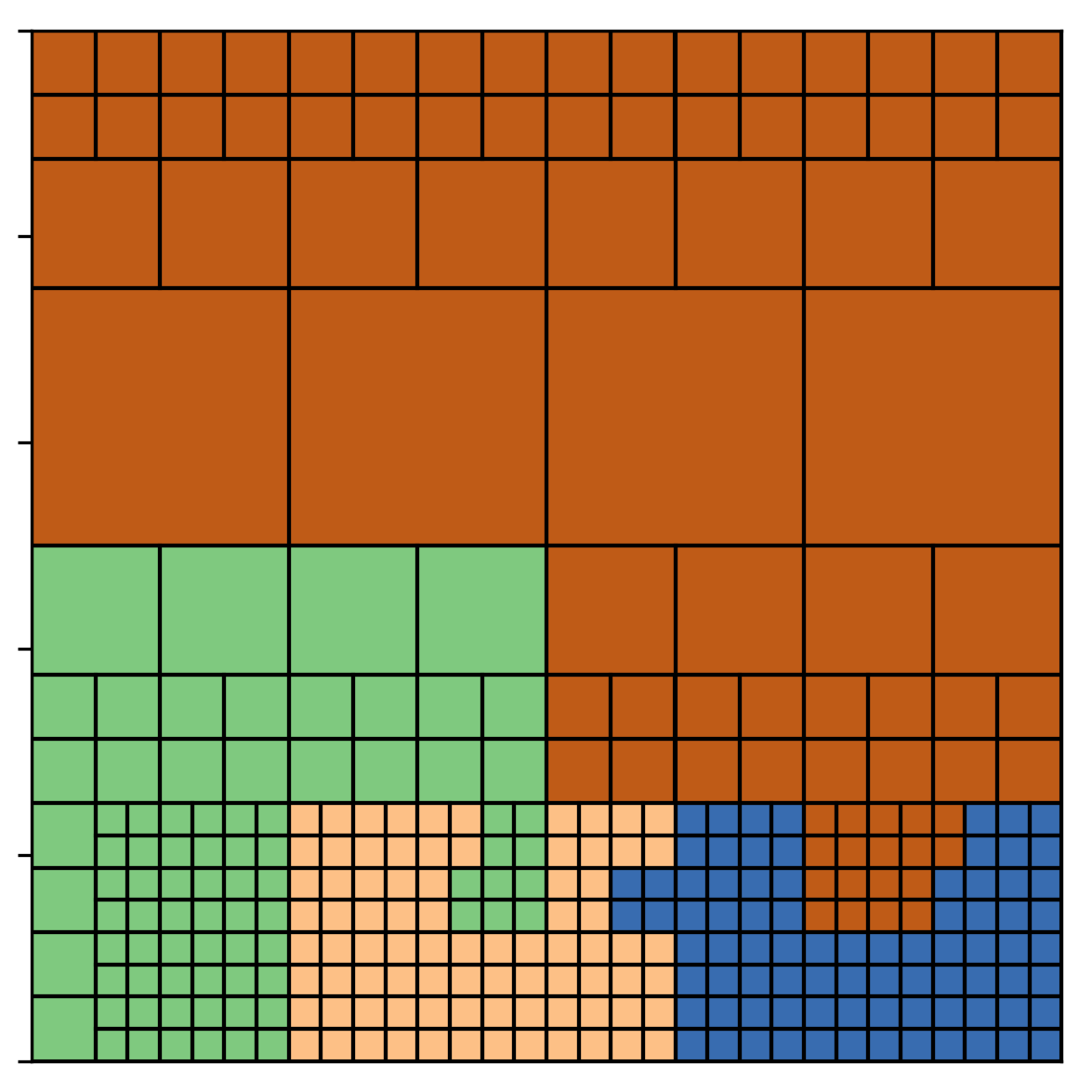}%
          \includegraphics[width=0.45\linewidth]{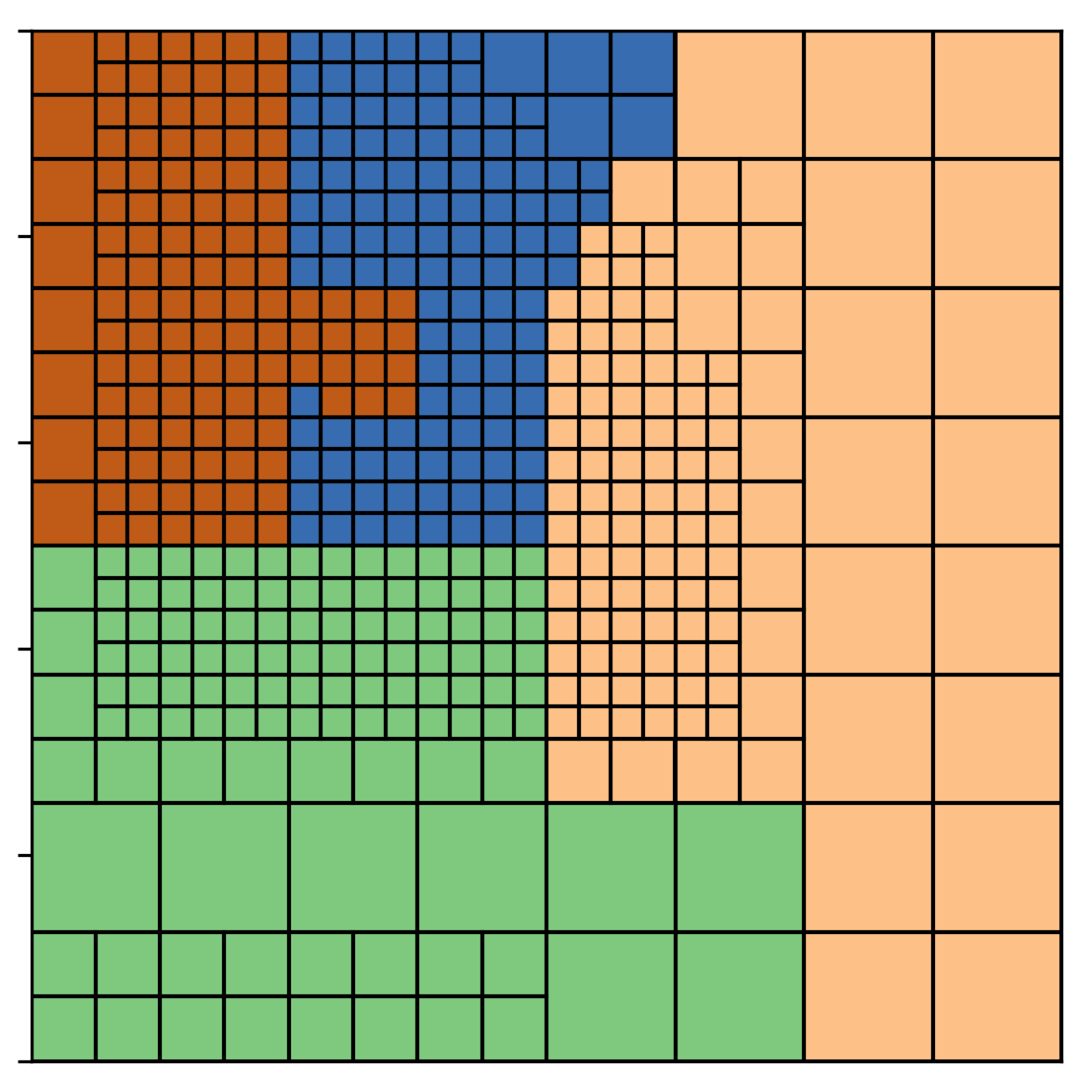}
          \caption{$u_i(\Omega_i)$ and $u_j(\Omega_j)$}
          \label{subfig:rawfields}
\end{subfigure}%
    \begin{subfigure}[c]{0.35\textwidth}
    \centering
	  \includegraphics[width=0.45\linewidth]{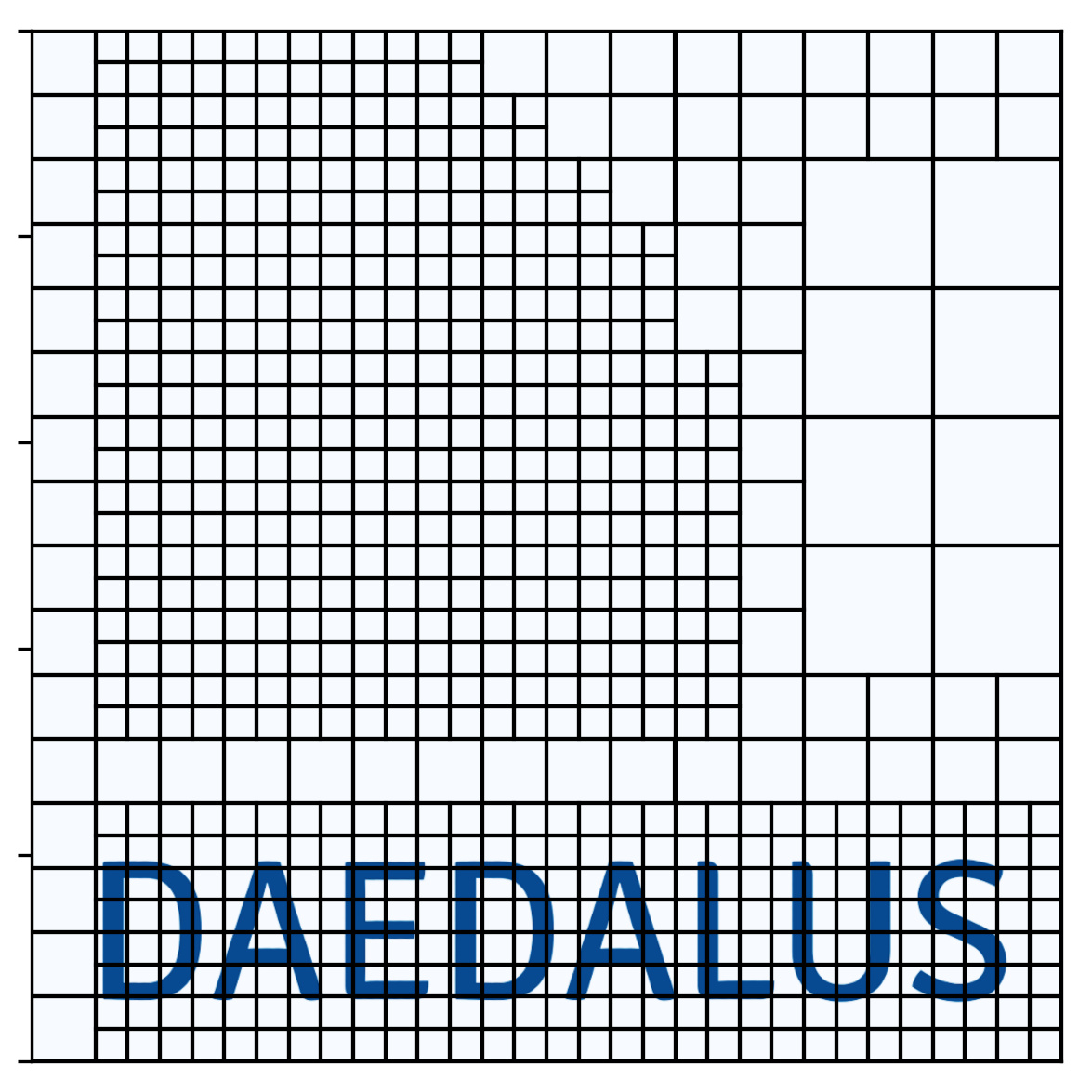}%
	  \includegraphics[width=0.45\linewidth]{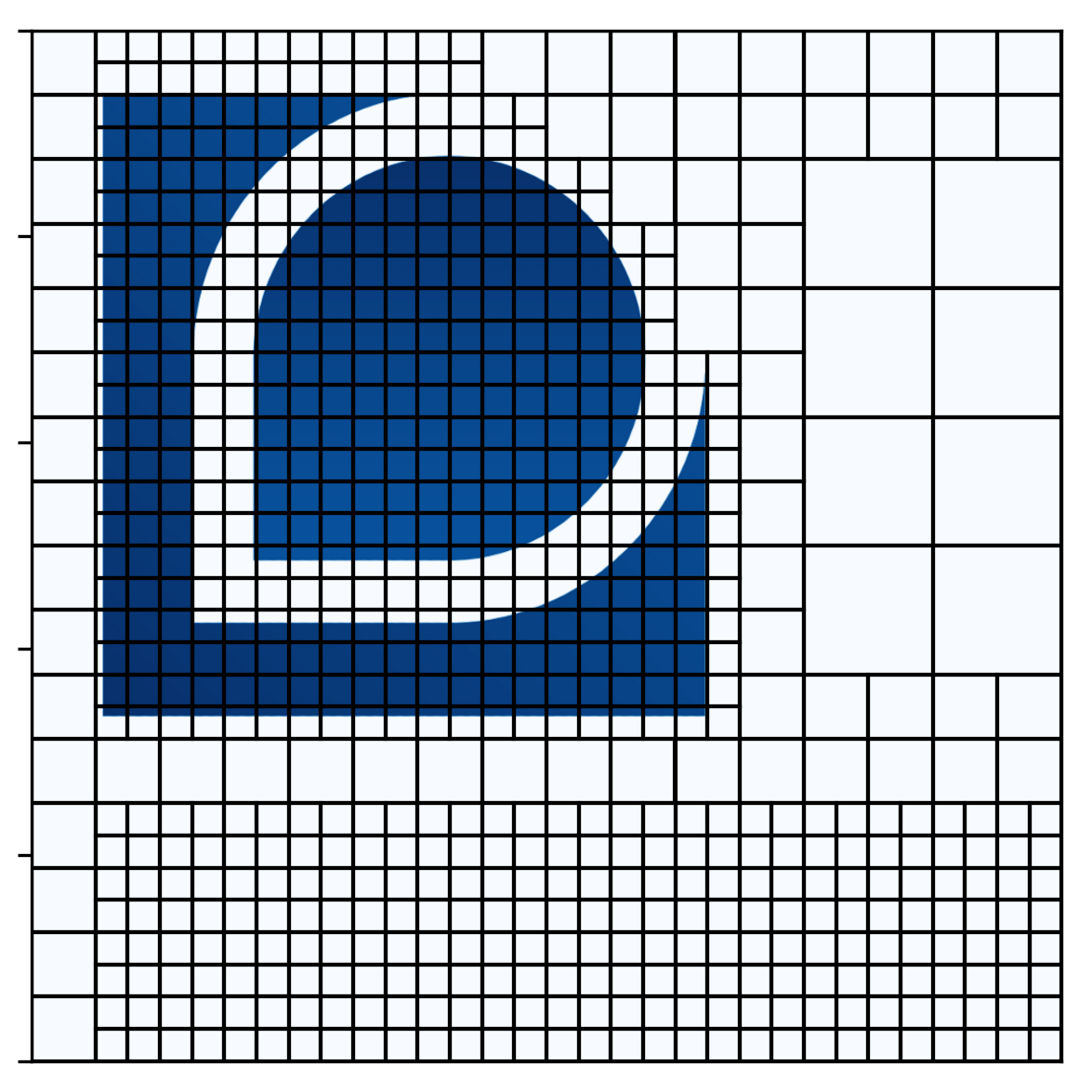}\\
	  \includegraphics[width=0.45\linewidth]{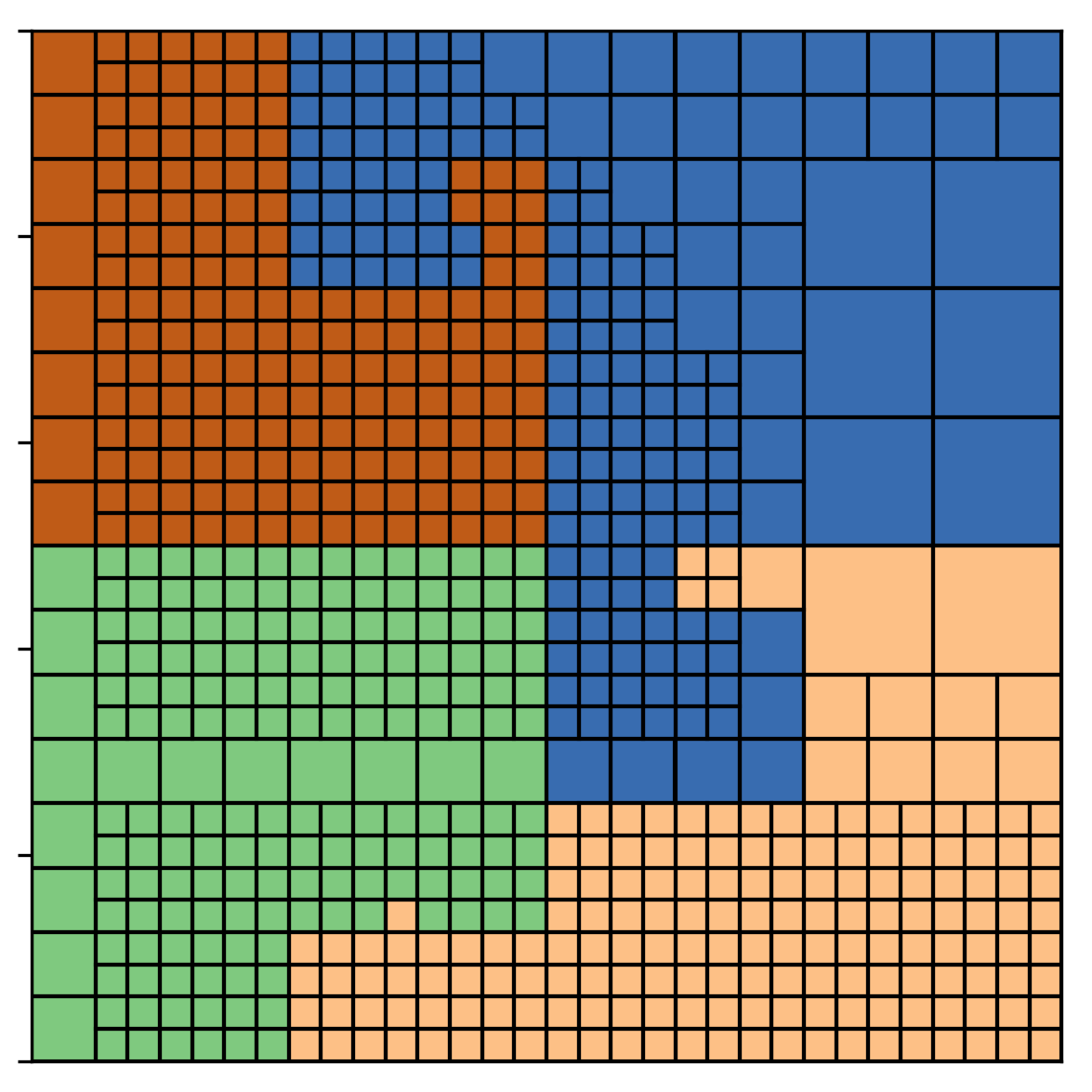}%
	  \includegraphics[width=0.45\linewidth]{dsymbol_refineprocs.png}
          \caption{$u_i(\Omega_{ij})$ and $u_j(\Omega_{ij})$}
          \label{subfig:refinedfields}
\end{subfigure}%
    \begin{subfigure}[c]{0.25\textwidth}
    \centering
	 \includegraphics[width=0.63\textwidth]{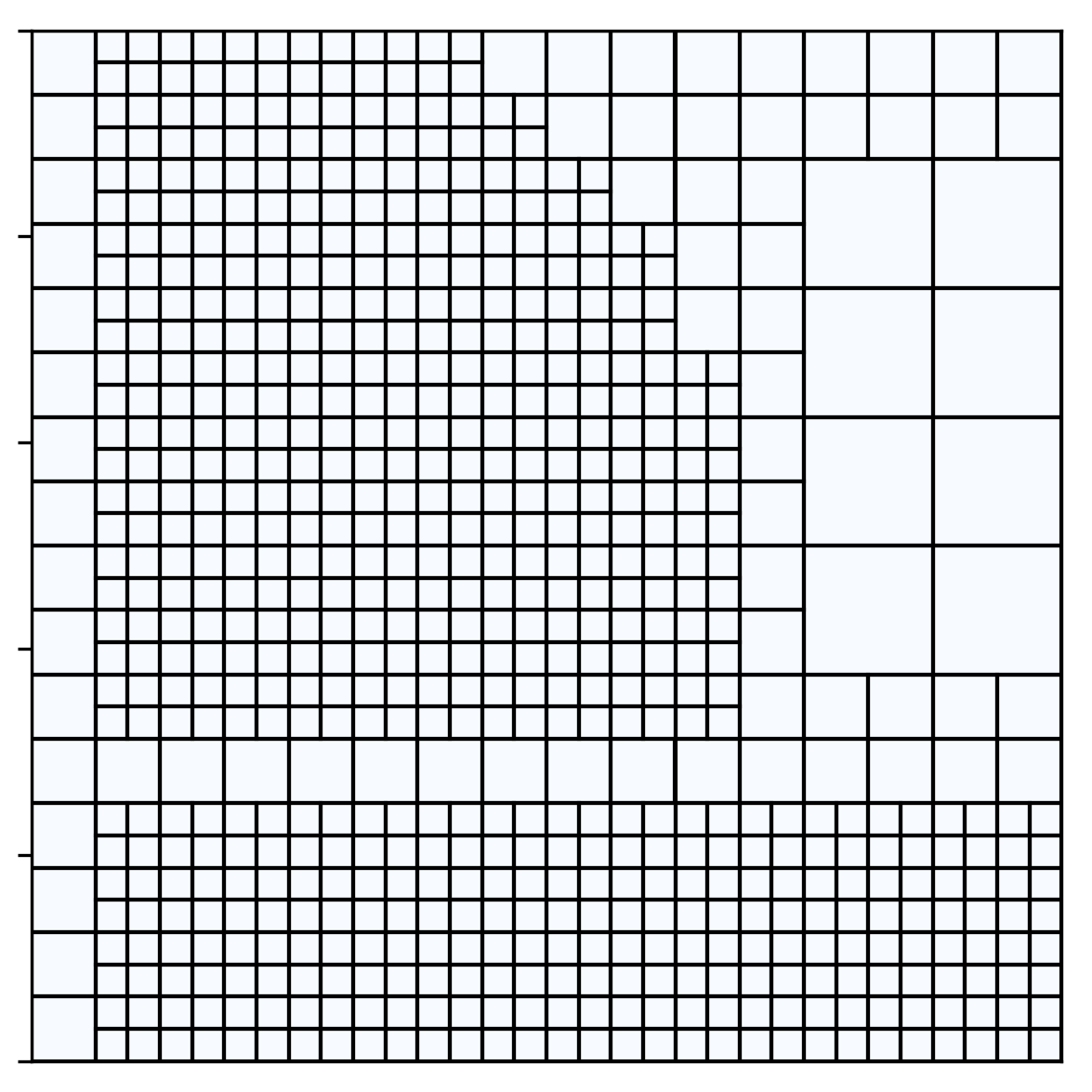}\\
	 \includegraphics[width=0.63\textwidth]{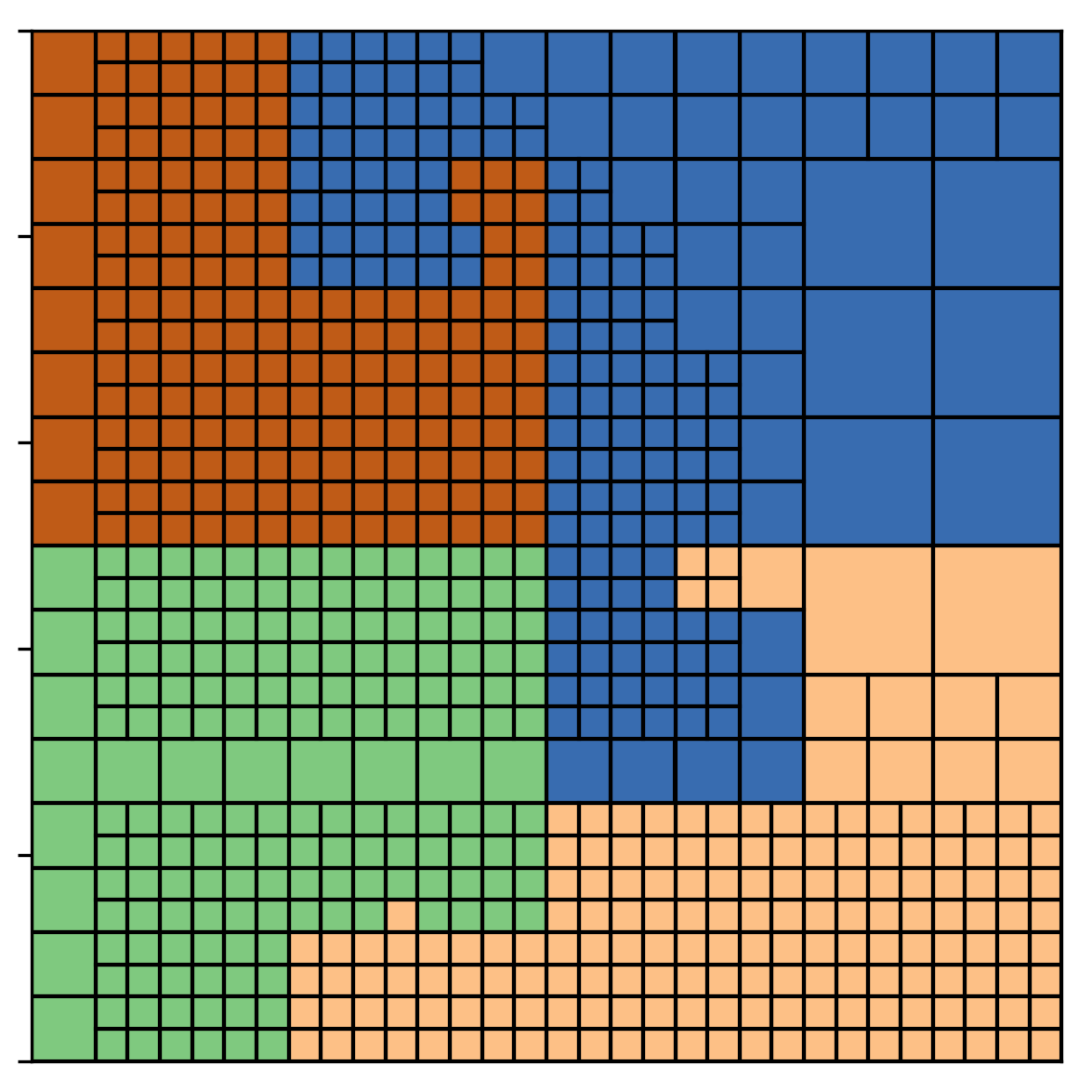}
         \caption{$(u_i \times u_j)(\Omega_{ij})$}
          \label{subfig:operatedfields}
\end{subfigure}%
  \caption{Visualization of pointwise operations on multiresolution grids
  $\Grid_i$ and $\Grid_j$. On the top row the scalar fields $u_i$ and $u_j$ are shown, where blue
  colors represent 1 and white colors 0. Below each field
the corresponding processor distribution grid is displayed. Each color of the processing grid represents one of
the four processing units.  Initial trees (\Cref{subfig:rawfields}), unified tree structures (\Cref{subfig:refinedfields})
and the processed field (\cref{subfig:operatedfields}).}
  \label{fig:pointoperation}
\end{figure}

In the wPOD algorithm, we follow the same approach as in the method of snapshots.
explained in the previous section (\cref{sec:snapshot_POD}). However, we use the sparsity of our wavelet block-based data representation to allow for an memory
efficient computation of the POD basis.
In contrast to the representation in terms of the snapshot matrix, as used by the SVD, our data is represented in terms of a forest $\Forest$  or a collection of trees (see \cref{sec:GridAndImplementation}). Here each snapshot $\u_i(\vec{x})\defeq\u(\vec{x},\mu_i)$ is associated with a tree $\Tree_i$ on a hierarchical structured multiresolution grid $\Grid_i$, $i=1,\dots,\NSnapshots$.
The leaves of the tree correspond to blocks, where each block $p$ stores coefficients $\{u^j_p[k_1,k_2]\}_{k_1,k_2}$ of the underlying basis $\{\phi^j_{k_1,k_2}\}_{k_1,k_2}$ at tree level $j$.
The interpolating basis allows to represent the data in a continuous form, when summing over all blocks $p\in \Lambda_i^j$ of each tree level j:
\begin{equation}
	\label{eq:sum_blocks}
	\u_i(\vec{x}) = \sum_{j=1}^{\Jmax}\sum_{p\in \Lambda_i^j} \u^{(p)}_i(\vec{x}) \quad\text{for}\quad i=1,\dots,\NSnapshots
\end{equation}
 We can thus introduce the \textit{snapshot set}:
\begin{align}
  \mathcal{U} &=\{\u( \vec{x}, \mu_1),\dots, \u( \vec{x}, \mu_{\NSnapshots})\mid \vec{x} \in \Domain\}\,,
\end{align}
as the continuous counterpart of the snapshot matrix $\mathbf{U}$.
As mentioned earlier, our algorithm is capable of handling 2D and 3D data fields on
a rectangular domain
$\Domain \subset \mathbb{R}^d$, $d\in\{2,3\}$.
Note that with the new data representation in terms of functions in the $L^2$ Hilbert space, the inner product in the POD formulation has changed to
\begin{equation}
	\label{eq-def:scalar-product}
	\sprod{\u_i}{\u_j} \defeq\; \int_{\Domain}  \u^T_i(\vec{x}) \u_j(\vec{x})\, \d{\vec{x}}\,.
\end{equation}
However, for inner products or any pointwise operation (+,-) between snapshots $\u_i$,$\u_j$, represented on locally different grids $\Omega_i, \Omega_j$,  it is required that both coefficient vectors $\ucoef_i,\ucoef_j$ are of same length, i.e. expressed in the same basis.
In contrast to the discussed FEM methods \cite{UllmannRotkvicLang2016,GrassleHinzeLangUllmann2019}, this can be achieved very efficiently, because the
hierarchical grid definition allows to merge two grids by the union
$\Omega_{ij} = \Omega_i \cup \Omega_j$ for the two snapshots involved.
\Cref{fig:pointoperation} visualizes the grid merging procedure.
In \cref{subfig:rawfields}, the initial grids $\Grid_i$ and $\Grid_j$ are displayed together with their processor distribution.
In this example both trees have a fundamentally different tree structure and processor
distribution. In areas where $\Grid_i$ has large details, $\Grid_j$ does not and vice versa.
The aim of the union of both grids is to merge them, such that no detail of both
trees gets lost. This implies that merging both trees only involves refinement operations, which are cheap when using wavelet up-sampling.
As explained in \cref{sec:GridAndImplementation} trees with
identical tree structure have identical processor distribution because of our load
balancing strategy by space filling curves. The \texttt{hvy-data}, i.e. grid quantities of $\Grid_i$ and $\Grid_j$
are therefore on the same processor (see \cref{subfig:refinedfields}) after unification.
The unification enables us to calculate the inner product in \cref{eq-def:scalar-product} as a weighted inner product
\begin{equation}
	\label{eq:weighted-inner-product}
		\sprod{\u_i}{\u_j}=\sprod{\ucoef_i}{\ucoef_j}_{\boldsymbol{M}}=\ucoef_i^T\boldsymbol{M}\ucoef_j,
\end{equation}
where $\ucoef_i$ represent the vectorized entries of tree $i$ and $\boldsymbol{M}$ is a positive definite and symmetric matrix (explicitly given in \cref{sec:weightedInnerProduct-appendix}). In FEM literature this
matrix is often called mass matrix and has been used by \cite{UllmannRotkvicLang2016,GrassleHinzeLangUllmann2019} in a similar approach.
With \cref{eq:weighted-inner-product} we are able to calculate the inner product
and the associated norm of our block-based multiresolution fields. Replacing the $L^2$ inner products by weighted inner products therefore generalizes the POD minimization problem in \cref{eq:minPOD} to multiresolution fields.
With this intermediate result we are able to go one step further to combine wavelet
adaptation and POD truncation.

\subsection{The wPOD Algorithm}
\label{sec:wPODAlgo}
The wPOD algorithm proceeds in the following steps:
\begin{enumerate}
    \item \textbf{Read and Coarsen Data} $\u_i^\epsilon \leftarrow \texttt{adapt}(\u_i,\epsilon,\Jmin,\Jmax)$\\
In the first step of the algorithm we read all block decomposed snapshots in $\mathcal{U}$ and coarsen them for a given threshold $\epsilon$ using the wavelet adaptation scheme, if the input fields are not already adapted.
The adapted fields are denoted by $\u^\epsilon$. This part of the algorithm is essential
because it allows to keep only the most relevant information (see \cref{sec:bbwadapt}) of the input data using wavelet compression and therefore
makes handling of large data feasible (see \cref{sec:numerical-results}).

\item \textbf{Computation of Correlation Matrix} $(\mathbf{C})^\epsilon_{ij}=\frac{1}{\NSnapshots V} \sprod{\u_i^\epsilon }{\u_j^\epsilon}$\\
The main computational effort of the algorithm is the construction of all elements of
 the correlation matrix $\mathbf{C}^\epsilon\in\mathbb{R}^{{\NSnapshots}\times{\NSnapshots}}$, for which the inner product of locally different resolved snapshots needs to be computed. For any pairwise operation ($+,-,\sprod{\cdot}{\cdot}$), we refine to a union of both grids as shown in \cref{fig:pointoperation}. After the operations (+,-) the resulting field is adapted again to the predefined threshold $\epsilon$.

\item\textbf{Solving the Eigenvalue Problem} $\mathbf{C}^\epsilon \vec{v}^\epsilon_k = \lambda_k^\epsilon \vec{v}^\epsilon_k$\\
After the correlation matrix $\mathbf{C}^\epsilon$ is constructed, we diagonalize it with Jacobi's method
for real symmetric matrices \texttt{DSYEV} implemented in \texttt{LAPACK}
\cite{Anderson1990}. As described in \cite{Anderson1990}
sec.~8, the chosen method computes all the eigenvalues and eigenvectors
close to machine precision. We therefore neglect errors made during the diagonalization. In contrast to the construction
of $\mathbf{C}^\epsilon$, the computational effort needed for diagonalization is relatively small.

\item \textbf{Construction of POD Modes} $\Mode^\epsilon_k = \frac{1}{\sqrt{\lambda_j^\epsilon \NSnapshots }}\sum_{i=1}^{\NSnapshots} (\vec{v}^\epsilon_k)_{i} \u_i^\epsilon$\\
The elements of the correlation matrix are the inner products between two snapshots, i.e. the $i$-th row/column contains the coefficients of $\u_i$ represented
by a linear combination of all snapshots in $\mathcal{U}$.
Diagonalizing $\mathbf{C}^\epsilon$ means finding a basis of coefficient vectors $\vec{v}^\epsilon_k\in\mathbb{R}^{\NSnapshots}$ which
generate an optimal representation of $\mathcal{U}$. The representation in terms of orthonormal modes $\{\vec{\Mode}^\epsilon_k\}$ is
computed according to \cref{eq:modes}. The summation in \cref{eq:modes} proceeds in multiple steps. In the first step we copy $\Mode^\epsilon_k\leftarrow (\vec{v}_k^\epsilon)_{1} \u_1^\epsilon$, after which we iteratively sum up $\Mode^\epsilon_k\leftarrow \Mode^\epsilon_k + (\vec{v}^\epsilon_k)_{i} \u_i^\epsilon$ for $i=2,\dots,\NSnapshots$ in the second step and divide by the normalization factor.    

\item \textbf{Computation of POD Modes Coefficients}
$a_{ki}^\epsilon = \frac{1}{V}\sprod{\Mode^\epsilon_k}{\u_i^\epsilon}$\\
The computation of the modes coefficients involves again a computation of the inner product between the orthonormal modes $\Mode^\epsilon_k$ and the snapshots $\u_i^\epsilon$. In most cases, this step needs less evaluations of the scalar product since the number of modes $r$ should be small, $r\ll\NSnapshots$.

\end{enumerate}

In summary, our algorithm generates sparse modes $\Mode^\epsilon_k$, $k=1,\dots,r$, with amplitudes $a^\epsilon_{ki}$ to approximate any of the snapshots $\u_i\in\mathcal{U}$ in terms of a linear subspace
\begin{equation}
	\label{eq-def:wPODsubspace}
  \u_i(\vec{x})\approx\tilde{\u}_i^\epsilon(\vec{x})= \sum_{k=1}^r a^\epsilon_{ki}\Mode^\epsilon_k(\vec{x})\quad\text{for}\quad i = 1,\dots,\NSnapshots\,.
\end{equation}
In this notation, the upper index $\epsilon$ denotes the quantities, which are indirectly affected by the wavelet threshold (e.g. $a_{ki}^\epsilon,C_{ij}^\epsilon$) or directly expressed as a truncated wavelet series (e.g. $\tilde{\u}_i^\epsilon(\vec{x}),\Mode_k^\epsilon(\vec{x})$). Furthermore, $\tilde{\u}$ denotes the truncation after the $r$-th mode.
  

%% file: errors.tex
\subsection{Error Estimation}
\label{sec:ErrorAnalysis}
Here, we discuss the dependency of the approximation error on the wavelet threshold $\epsilon$ and the truncation rank $r$. 
Additionally, we explain how to choose both values in order to obtain a given accuracy.

Therefore,  we provide an error estimate of the approximation $\tilde{\u}_i^\epsilon$ in
\cref{eq-def:wPODsubspace}. The approximation projects our data $\mathcal{U}=\{\u_1,\dots,\u_{\NSnapshots}\}$ onto a linear subspace spanned by a set of $r$ orthonormal modes $\{\Mode_k^\epsilon\}_{k=1,\dots,r}$.
The sparsity of the modes is determined by the wavelet threshold $\epsilon$ (see \cref{sec:bbwadapt}) and the dimension of the subspace $r$ shall be much smaller than the number of snapshots: $r\ll\NSnapshots$.
For given $r,\epsilon$ we define the relative error of our approximation in the $L^2$-norm,
	\begin{align}
	  \label{eq-def:errWPOD}
	  \errWPOD(r,\epsilon)\defeq \frac{\sum_{i=1}^{\NSnapshots}\norm{\u_i(\vec{x}) - \tilde\u_i^\epsilon(\vec{x})}^2}
	  { \sum_{i=1}^{\NSnapshots}\norm{\u_i(\vec{x})}^2}
	\end{align}
which can be split into two contributions, i.e. compression and truncation errors. We thus have,
\begin{align}
   \sum_{i=1}^{\NSnapshots}\norm{\u_i(\vec{x}) - \tilde\u_i^\epsilon(\vec{x})}^2 &=
  \sum_{i=1}^{\NSnapshots}\norm{\u_i(\vec{x}) - \u_i^\epsilon(\vec{x}) + \u_i^\epsilon(\vec{x}) - \tilde\u_i^\epsilon(\vec{x})}^2 \\
        & \le \, \underbrace{\sum_{i=1}^{\NSnapshots}\norm{\u_i(\vec{x}) - \u_i^\epsilon(\vec{x})}^2}_{\text{compression error}}
        +\underbrace{\sum_{i=1}^{\NSnapshots}\norm{\u_i^\epsilon(\vec{x}) - \tilde\u_i^\epsilon(\vec{x})}^2}_{\text{POD truncation error}}\,.\label{eq:errorsum}
\end{align}
Hereby, the relative error arising from thresholding details is defined by 
\begin{equation}
  \label{eq-def:waveleterror}
  \errWavelet(\epsilon)\defeq\frac{\norm{\u_i(\vec{x}) - \u_i^\epsilon(\vec{x})}}{\norm{\u_i(\vec{x})}}
\end{equation}
and the relative error due to the truncation after the $r$-th POD mode in \cref{eq-def:wPODsubspace} is:
\begin{equation}
  \label{eq-def:PODerror}
  \errPOD(\epsilon,r)\defeq\frac{\sum_{i=1}^{\NSnapshots}\norm{\u_i^\epsilon(\vec{x}) - \tilde\u_i^\epsilon(\vec{x})}^2}
  {\sum_{i=1}^{\NSnapshots}\norm{\u_i^\epsilon(\vec{x}) }^2}
  = \frac{\sum_{k=r+1}^ {\NSnapshots}\lambda^\epsilon_{k}}{\sum_{k=1}^ {\NSnapshots}\lambda^\epsilon_{k}}
\end{equation}
Further details can be found in \cref{sec:wPODAlgo}.
Using \cref{eq:errWavelet,eq-def:waveleterror,eq-def:PODerror,eq:errorsum} one obtains for the total
relative error of the wPOD (see \cref{sec:deriv-eq23}):
\begin{align}
  \label{eq:errWPOD}
  \errWPOD(\epsilon,r)
    \le &\; \errPOD(0,r) +\mathcal{M}_r \epsilon + \epsilon^2
  \approx   \errPOD(0,r) + \epsilon^2 \,,\\
  \text{where}\quad &\mathcal{M}_r = \frac{\sum_{k=r+1}^{\NSnapshots} l_k}{\sum_{k=1}^{\NSnapshots}\lambda_k}\,,
\end{align}
when assuming perturbed eigenvalues $\lambda^\epsilon_k=\lambda_k+l_k\epsilon$, with perturbation $l_k\in\mathbb{R}$. An error bound for the perturbation of the eigenvalues is given for a simplified correlation matrix in \cite{CastrillonAmaratunga2002}.
$ \mathcal{M}_r $ is often small, in which case it can be neglected. 
This is further discussed in the numerical example section. 
Note that in the limit $\epsilon \to 0$ the wPOD error yields exactly the POD error. In
the limit $r \to \infty$ the POD error vanishes and we we are left with the wavelet compression error.
These limits are visualized for our numerical studies in \cref{fig:vort-PODerror,fig:bumblebee-wPODerror}.

In most of the applications the wavelet threshold $\epsilon$ will be chosen according to the available memory of the hardware. If memory limitations are not an issue,
it is advantageous to balance the wavelet and POD truncation error for better efficiency.
Assuming that the approximation in \cref{eq:errWPOD} holds $\errPOD(\epsilon,r)\approx\errPOD(0,r)$ (i.e. $\mathcal{M}_r\le \epsilon$), we can treat the two errors $\errWavelet,\errPOD$ independently.
For a predefined error $\err^*$, we
first fix the compression error choosing $\epsilon^*\le \sqrt{\err^*/2}$ and adjust the truncation
error by \texttt{err}$=\err^*-(\epsilon^*)^2$. The truncation rank $r^*$ is then
chosen to compensate the additional error introduced by the wavelet compression.
In the case when $\mathcal{M}_r$ is expected to be larger than $\epsilon$, the errors cannot be balanced without having an estimate of the POD eigenvalues. Eigenvalues  $\lambda^\epsilon_k$, which are smaller then the compression error are not reliable. Therefore we recommend the conservative setting, choosing $\epsilon^*<\err^*$, for a first estimate.

%% file: results.tex
\section{Numerical Results}
\label{sec:numerical-results}

In this section, we test the wPOD algorithm, outlined in \cref{sec:wPODAlgo}, on 2D and 3D numerical data and assess its efficiency and precision.
The algorithm is integrated into the open source software package \Softwarename{WABBIT}~\cite{WABBIT_github} and can be called as a post processing
routine\footnote{
  The program has several options: \texttt{wabbit-post --POD --  nmodes=<}$r$\texttt{> --error=<err> --memory=<RAM> --adapt=<}$\epsilon$\texttt{> --components=<}$K$\texttt{> --list=<ComponentList1> }$\cdots$\texttt{-- list=<ComponentListK>}
  in which the number of modes $r$ or the truncation error \texttt{err} can be specified. For the latter $r$ is automatically chosen from the error criterion given in \cref{sec:wPODAlgo}. The algorithm
  requires specifying the number of components $K$ together with $K$ lists of
  files which stores the snapshots of each component in a HDF5 format. Furthermore,
  the memory and adaptation level should be chosen in accordance with given resources.
}.

We provide two types of case studies:
A synthetic test case (see \cref{sec:SyntheticTestCase}) in 2D, which is used to
benchmark our code. We also compare it to the randomized singular value decomposition  (rSVD), outlined in \cref{sec:snapshot_POD},
and case studies for 2D and 3D data obtained by  numerical simulation of
the incompressible Navier-Stokes equations in \cref{subsec:directNumericalandAdaptive}.

\subsection{Synthetic Test Case}  
\label{sec:SyntheticTestCase}
For the synthetic test case we define a combination of dyadic structures, inspired by \cite{MendezBalabaneBuchlin2018}:
\begin{align*}
  u(x,y,t ) = \sum_{k=1}^R a_k(t)\Psi_k(x,y)\,,
\end{align*}
of $R=15^2$ orthogonal modes $\Psi_k: [0,30]^2 \rightarrow \mathbb{R}$ and temporal amplitudes
$a_k: [0, 2\pi] \rightarrow \mathbb{R}$.
The modes are smooth, two-dimensional bumps
\begin{align}
  \Psi_{m+15n+1}(x,y)
        &=b(\sqrt{(x-x_m)^2+(y-y_n)^2}) \quad\text{for}\quad n,m = 0,\dots,14\,.
        \label{eq:modes_bump}\\
   b(x) &=
   {\begin{cases}
    \exp \left(-{\frac {1}{1-x^{2}}}\right),&x\in(-1,1)\\
    0,&{\mbox{otherwise}}
  \end{cases}}
 \end{align}
placed at $(x_m,y_n)=(1+2m,1+2n)$ in a checkerboard pattern.
Note that the modes are orthogonal because of their non-overlapping support.
Furthermore we choose oscillating amplitudes:
\begin{equation}
  \label{eq:amplitudes_bump}
  a_k(t)=e^{-k/\Delta\lambda}\sin(\pi f_k t) \quad\text{for}\quad k = 1,\dots,15^2
\end{equation}
with randomly shuffled frequencies $f_k\in\{1,\dots,15^2\}$ and moderate decrease
in magnitude: $\Delta\lambda=3$. We choose
$\NSnapshots=2^7$ equally spaced snapshots on a $N_x\times N_y=1024\times1024$ initial grid.
Before starting the algorithm the initial grid has to be partitioned into blocks.
This is done using a python
routine available in the \texttt{WABBIT} software package \cite{WABBIT_pythongithub}.
In our studies we block-decompose the initial grid in three different configurations to compare
the effect of different block sizes. The sizes of the blocks are $\Bs=N_x/2^{\Jmax}+1=17,33,65$ with $\Jmax=6,5,4$, respectively.

\subsubsection*{Wavelet Compression}
\begin{figure}[htp!]
  \centering
  \includegraphics[width=0.32\linewidth]{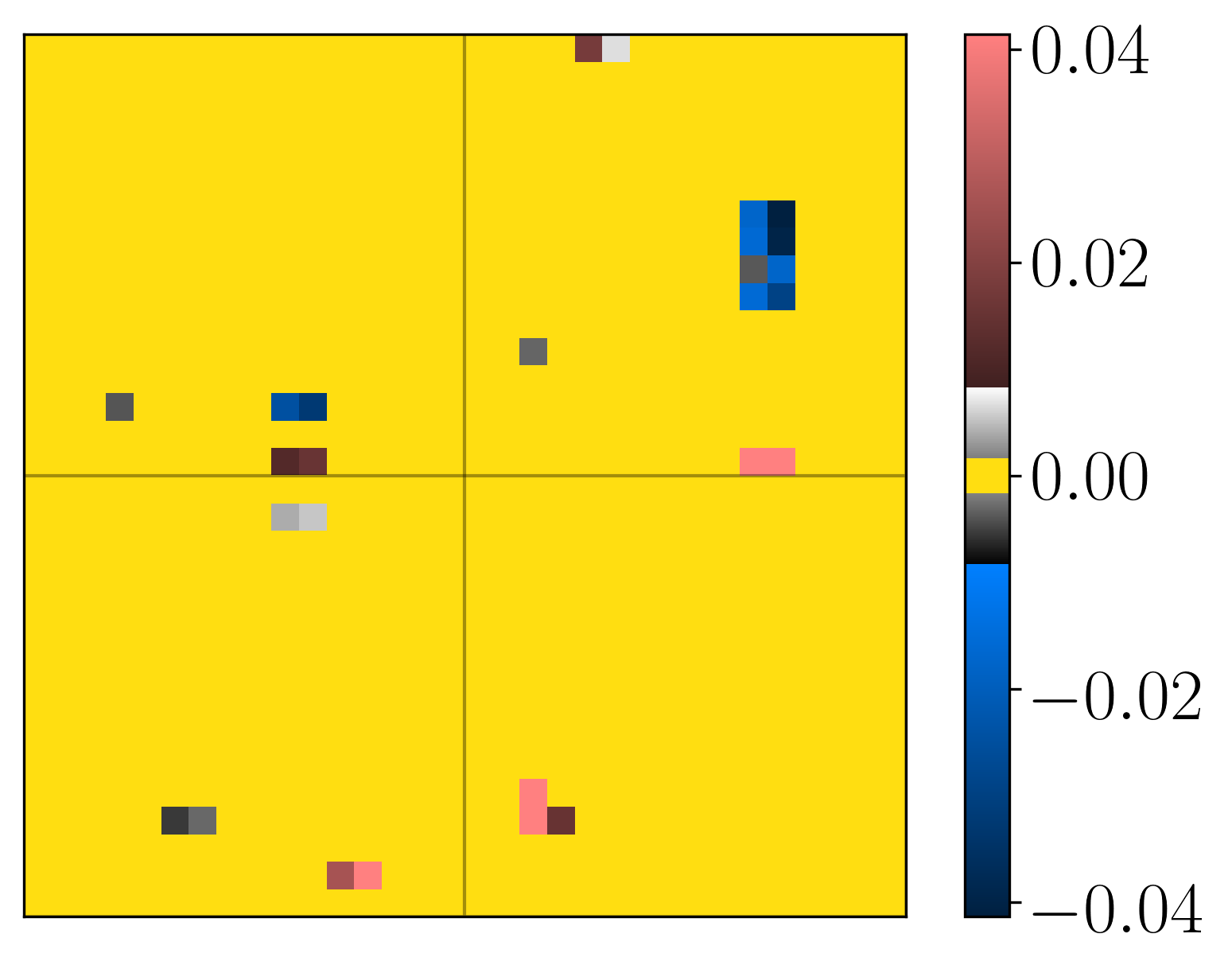}
  \includegraphics[width=0.32\linewidth]{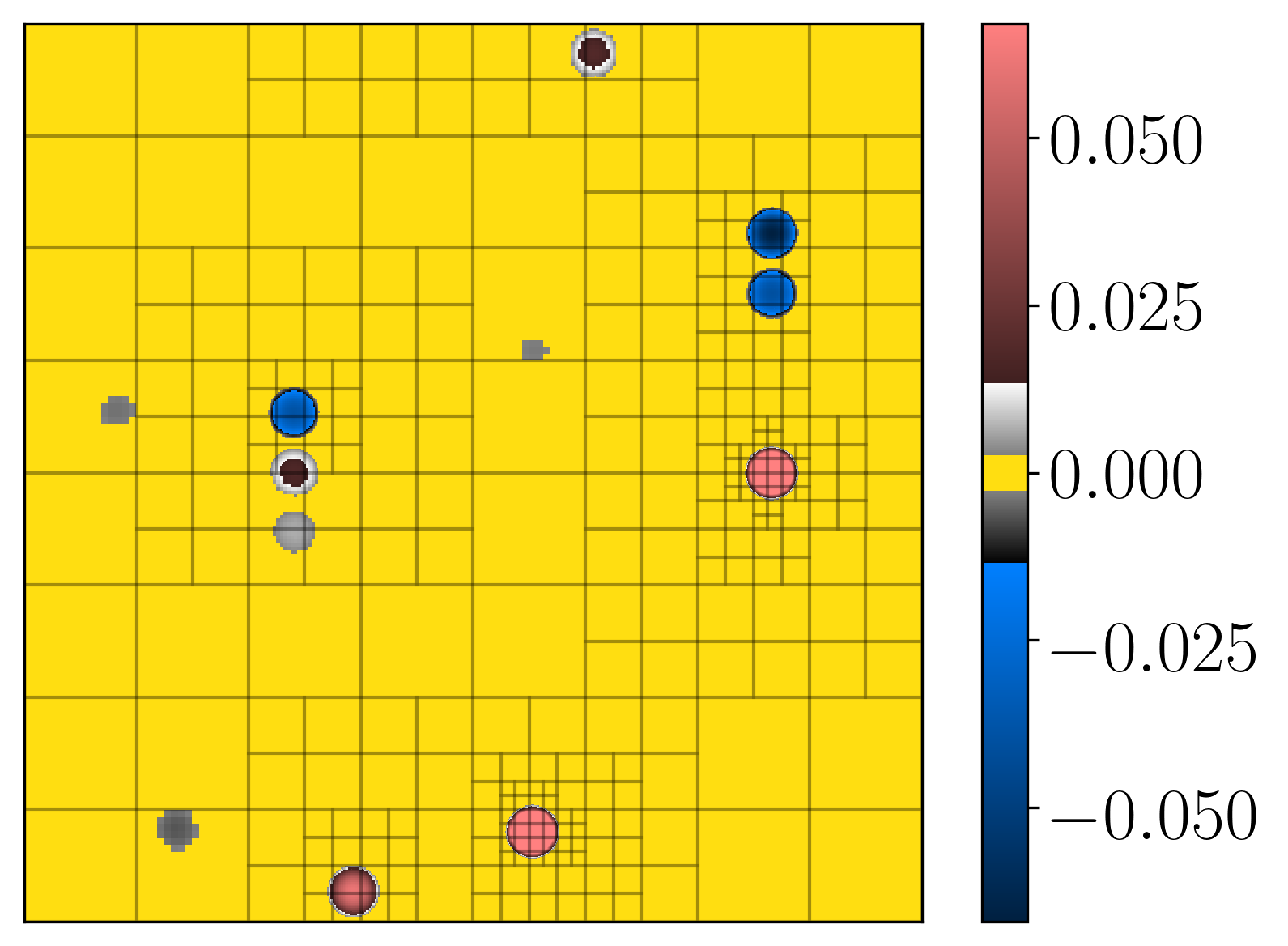}
  \includegraphics[width=0.32\linewidth]{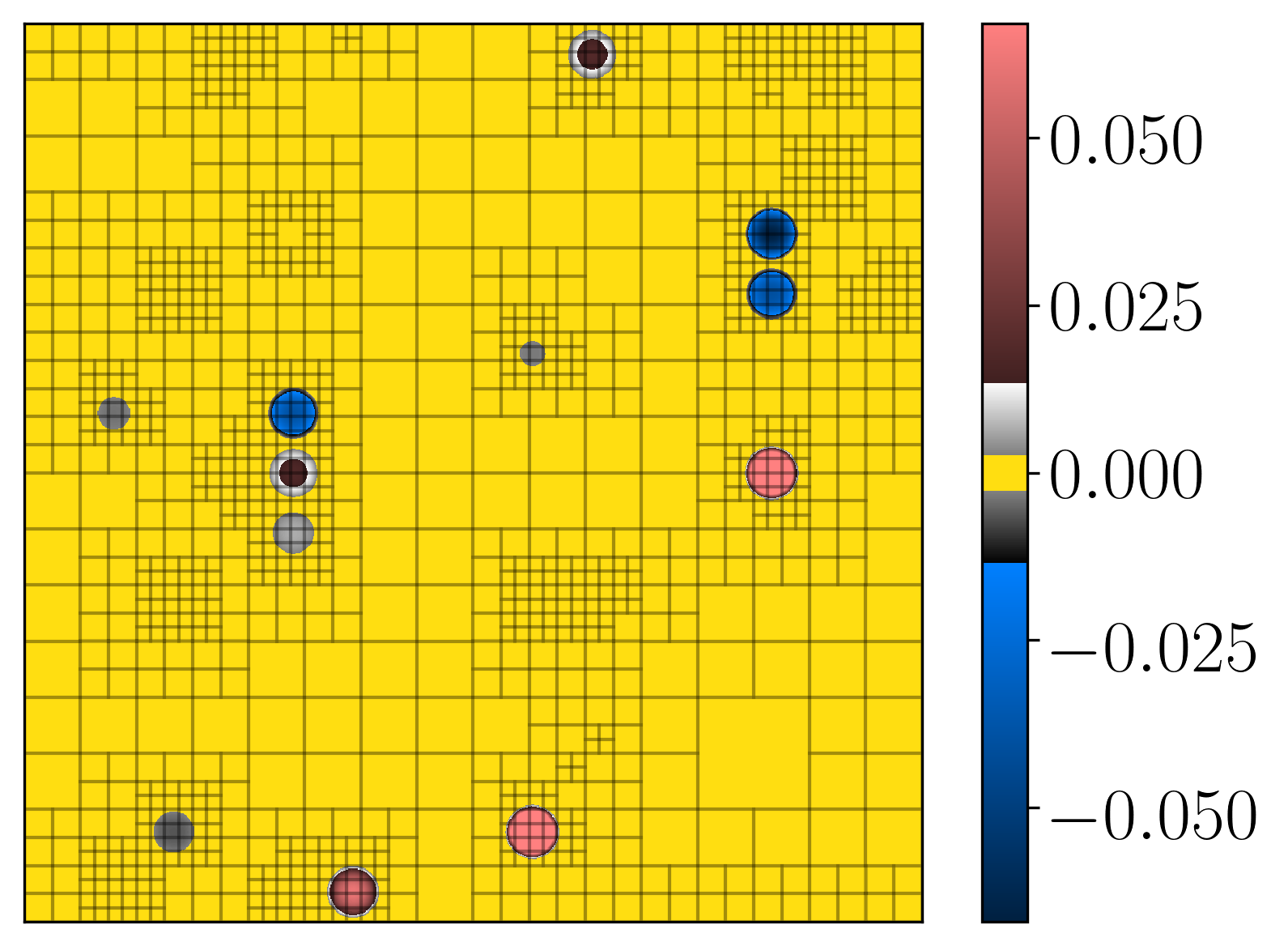}
  \caption{
  Block-based adaptation of $u^\epsilon(x,y,t)$ at $t=42\Delta t$ for $ \epsilon=1.0,\num{2.2e-2},\num{1.0e-5}$ (from left to right) and with $\Bs=17$.}
  \label{fig:bumpsnapshot}
\end{figure}

\begin{figure}[htpb]
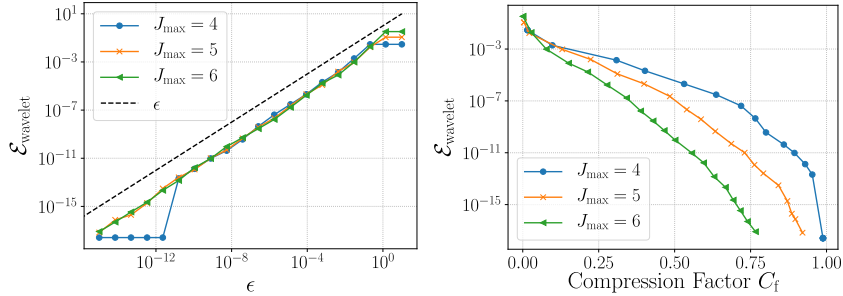

    \centering
    \includesvg[width=0.45\textwidth]{bump_CDF44c_02_compression_err4th}
    \includesvg[width=0.45\textwidth]{bump_CDF44c_02_compression_err_vs_rate}
    \caption{Compression error $\errWavelet$ (left) defined in \cref{eq-def:waveleterror} and compression factor $C_\mathrm{f}$ (right) of the bump test case using different block sizes $\Bs=N_x/2^{\Jmax}+1$, with maximal refinement levels $\Jmax=4,5,6$. The compression factor is the fraction between the sparse and
    dense number of grid points / blocks.}
    \label{fig:bumpscompress}
\end{figure}

First we examine the compression of the data with varying block size and thresholds $\epsilon$ with $10^{-15}\le\epsilon\le10$.
\Cref{fig:bumpsnapshot} shows the adaptation of a single snapshot for $ \epsilon=1.0,\num{2.2e-2},\num{1.0e-5}$. For larger $\epsilon$ the number of blocks decreases, leading to
stronger compression of the data and increasing compression errors.
This behavior
is quantified for varying block size $\Bs$ and $\epsilon$ in \cref{fig:bumpscompress}.
Here we plot the relative compression error
$\errWavelet$ and compression factor $C_\mathrm{f}\le1$, i.e. the fraction between the number of blocks at a given threshold and the total number of blocks available needed for the full grid.
As can be seen from \cref{fig:bumpscompress},
a higher maximal refinement level $\Jmax$, i.e. smaller
blocks, corresponds to smaller overall compression factor, while the compression error
$\errWavelet$ is approximately the same.
This observation is expected, because smaller blocks enable better resolution
of local structures, however increasing the data handling effort. For all the compression curves in the numerical examples in \cref{fig:bumpscompress,fig:compressVortStreet} we see
the classical saddle shaped error curve:
with rapid error decay for small $C_\mathrm{f}\lesssim 0.05$  (i.e. large $\epsilon\gtrsim 10^{-2}$)
until a plateau is reached with
a saddle point from which it begins to decay again.
Regardless of the final error of our algorithm, it is recommended to set $\epsilon$
at the onset of the plateau, since after the plateau is reached only little gain
in precision is achieved.

Furthermore we emphasize that the compression error scales linear in $\epsilon$, independent from the chosen block size, which is an important property for the
error control of our algorithm. The sudden drop of the error for $\Jmax=4$ is
due to the fact, that after $\epsilon\lesssim 10^{-11}$ the blocks are refined
to the maximal level.

\subsubsection*{POD Truncation}

\begin{figure}[htp!]
  \centering
  \includegraphics[width=0.32\linewidth]{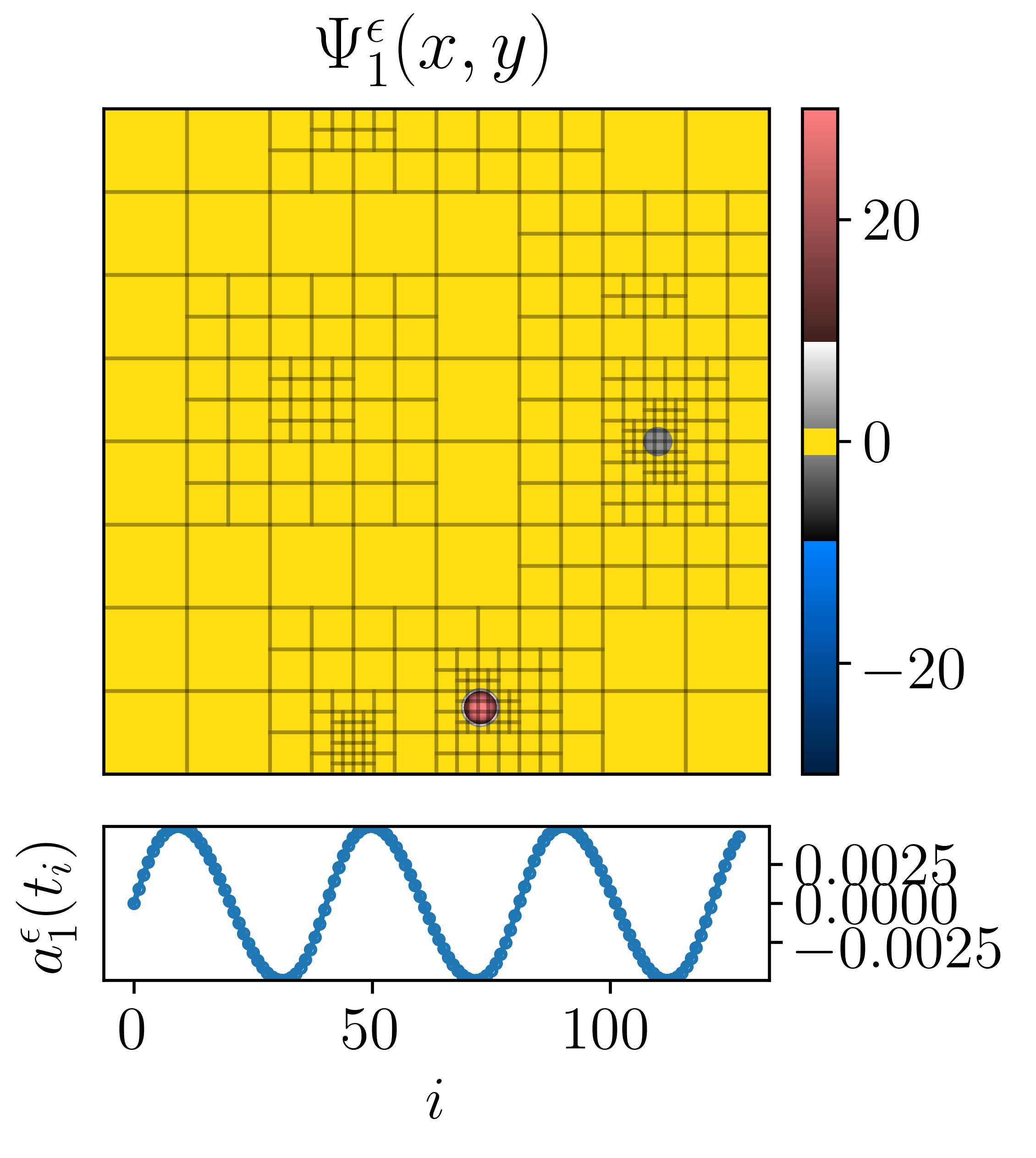}
  \includegraphics[width=0.32\linewidth]{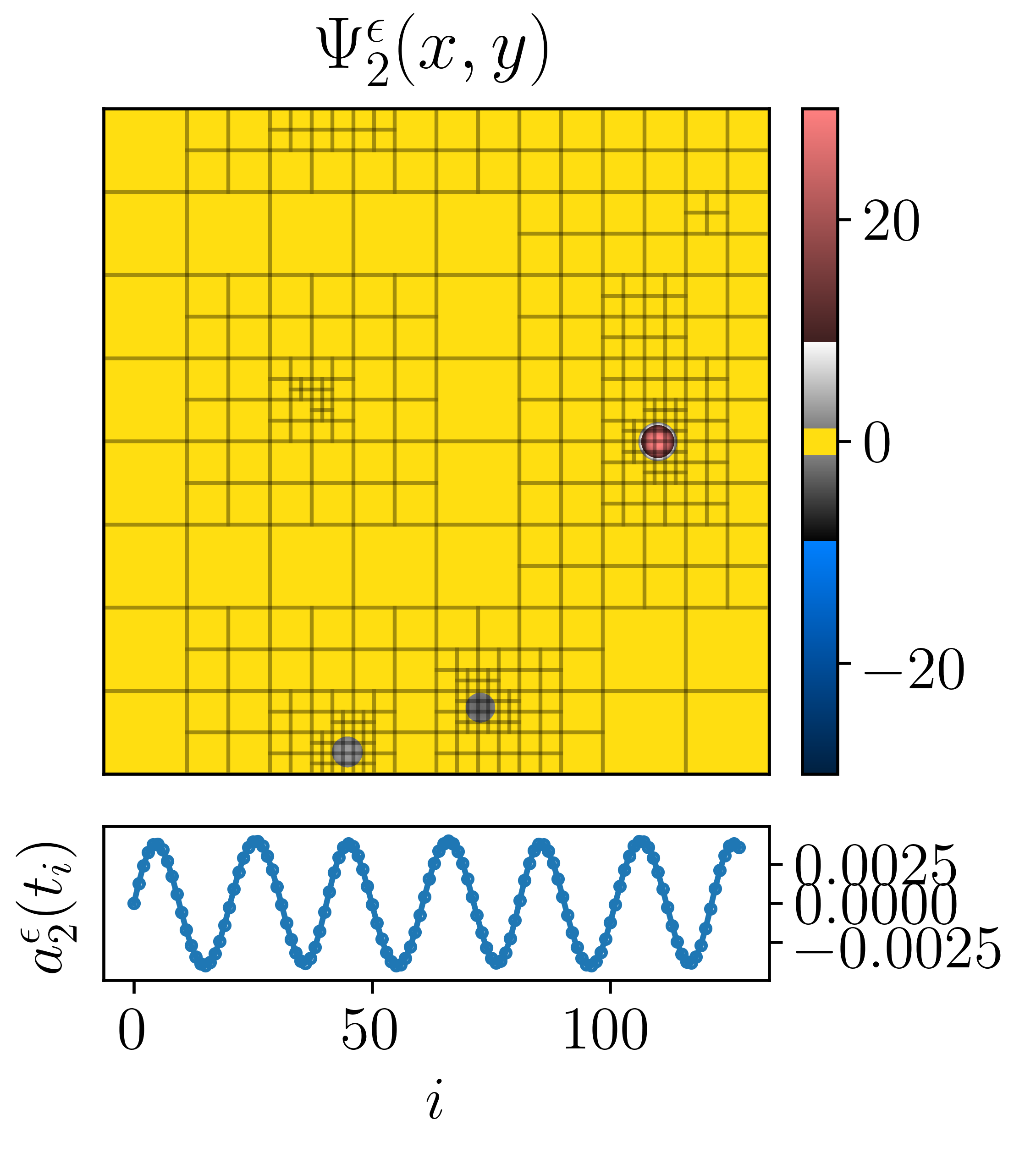}
  \includegraphics[width=0.32\linewidth]{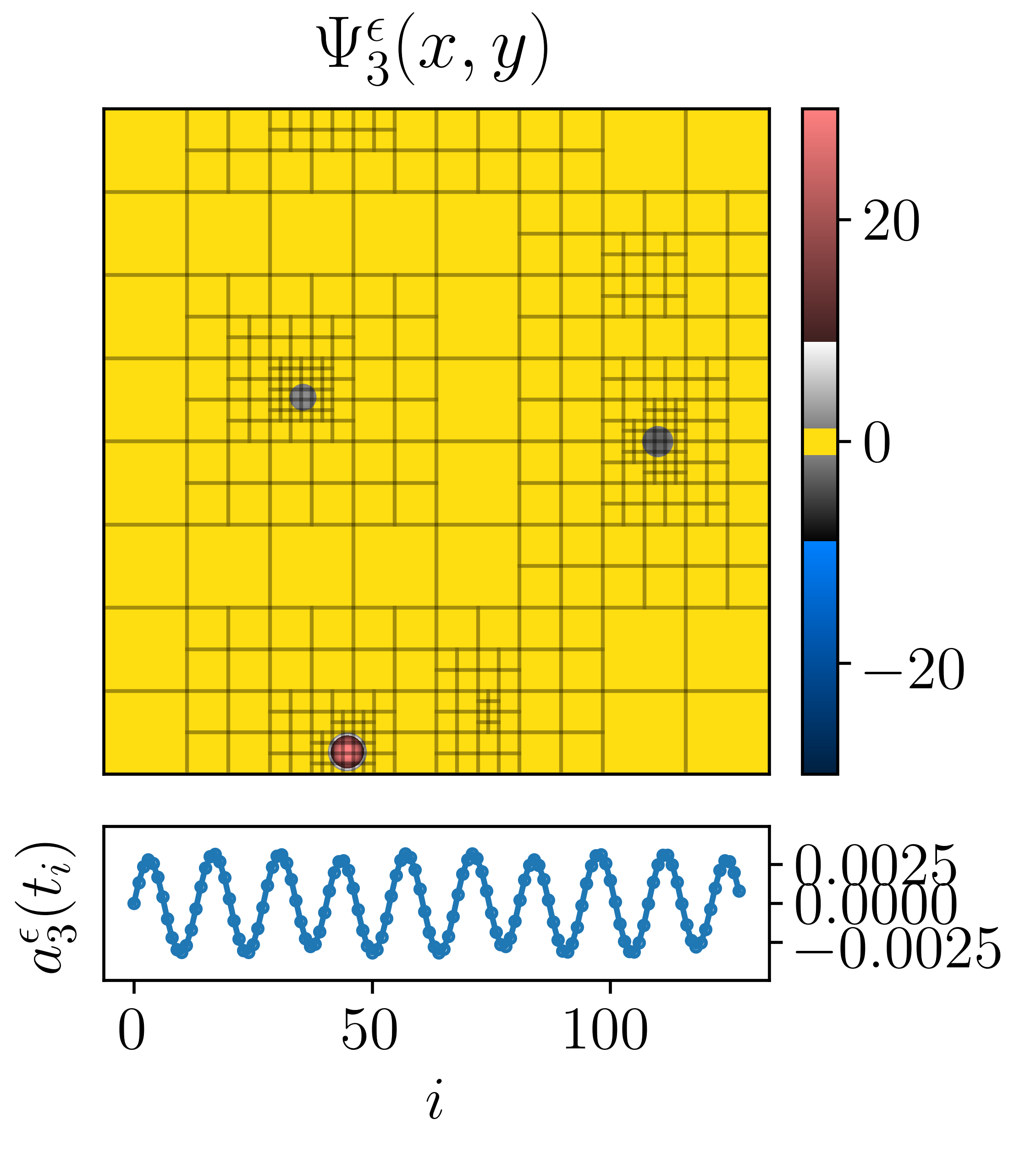}
  \caption{First three modes $\Psi^\epsilon_k$ and their amplitudes $a^\epsilon_k(t_i)$, $k=1,2,3$, for $\epsilon=10^{-5}$. Blocks are of size $\Bs=17$.}
  \label{fig:bumpmodes}
\end{figure}

Next we study the impact of the wavelet compression on the computed POD modes
and the overall approximation error.
For $\epsilon=10^{-5}$, \cref{fig:bumpmodes} visualizes the first three modes and the corresponding amplitudes obtained with
the wPOD algorithm. Note that the modes and amplitudes of the POD problem \cref{eq:minPOD} are only unique up
 to an orthogonal transformation. Hence the initial input structures $\Psi_k,a_k$, defined in \cref{eq:amplitudes_bump,eq:modes_bump}, are not exactly recovered by the wPOD.
 However, we see that the magnitude of the amplitudes decreases and its frequency increases with increasing mode number. Moreover one observes that the computational grid is nicely adapted to the structure of the modes.

\begin{figure}[htpb]
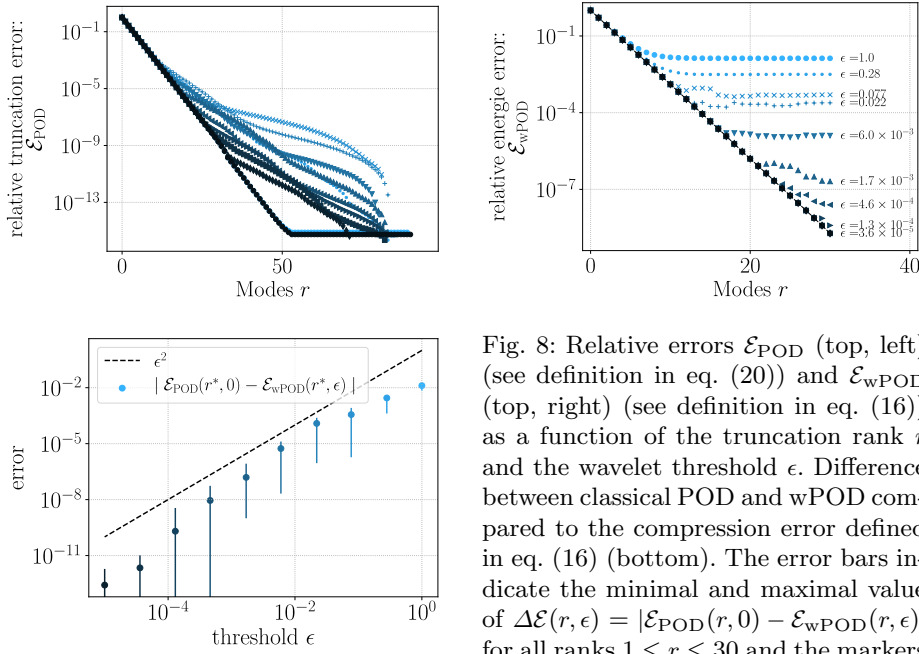

    \centering
    \begin{subfigure}[t]{.48\textwidth}
        \centering
        \includesvg[width=1\textwidth]{bump_CDF44c_02_PODerror_J5}
        \label{subfig:bumpPODerr}
    \end{subfigure}\hfill%
      \begin{subfigure}[t]{.48\textwidth}
        \centering
        \includesvg[width=1\textwidth]{bump_CDF44c_02_wPODerror_J5}
            \label{subfig:bumpwPODerr}
    \end{subfigure}
      \begin{subfigure}[t]{.48\textwidth}
        \centering
        \vspace{0pt}
        \includesvg[width=1\textwidth]{bump_CDF44c_02_deltaPODerror_J5}
    \end{subfigure}\hfill%
    \begin{minipage}[t]{.48\textwidth}
    \caption{Relative errors $\errPOD$ (top, left) (see definition in \cref{eq-def:PODerror}) and $\errWPOD$ (top, right) (see definition in \cref{eq-def:errWPOD}) as a function of the truncation rank $r$ and the wavelet threshold $\epsilon$.  Difference between classical POD and wPOD compared to the compression error
    defined in \cref{eq-def:errWPOD} (bottom). The error bars indicate the minimal and maximal value of $\Delta \mathcal{E}(r,\epsilon)=\abs{\errPOD(r,0)-\errWPOD(r,\epsilon)}$ for all ranks $1\le r\le30$ and the markers the mean $\overline{\Delta \mathcal{E}}(\epsilon)=1/30\sum_{r=1}^{30} \Delta \mathcal{E}(r,\epsilon)$.  The colors vary from bright blue at $\epsilon=1.0$ to black at $\epsilon=0.0$. Blocks are of size $\Bs=17$.}
    \label{fig:bump-wPODerror}
    \end{minipage}
\end{figure}

Additionally, we estimate the truncation error $\errPOD(\epsilon,r)$ and the total error $\errWPOD(\epsilon,r)$ in the $L^2$-norm, as defined in \cref{eq-def:errWPOD}. We compare
the errors for $10^{-5}\le\epsilon\le\num{1.0},r\le30$ in \cref{fig:bump-wPODerror}.
In these plots the impact of the wavelet adaptation, corresponding to blue lines with $\epsilon>0$, is visualized and compared to the classical snapshot POD procedure corresponding to $\epsilon=0$, which is drawn in black.
The exponential decay of the eigenvalue spectra, given by the magnitude of the input modes $\abs{a_k}\sim\exp(k/\Delta\lambda)$, is nicely recovered in the $\epsilon=0$ case up to values $r\lesssim 60$ below machine precision. For increasing $\epsilon$ we see that the eigenvalues become increasingly distorted, as expected. However, the eigenvalue distortion does not influence the overall approximation error if epsilon is chosen with care (see \cref{sec:ErrorAnalysis}).
In fact the total approximation error converges approximately with $\epsilon^2$ to the exact values as the difference in \cref{fig:bump-wPODerror} (bottom) shows. The presented results are independent of the chosen block size $\Bs$, although only $\Bs=17$ is used in the shown figures.

\subsubsection*{Comparison to Randomized SVD}

\begin{figure}[htp!]
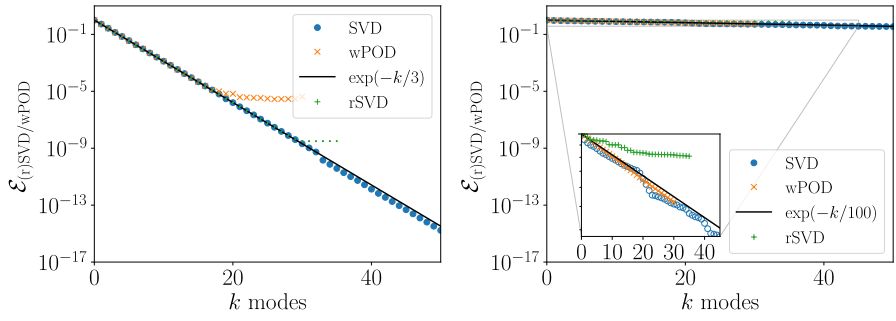

  \centering
  \includesvg[width=0.48\linewidth]{bump_CDF44c_02_error_wPODvsrSVD}
  \includesvg[width=0.48\linewidth]{bump_CDF44c_01_error_wPODvsrSVD01}
  \caption{Comparison of the total approximation errors  $\mathcal{E}_\text{(r)SVD},\errWPOD$  for $\Delta\lambda=3$ (left) and $\Delta\lambda=100$ (right). As reference we plot the exact values $\exp(-k/3)$ and $\exp(-k/100)$ indicated by a black line. The rSVD is computed with 35 random samples of $\Smat$ and the wPOD set up with $\epsilon=6\times 10^{-3}$ and $\Jmax=5$, to ensure approximately the same memory consumption. The inset shows a zoom.}
  \label{fig:rSVD-wPOD-slow}
\end{figure}

The overall purpose of our method is to be able to fit the snapshot data into
the fast memory by tuning $\epsilon$, in order to compute the POD without having to address the slow memory again, after the data has been compressed. This enables us to cope with large data sets. Therefore, a comparison to other methods suited for large data like the randomized SVD suggests itself.
We follow the algorithm outlined in \cref{sec:snapshot_POD} taken from \cite{HalkoMartinssonTropp2011}.
For a fair comparison we refrain from power iterations, which would need additional passes over the slow memory and we use $n=5$ extra random samples
as suggested in \cite{HalkoMartinssonTropp2011}. Hence for a target rank of $r^*=30$ we take $q=35$ random samples of our snapshot matrix $\Smat\in\mathbb{R}^{M\times\NSnapshots}$, $M=1024^2,\NSnapshots=2^7$ to compute an orthogonal matrix $\mathbf{Q}\in\mathbb{R}^{M\times q}$, which approximates the column rank of $\Smat$. The computation of $\mathbf{Q}$ however is only feasible if the random samples fit in the fast memory. Therefore, the minimum amount of memory needed by the rSVD is given by $S_\text{rSVD}=Mq$ in units of the floating-point arithmetic. In contrast the wPODs memory requirements in units of the floating-point arithmehtic: $S_\text{wPOD}=N^\text{tot}_\text{blocks}\Bs^2$ depend on the total number of blocks $N^\text{tot}_\text{blocks}$ in the snapshot set. Nevertheless $S_\text{wPOD}$ is tunable with $\epsilon$, but increasing $\epsilon$ also increases the compression error. To compare both methods we estimate $\errWPOD(r,\epsilon)$ as before in a range from $10^{-5}\le\epsilon\le1$ and $r$ up to $r^*=30$ together with the relative error
\begin{equation}
  \label{eq:frob_error}
  \mathcal{E}_\text{(r)SVD}(r)=\norm{\mathbf{U}-\mathbf{\tilde\Psi}\mathbf{\tilde\Sigma}\mathbf{\tilde V}^T}_F/\norm{\mathbf{U}}_F
\end{equation}
 of the truncated (r)SVD in the Frobenius norm. Here $\mathbf{\tilde\Psi}\in\mathbb{R}^{M\times r}, \mathbf{\tilde V}\in\mathbb{R}^{\NSnapshots \times r}$ are matrices, with columns composed of all spatial modes as orthonormal vectors $\Mode_k\in\mathbb{R}^{M}$ and the corresponding temporal coefficients $\vec{v}_k\in\mathbb{R}^{\NSnapshots}$ and $\mathbf{\tilde\Sigma}\in\mathbb{R}^{r\times r}$ contain the POD eigenvalues $\lambda_k$ as singular values
 $\sigma_k=\sqrt{\lambda_k}$ on the diagonals.
 A direct comparison of the total approximation errors $\mathcal{E}_\text{(r)SVD},\errWPOD$ defined in \cref{eq:frob_error,eq-def:errWPOD}, respectively, is shown in the left of \cref{fig:rSVD-wPOD-slow}. In both figures the wPOD was set up with $\epsilon=3.6\times 10^{-5}$ and $\Jmax=5$, such that $S_\text{wPOD}\approx S_\text{rSVD}$.

\begin{figure}[h]
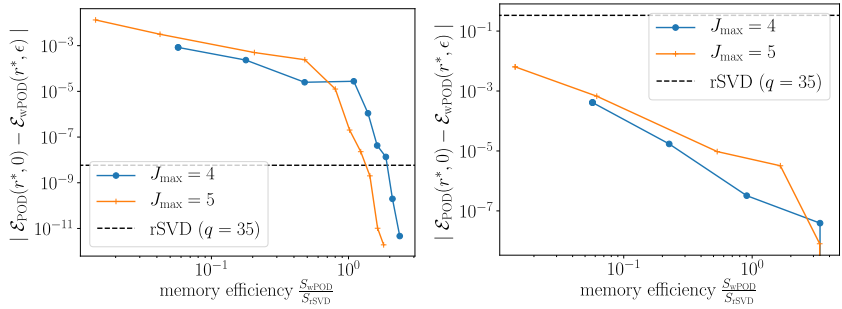

  \centering
  \includesvg[width=0.45\linewidth]{bump_CDF44c_02_memory_wPODvsrSVD}
  \includesvg[width=0.45\linewidth]{bump_CDF44c_01_memory_wPODvsrSVD01}
  \caption{Memory consumption vs. deviation from the exact POD values. Comparison between rSVD and wPOD. On the left we show the test case with $\Delta \lambda =3$ and on the right $\Delta \lambda =100$.}
  \label{fig:memory_rSVD-wPOD}
\end{figure}

One important aspect of this comparison has to be highlighted first.
 The performance of the rSVD and wPOD strongly depends on the data.
 The wPOD will always benefit from data with localized smooth structures, but it will be less efficient than the rSVD for cases which are distorted by random noise. In our studies the oscillating bump structures are very localized, therefore the wPOD has an advantage over the rSVD. In fact wPOD needs less memory than the rSVD to achieve the same overall error as shown in the studied parameter range, see \cref{fig:memory_rSVD-wPOD}.

A very interesting case, that cannot be efficiently dealt with by the rSVD, is the case
of slowly decaying singular values. Here, random sampling can hardly capture the column rank
of $\mathbf{U}$, since all columns are nearly equally important. Even the SVD algorithm implemented in pythons numpy package (using LAPACK, the implicit zero-shift QR
algorithm after reducing $\mathbf{U}$ to bidiagonal form) exhibits instabilities, as can be seen from the blue dotted markers in \cref{fig:rSVD-wPOD-slow}.
This is not the case for the wavelet POD, as shown in \cref{fig:rSVD-wPOD-slow}. However small deviations from the expected error decay $\sim \exp(-k/100)$ are visible.
For this case we have chosen all parameters as before, except that the
magnitude of the modes in \cref{eq:amplitudes_bump} decays much slower with $\Delta\lambda=100$.
This result demonstrates that our method can be used in the combination with
large model order reduction problems, which suffer under slowly decaying eigenvalues or singular values.
We would like to highlight that especially in large transport dominated systems, as they often occur in fluid dynamics, MOR is negatively impacted by slowly decaying singular values as
reported in \cite{OhlbergerRave2015}. Therefore, our method could be very useful in the treatment
of such problems. We will come back to this in \cref{sec:3DInsectsFlight}, where
we compute a POD of a bumblebee in forward flight.

%

\subsection{Application to 2D and 3D numerical flow data}
\label{subsec:directNumericalandAdaptive}
In this section we compute a sparse POD basis of 2D and 3D flow data, computed with
\Softwarename{WABBIT}. In the first study, we use data from a numerical simulation with an equidistant grid of a 2D flow past a cylinder.
In the second application, we apply the algorithm to highly
resolved 3D
data of a flapping flight simulation of a bumblebee \cite{EngelsSchneiderReissFarge2019}.
\begin{table}[hbp!]
  \centering
  \caption{Parameters of the numerical test cases. Here $\epsilon^{\infty}$ is the wavelet threshold with CDF4,4 wavelets being normalized in $L^{\infty}$. }
  \label{tabl:vortexParams}
  \begin{tabular}{ll}
    \toprule
    \textbf{Von Kármán Vortex Street}\\
    Parameter & Value\\
    \midrule
    Number Snapshots $\NSnapshots$ & 225\\
    Resolution $N_1\times N_2$ & 4096$\times$1024\\
    Domain size $L_1\times L_2$ & 64 $\times 16$\\
    Reynolds Number $\mathrm{Re}$ & 200\\
    \bottomrule
    \end{tabular}%
    \begin{tabular}{|ll}
      \toprule
      \textbf{Bumblebee} \\
      Parameter & Value\\
      \midrule
      Number Snapshots $\NSnapshots$ & 41\\
      Resolution  $(\Jmax,B_\alpha,\epsilon^\infty)$ & (7,\,23,\,0.01)\\
      Domain size $L^3$ & $8^3$\\
      Reynolds Number $\mathrm{Re}$ & 2000\\
      \bottomrule
    \end{tabular}
  \end{table}
\subsubsection{2D Case - Von Kármán Vortex Street}
\label{sec:vortex_street}
In this first example, our dataset $\mathcal{U}$ results from a 2D simulation of an incompressible flow past a cylinder using the artificial compressibility method computed with \texttt{WABBIT}.
Details about the software can be found in \cite{EngelsSchneiderReissFarge2019}.
For this study the block structured grid of \texttt{WABBIT} is fixed at the resolution $\Jmax=6$, with $(\Bs_1,\Bs_2)=(65,17)$ and domain size $\Domain\defeq[0,64]\times[0,16]$.
This is equivalent to an equidistant grid of 4096$\times$1024 grid points.
Each snapshot has three components $\vec{u}=(\vec{v},p)$, where $\vec{v}$
denotes the velocity and $p$ the pressure.
For this case study we chose the vortex street for a Reynolds number $\mathrm{Re}=200$,
because it is known to have fast decaying POD truncation errors. The Reynolds number $\mathrm{Re}=2v_\infty R/\nu$ is based on the cylinder radius $R=1$, freestream velocity $v_\infty=1$ and kinematic viscosity $\nu=0.01$.
The simulation was run until a stable Kármán Vortex shedding was achieved.
From this point onwards, we sample the solution in the time interval $500\le t\le 612$ with $\Delta t = 0.5$,
resulting in $\NSnapshots=225$ snapshots.
The relevant parameters for our algorithm are summarized in \cref{tabl:vortexParams}.
For concise overview we only show the scalar-valued vorticity $\omega=\nabla \times \vec{v}$ computed
from the velocity components of the state vector in \cref{fig:reconstruct_vort,fig:threshold-vorx}
and the curl of the two velocity components of the POD-basis in \cref{fig:vort_modes}.

\subsubsection*{Wavelet Compression}
\begin{figure}[htpb]
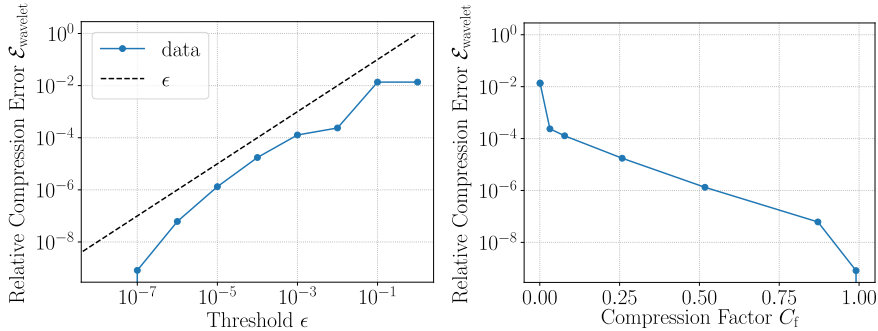

  \centering
  \includesvg[width=0.47\linewidth]{vortex_error_compression_err4th_cyl}
  \includesvg[width=0.47\linewidth]{vortex_error_compression_err_vs_rate_cyl}
  \caption{Compression error and compression factor $C_\mathrm{f}$ of the vortex street. Left: The compression error in the $L^2$-norm is bounded by $\epsilon$, drawn with dashed line
  (\dashed). Right: Relative error in the $L^2$-norm vs. compression factor. For direct comparison the vertical axis limits of both figures are identical.}
  \label{fig:compressVortStreet}
\end{figure}

As in the synthetic test case in \cref{sec:SyntheticTestCase}, we first study the compression of the flow data.
To this end, we sample $\errWavelet(\epsilon)$ of one representative snapshot $\u$ at $t=550$ with  different $\epsilon\in\{10^{-8}, 10^{-7}, \dots, 1\}$. In \cref{fig:compressVortStreet} we plot the  relative compression error $\errWavelet$ against the wavelet threshold $\epsilon$ (left) and against the compression factor $C_\mathrm{f}$ (right).
In contrast to the scalar field in the synthetic example, the thresholded quantity is vector
valued. Therefore, we normalize each state vector component before thresholding. All norms in the plots are vector norms.
Note, that for the threshold $\epsilon=0.1$ the grid is on the coarsest resolution ($j=1$) with 4 blocks only (corresponding blocks are shown in \cref{fig:threshold-vorx}). This explains why the compression errors beyond $\epsilon=0.1$ do not differ. Similarly $\epsilon=\num{1e-8}$ corresponds to the finest resolution at grid level $j=\Jmax$ and the error in \cref{fig:compressVortStreet} drops to zero.
However, as mentioned earlier, for maximal performance the wavelet threshold $\epsilon$ should be located at the onset of the plateau,
which is reached at $\sim \num{1e-2}$ (i.e. $C_\mathrm{f}\sim 0.035$).
Comparing this observation with the block distribution in \cref{fig:threshold-vorx}, one observes that smaller values of $\epsilon$ produce denser grids, with only little gain in precision.
It is remarkable that according to \cref{fig:compressVortStreet} less then $3.5\%$ of the actual data is needed to represent the full
data with an $L^2$-error less then 0.5$\permil$. At this compression level we have
compressed the full data with $225$ snapshots from $\sim 24$GB to $\sim 0.84$GB, which
makes it easily manageable for most laptops.

\subsubsection*{POD Truncation}
\begin{figure}[ht!]
  \centering
  \includegraphics[width=1\textwidth]{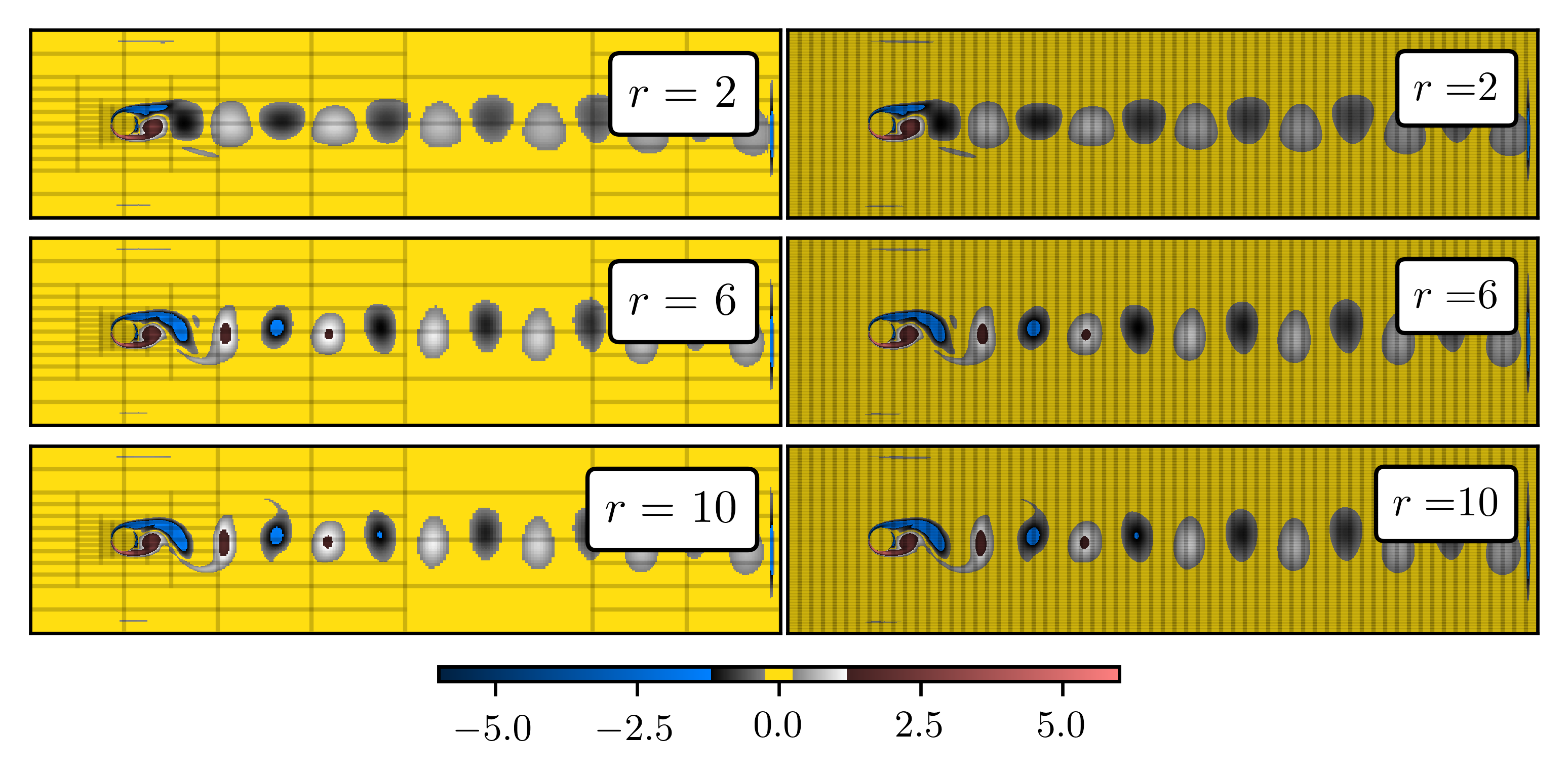}
  \caption{Direct comparison between sparse reconstruction $\tilde \u^\epsilon$ with $\epsilon =10^{-2}$ (left) and dense reconstruction with $\epsilon=0.0$ (right) using $r=2,6,10$ modes (from top to bottom). In both columns we display the vorticity $\tilde\omega^\epsilon=\partial_x \tilde v_y^\epsilon - \partial_y \tilde v_x^\epsilon$ of the reconstructed velocity components $(\tilde v_x^\epsilon,\tilde v_y^\epsilon)$.}
  \label{fig:reconstruct_vort}
\end{figure}

\begin{figure}[htpb]
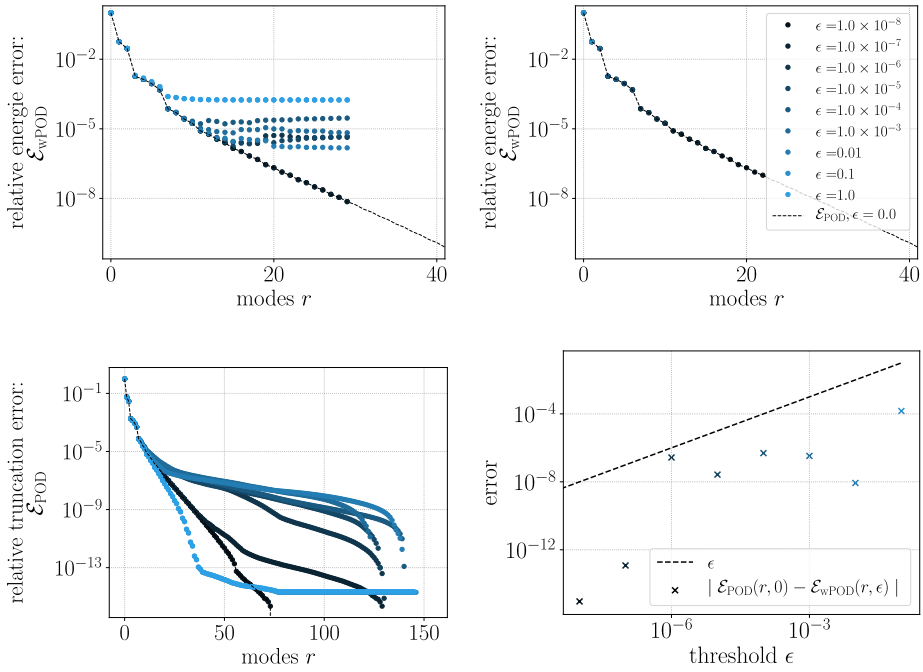

    \centering
          \begin{subfigure}[t]{.49\textwidth}
        \centering
        \includesvg[width=1\linewidth]{vortex_error_wPODerror_J6}
            \label{fig:vort-PODerror_c}
    \end{subfigure}\hfill%
      \begin{subfigure}[t]{.49\textwidth}
        \centering
        \includesvg[width=1\linewidth]{vortex_error_wPODerror_J6_conservative}
            \label{fig:vort-PODerror_d}
    \end{subfigure}
     \begin{subfigure}[t]{.49\textwidth}
        \centering
        \includesvg[width=1\linewidth]{vortex_error_PODerror_J6}%
        \label{fig:vort-PODerror_a}
    \end{subfigure}\hfill%
       \begin{subfigure}[t]{.49\textwidth}
         \centering
         \includesvg[width=1\linewidth]{vortex_error_deltaPODerror_J6_conservative}
     \end{subfigure}\hfill%
    \caption{
     Relative errors $\errPOD$ (lower left) and $\errWPOD$ for the vortex street without error control (upper left) and with conservative error control (upper right).
    In the lower right the absolute difference between the classical POD results and the wPOD:  $\Delta \mathcal{E}(r,\epsilon)=\abs{\errPOD(r,0)-\errWPOD(r,\epsilon)}$ are shown in the conservative error setting.
     The colors vary from bright blue at $\epsilon=1$ to black at $\epsilon=0$.
    }
    \label{fig:vort-PODerror}
\end{figure}

We will now study the error behavior of the wPOD numerically for fixed threshold $\epsilon$. To this end, we compute $\errPOD(\epsilon,r), \errWPOD(\epsilon,r)$ for
$\epsilon\in\{10^{-8}, 10^{-7}, \dots, 1\}$. Three calculated modes and their temporal coefficients are
visualized with the corresponding block structure in \cref{appxC},  \cref{fig:vort_modes}.
Furthermore, we compare the reconstruction $\tilde\u^\epsilon$ in \cref{fig:reconstruct_vort} for the dense case with $\epsilon=0$ (right column) and one adaptive case $\epsilon=10^{-2}$ (left column) using $r=2,6,10$ modes. The comparison shows that with increasing number of modes the typical vortex structure is recovered. No qualitative differences, except local changes in resolution, can be seen, when comparing the adaptive and non adaptive results.
For a quantitative analysis we have plotted the total $L^2$-error and the truncation error in \cref{fig:vort-PODerror}.
In both plots the impact of the wavelet adaption (corresponding to $\epsilon > 0$, blue lines) is visualized and compared to the results of the classical POD procedure (corresponding to $\epsilon = 0$, black line).
The numerical data show the behavior stated in \cref{sec:ErrorAnalysis}:
With the wavelet adaption of the snapshots, errors are introduced, which lead to distortion of
the POD eigenvalue problem and compression errors in the POD basis.

The perturbation $l_k$ of the eigenvalues is especially visible in $\errPOD$ (lower left plot in \cref{fig:vort-PODerror}) for the low energy modes ($r>30$), corresponding to small eigenvalues. In this regime
the distorted eigenvalues $\lambda_r^\epsilon = \lambda_r + l_r \epsilon $ fluctuate around the exact value $\lambda_r$ (black markers).
For the smallest eigenvalues the relative fluctuation can lead to a total failure of the algorithm, with {even} negative eigenvalues. This regime, however, can be ignored,
since the total error $\errWPOD$ will be dominated by the compression effects for large $r$.

Figure \ref{fig:vort-PODerror} (upper left) shows that the error behavior of $\errWPOD$ is
dominated by the truncation error $\errPOD$ for small number of modes $r$.
With increasing $r$ the truncation error decreases, while the error introduced by the compression
remains constant. This leads to a saturation of the total error, as soon as the truncation error is smaller than the error due to compression. This saturation effect is not a special case of the chosen wavelet compression scheme, as it also appears in finite element schemes as well, see for instance  \cite{UllmannRotkvicLang2016,GraessleHinze2018}.
However, the wavelet basis has a major
advantage over finite element schemes in this setting, since the grids are
hierarchically structured, which is easier to handle and computationally efficient. In fact, no
additional computations for \textit{"(i) collision detection, (ii) mesh intersection (detect intersection interface) and (iii) integration of complex polyhedra"} \citep[p.9]{GraessleHinze2018} or special vertex bisection triangulation, as discussed in \cite{UllmannRotkvicLang2016}, are needed.

If we choose the conservative error setting $\epsilon^*<\mathcal{E}^*$, recommended in \cref{sec:ErrorAnalysis}, with for example $\epsilon^*=0.1\mathcal{E}^*$ and truncate all modes for which $\errPOD(\epsilon^*,r)\le \mathcal{E}^*$ we obtain the results shown in the upper right \cref{fig:vort-PODerror}. This is essentially the same figure as shown in the left, but all points are excluded which do not fulfill the conservative error criterion. With the help of the conservative error setting, we are able to control the errors introduced by the perturbation of the eigenvalues. Therefore, the difference $|\errWPOD(\epsilon,r) - \errPOD(0,r)|$, shown in the lower right of \cref{fig:vort-PODerror}, stays below $\epsilon$.

\subsubsection{3D Case - Insect Flight}
\label{sec:3DInsectsFlight}
   The data come from a three dimensional, highly resolved block-based adaptive simulation of a bumblebee in forward flight using \Softwarename{WABBIT} \cite{WABBIT_github}. A summary of relevant parameters is given in \cref{tabl:vortexParams}. Additional details of the adaptive flight
  simulation can be found in \cite{EngelsSchneiderReissFarge2019}.

\begin{figure}[htp!]
    \centering
    \begin{subfigure}[t]{.48\textwidth}
        \centering
        \includegraphics[width=1\textwidth]{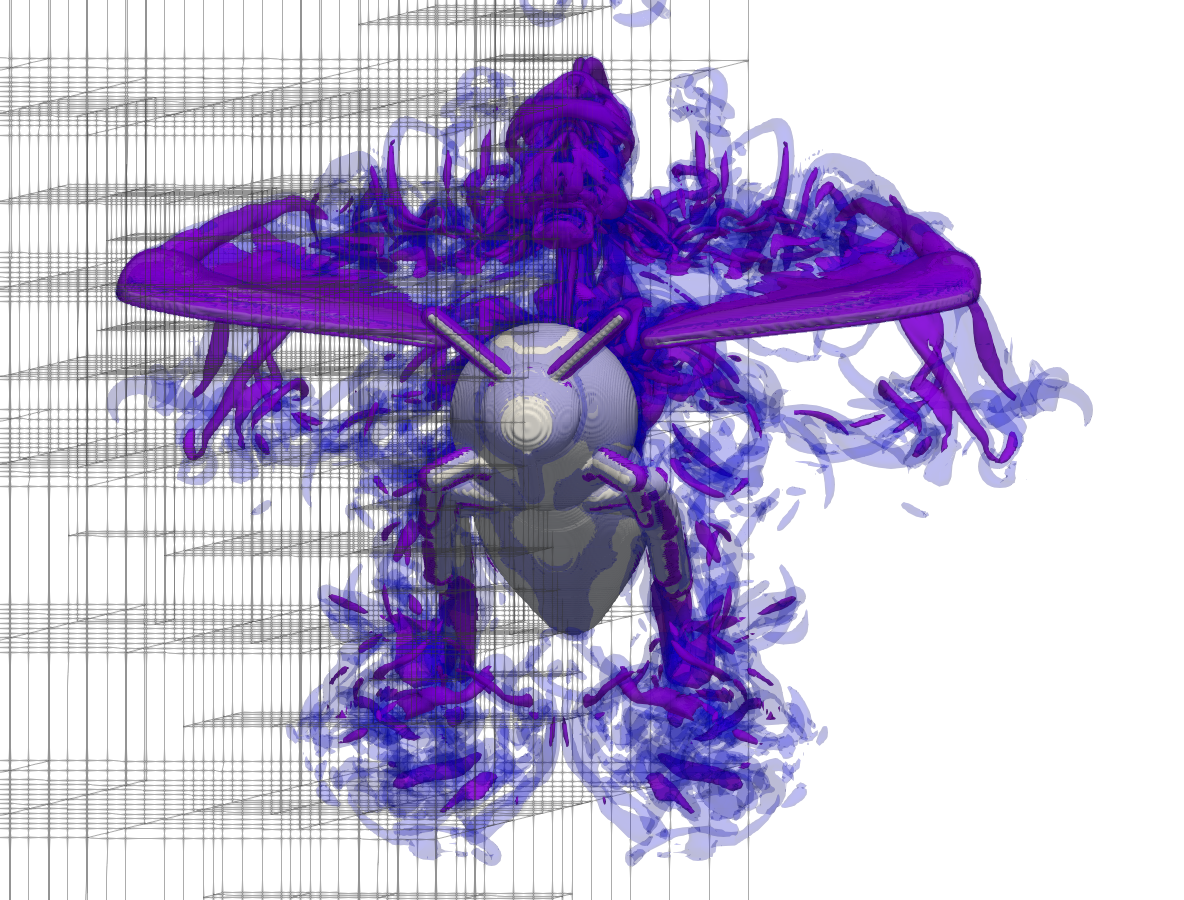}
        \caption{Reference snapshot}
        \label{fig:bubmlebee-reconst_full}
    \end{subfigure}\hfill%
     \begin{subfigure}[t]{.48\textwidth}
        \centering
            \includegraphics[width=1\textwidth,trim=150 100 100 100, clip]{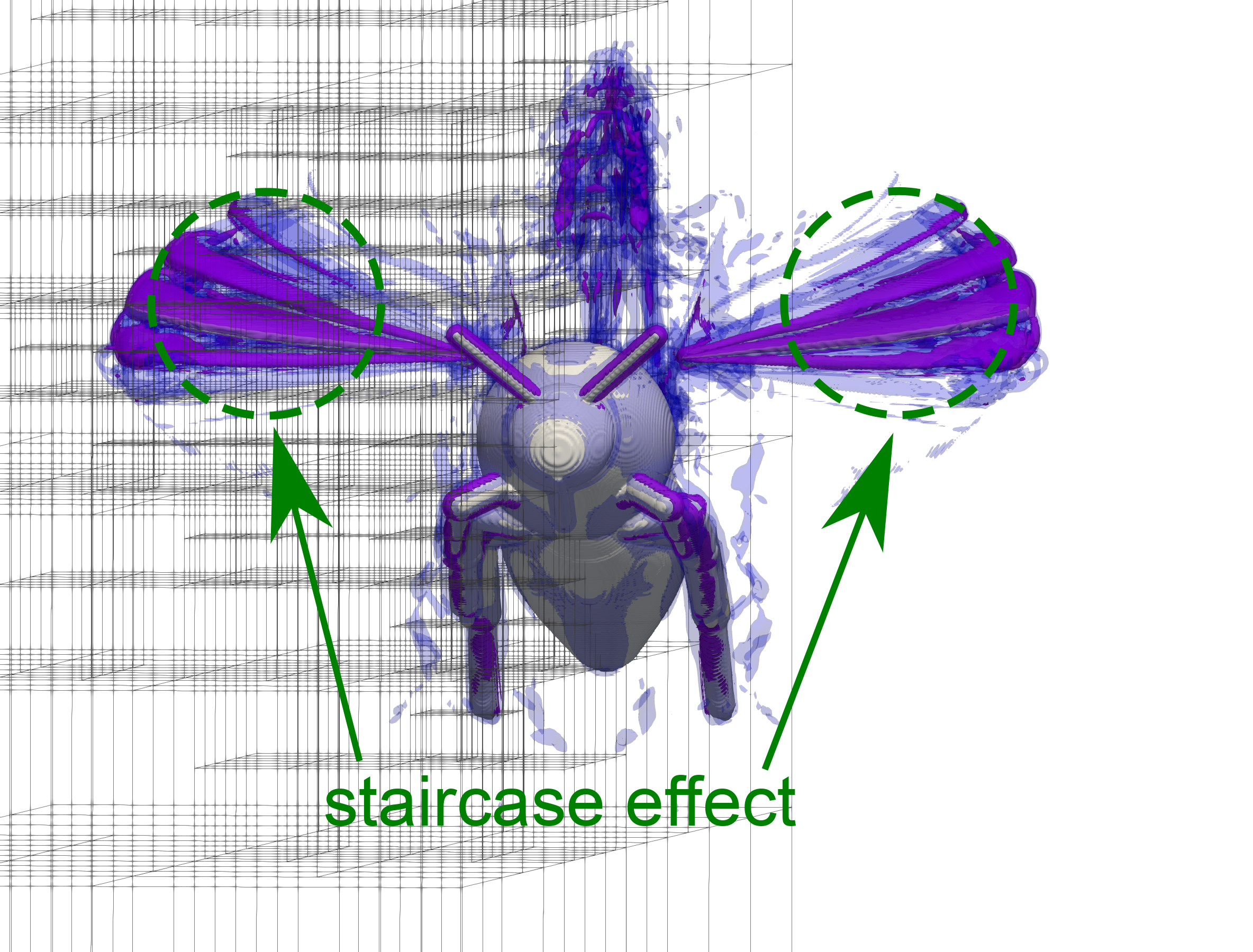}%
        \caption{Reconstruction with 5 modes}
        \label{fig:bubmlebee-reconst_5modes}
    \end{subfigure}
    \begin{subfigure}[t]{.48\textwidth}
        \centering
        \vspace{0pt}
        \includegraphics[width=1\textwidth,trim=150 100 100 100, clip]{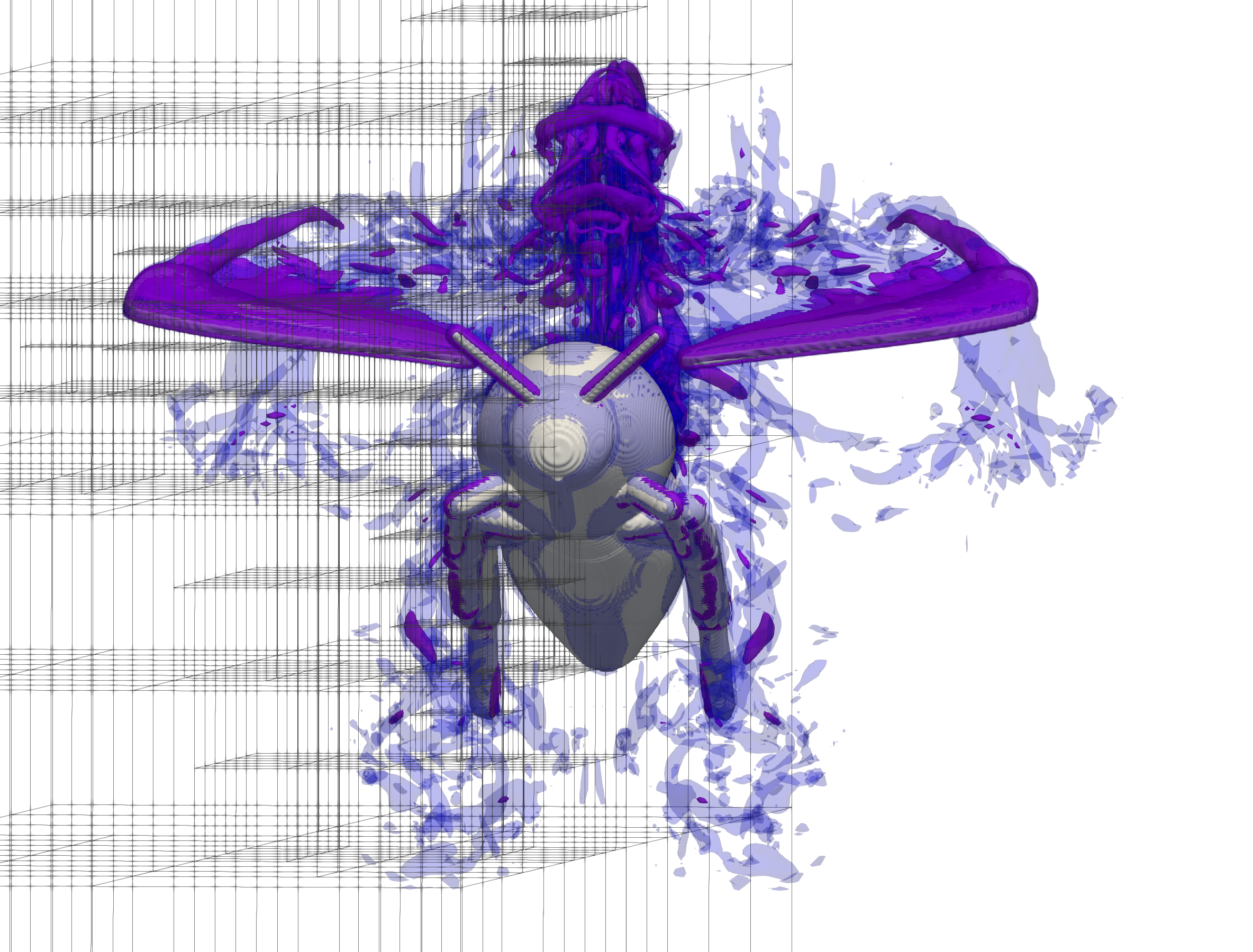}
            \caption{Reconstruction with 15 modes}
            \label{fig:bubmlebee-reconst_15modes}
    \end{subfigure}\hfill%
    \begin{minipage}[t]{.48\textwidth}
    \caption{Comparison between a bumblebee snapshot at time $t_i, i=13$ a)  and its POD reconstruction $\tilde\u^\epsilon_i$ using 5 modes b) and 15 modes c) with $\epsilon=0.01$.
    Shown are two isosurfaces of the magnitude of vorticity, i.e., $\Vert\curl(\vec{v})\Vert_2 =50$  and $100$.
    The rigid body of the bumblebee is displayed in gray. The grid structure is indicated behind. Staircase effects at the wings are indicated in \cref{fig:bubmlebee-reconst_5modes}. These artifacts appear when POD is applied to sharp structures or discontinuities which move.}
    \label{fig:bumblebee-reconst}
    \end{minipage}
  \end{figure}

\begin{figure}[htp!]
	  \centering
          \includegraphics[width=0.48\textwidth,trim=14cm 8cm 14cm 3cm,clip]{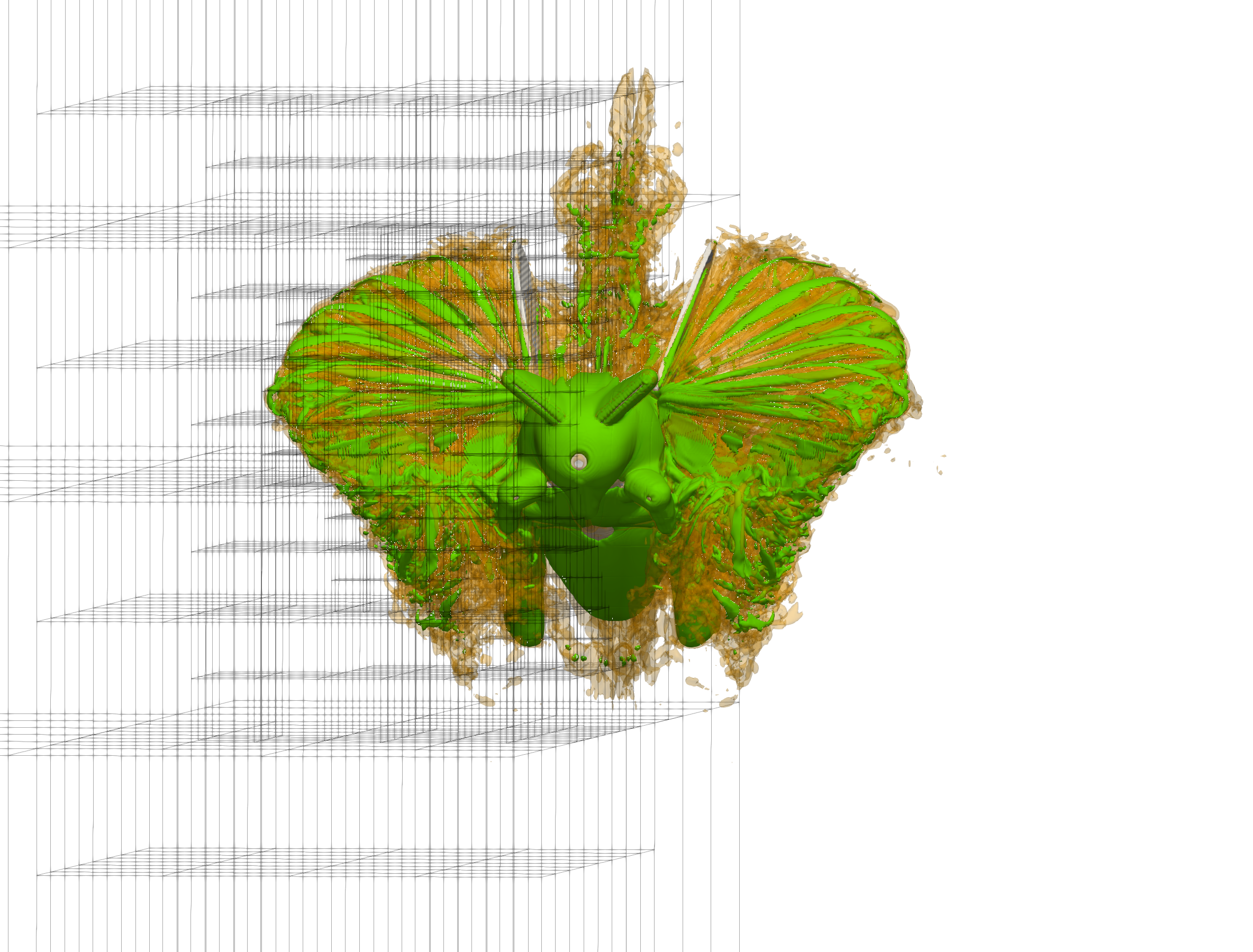}
          \includegraphics[width=0.48\textwidth,trim=14cm 8cm 14cm 3cm,clip]{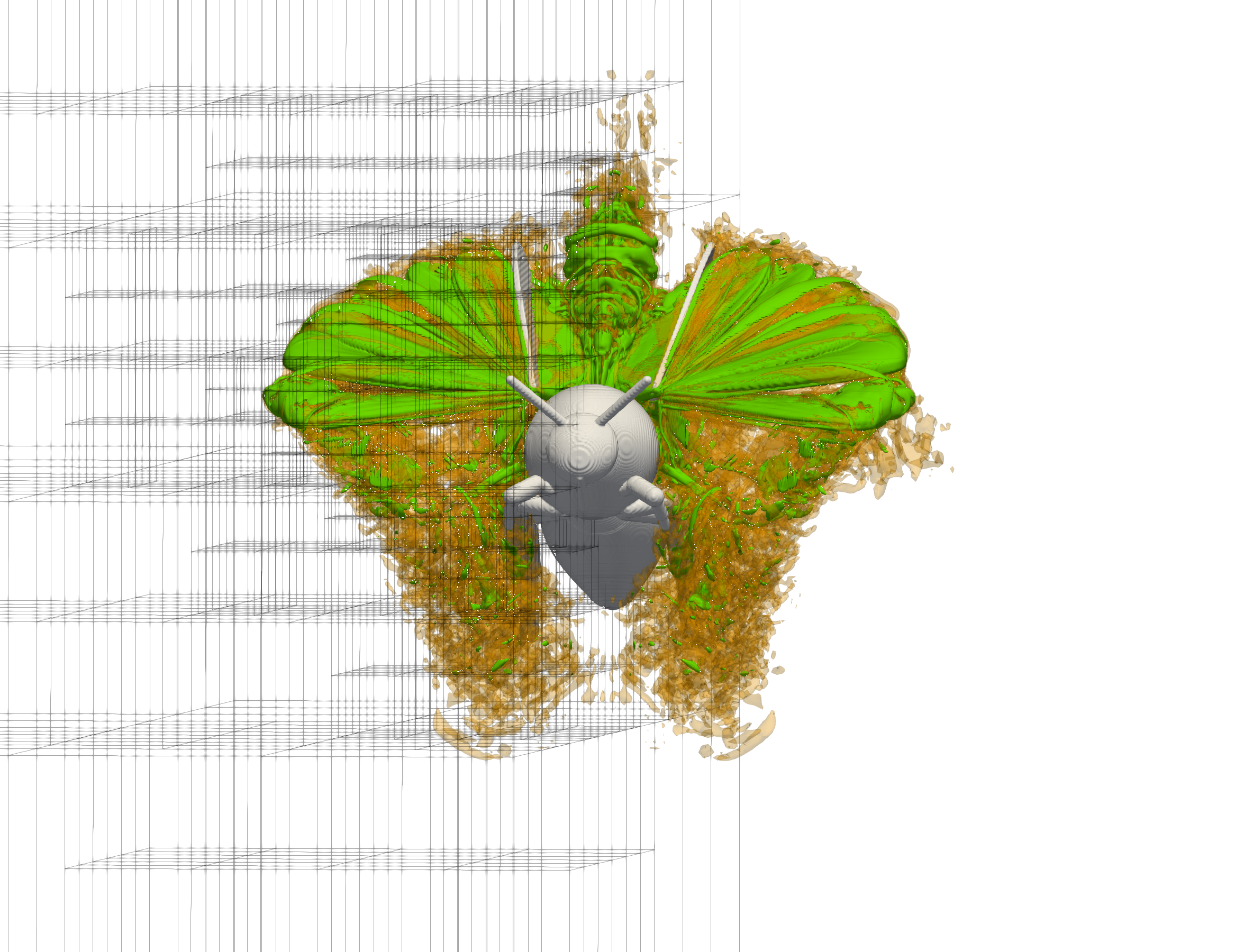}
          \includegraphics[width=0.48\textwidth,trim=14cm 8cm 14cm 3cm,clip]{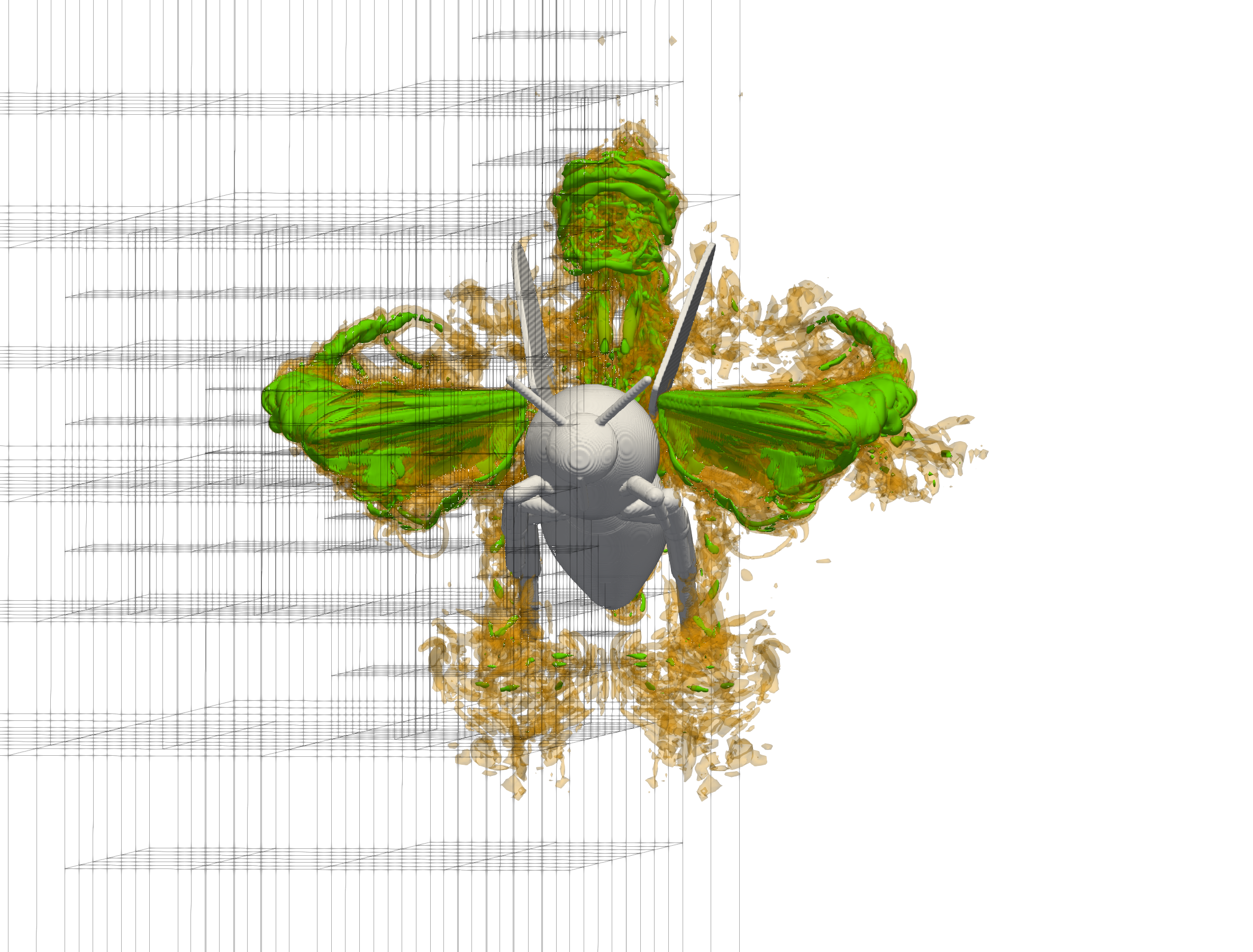}
          \includegraphics[width=0.48\textwidth,trim=14cm 8cm 14cm 3cm,clip]{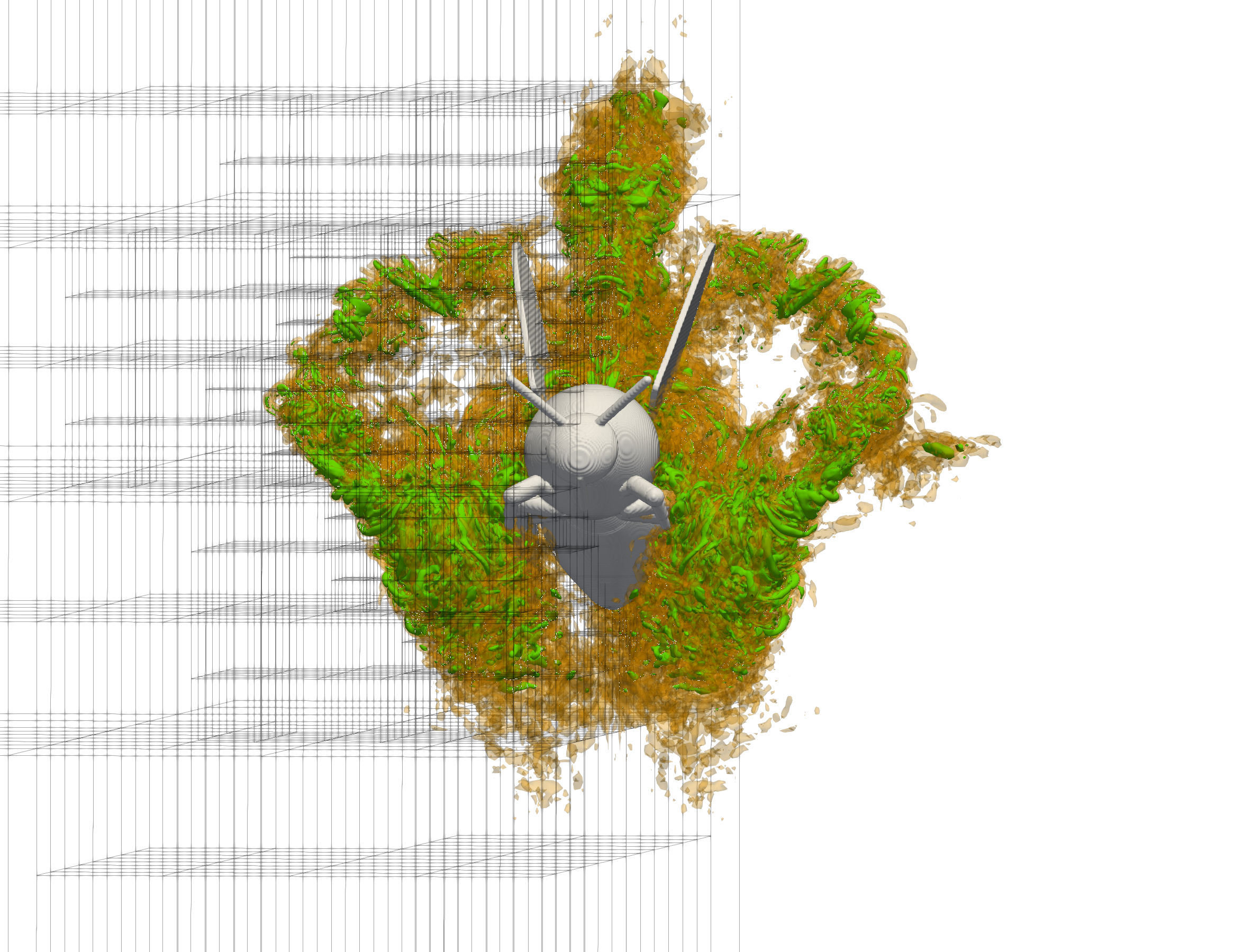}
          \caption{Bumblebee modes $\Mode_i^\epsilon$, with $i=1,9,18,27$ and $\epsilon=\num{1e-4}$ in row major order. The modes are visualized as isosurfaces of the magnitude $\norm{\Mode_i^\epsilon}_2=10,20$ with colors green and orange, respectively. For reference, the rigid body of the insect is shown in gray. The adaptive grid is only indicated on the left of each figure.}
          \label{fig:bumblebee-modes}
\end{figure}

  One representative snapshot is shown in \cref{fig:bubmlebee-reconst_full} together with the reconstruction of our algorithm using either 5 (\cref{fig:bubmlebee-reconst_5modes}) or 15 modes (\cref{fig:bubmlebee-reconst_15modes}). Additionally we plot some selected modes in \cref{fig:bumblebee-modes}. The wPOD algorithm is applied to the vorticity vector $\vec{w}=\curl(\vec{v})$, which is computed from the velocity $\vec{v}$. Two isosurfaces of the magnitude of the vorticity, $50$ and $100$, are shown in \cref{fig:bumblebee-reconst,fig:bumblebee-modes}.

\begin{figure}[htp!]
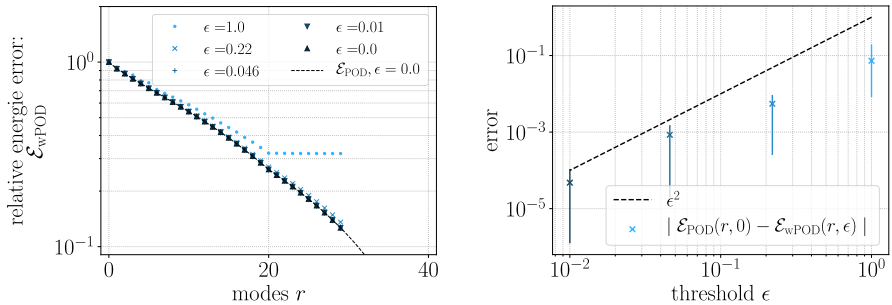

    \includesvg[width=0.51\linewidth]{bumblebee_wPODerror_J7}
    \includesvg[width=0.48\linewidth]{bumblebee_deltaPODerror_J7}
    \caption{Total relative error , $\errWPOD$, of the wPOD (left) and difference of the wPOD and POD error, $\Delta \mathcal{E}(r,\epsilon)=\abs{\errPOD(r,0)-\errWPOD(r,\epsilon)}$ (right).
    The error bars in the right figure indicate the minimal and maximal value of $\Delta \mathcal{E}$ for all ranks $1\le r\le30$ and the markers the mean $\overline{\Delta \mathcal{E}}(\epsilon)=1/30\sum_{r=1}^{30} \Delta \mathcal{E}(r,\epsilon)$.}
    \label{fig:bumblebee-wPODerror}
  \end{figure}

  The moving wing geometry causes large gradients of the flow field at the interfaces of the object. This large gradients move with the flapping wing and cause major problems to the POD, like staircase effects (see \cref{fig:bubmlebee-reconst_5modes}) of the reconstructed field with slowly decaying energy error (see \cref{fig:bumblebee-wPODerror}).
  This drawback of the POD is known for transport dominated fields with large gradients
  and is theoretically studied in \cite{OhlbergerRave2015,GreifUrban2019} with help of the Kolmogorov $n$-width. Although methods like \cite{KaratzasBallarinRozza2019} are available for
  parametric moving discontinuities, we obviate from using them here, since it is not within
  the scope of this work. However, wavelet adaptation reduces the amount of computational resources needed in favor of additional accuracy, like increased number of modes.
  In fact, for the data presented here ($\Jmax=7,\Bs=23,\epsilon^\infty=0.01,N_\text{blocks}\le 8000$) the factor in memory savings in comparison to the dense grid ($N_\text{blocks}=2^{3\Jmax}=2097152$) is larger than 260. This factor can be further increased, when increasing $\epsilon$. A full POD would be prohibitive 
  because of its tremendous
memory demand of approximately 31 TB ($N_\mathrm{b}=2^{3\Jmax}\NSnapshots$ Blocks with 0.4 MB each).
It should be further noted that, to the best of our knowledge, all previous results in the literature including \cite{UllmannRotkvicLang2016,GrassleHinzeLangUllmann2019,GraessleHinze2018,FangPainNavonPiggotGormanAllisonGoddard2009,CastrillonAmaratunga2002,KaratzasBallarinRozza2019} have been only applied to 1D or 2D cases.

 The statements about the error made in \cref{sec:ErrorAnalysis} also hold for the adaptive data:
 Here the slowly decaying eigenvalues are rather large compared to their perturbation. Hence, $\mathcal{M}_r$ is negligible and the total error behavior is mainly dominated by $\errPOD(r,0)$ and $\errWavelet(\epsilon)$ (see \cref{fig:bumblebee-wPODerror}, left).
 For the case of $\epsilon=1.0$, it can be nicely seen that the truncation error dominates the total error, after it falls below  the compression error plateau $\errWavelet(\epsilon)^2$ at $r\ge20$.
 {In \cref{fig:bumblebee-wPODerror} (right)} the difference between the total error $\errWPOD$ and the POD truncation error $\errPOD$ is shown. Note that for adaptive input data of the bumblebee $\errPOD(0,r)$ can be only assessed approximately, since wavelet details have been already discarded during the generation of the data. Hence for the adaptive case $\errPOD(0,r)$ means that no additional compression errors were introduced during the wPOD algorithm. The difference scales quadratically with the compression error $\errWavelet(\epsilon)^2\sim \epsilon^2$, drawn as dashed line in  \cref{fig:bumblebee-wPODerror} (right).
From these results we thus can conclude that our algorithm is able to reproduce the POD eigenvalue spectra of the original data, even for larger thresholds $\epsilon>0.01$ within a predefined precision given by the squared wavelet compression error.
As shown in our synthetic test case, this would be a challenge for the randomized SVD,
since the eigenvalues decay slowly.

%% file: conclusion.tex
\section{Conclusion and Outlook}
\label{sec:conclusion_and_outlook}

In this paper we presented a novel method to calculate the proper orthogonal decomposition for two or three-dimensional data, typically velocity or vorticity fields, given either on  equidistant or block-based adaptive grids, the later obtained with wavelet adaptation. The so-called wPOD algorithm endows the method of snapshots, proposed by {Sirovich \cite{SiL1987}} with nonlinear approximation for computing the POD efficiently. Our method makes use of a multiresolution framework called \Softwarename{WABBIT} to spatially adapt vector fields and thus generating sparse POD modes {using wavelet compression on block-based grids.}

The introduced compression errors of the POD basis are well controlled using nonlinear approximation by means of wavelet thresholding. While the compression error depends linearly on the chosen threshold $\epsilon$, the total energy error of the wPOD procedure scales with $\epsilon^2$. 
The wavelet compression and POD truncation errors can thus be balanced, assuming that the perturbation of the eigenvalues remains sufficiently small. 
{In comparison to the classical POD, the compression results in overall savings in memory and computational resources, while obtaining approximately the same error}. As a consequence, data from highly resolved 3D direct numerical simulation computed on massively parallel platforms can be processed on much smaller systems.

However, these savings in terms of computational effort are accompanied by additional
{overhead} for handling the tree-like data structure. Hence, the wPOD cannot yet compete with equivalent randomized techniques \cite{YuChakravorty2015, HalkoMartinssonTropp2011} in terms of CPU time, if the memory is not a limiting factor. Nevertheless, we were able to show that our method can handle data efficiently, when the singular values
decay slowly. In this case, the proposed wPOD  does indeed outperform the rSVD.
Furthermore, our method expresses the correlation matrix in terms of the underlying wavelet basis, similar to what has been done {in the context of} finite elements in \cite{GraessleHinze2018}. However, using the scaling relations of wavelets allows to express the inner products (\cref{eq-def:scalar-product}) effectively and thus avoids many problems associated with finite element schemes, cf. listed caveats in \cite[p.6]{GraessleHinze2018}.

The framework could be extended to non-periodic grids using boundary {adapted} wavelets (see e.g. \cite{SakuraiYoshimatsuSchneiderFargeMorishitaIshihara2017} for applications to channel flows) as well as to other algorithms like DMD {\cite{Schmid2010}, shifted POD \cite{ReS2018} or multiscale POD \cite{MendezBalabaneBuchlin2018}. Furthermore, our framework provides interesting perspectives for reduced order models introducing} adaptive POD-Galerkin simulations. It has the potential to become a practical tool for turbulence research when combining it with wavelet denoising for coherent vortex structures \cite{FargePellegrinoSchneider2001,FargeKevlahanSchneider1999} as we are able to split {the flow into coherent structures and incoherent background noise} before computing a POD basis.

Finally, in combination with wavelet denoising, the proposed wPOD technique could be likewise applied to
highly resolved flow images obtained experimentally, e.g. using particle image velocimetry or Schlieren imaging.

%% file: appendix.tex
\renewcommand{\ucoef}{\underline{{\mathsf{u}}}}
\addcontentsline{toc}{section}{Appendices}
\normalsize
\section*{Appendix}

\section{Block-Based Wavelet Adaptation}
\label{appx:block-based_wavelet_adaptation}

\subsection{Refinement and Coarsening}

For the wavelet adaptation scheme, here illustrated for the two-dimensional case, we assume real valued and continuous $L^2$-functions $u(x,y)$, such as the pressure or velocity component of a flow field.
The function is sampled on a block-based multiresolution grid with maximum tree level
$\Jmax$.
The sampled values on a block $\Block_p^j$ are denoted by:
\begin{align}
  &x^{j}_{p,k_1} = {x}_p + k_1 \Delta x^j &k_1=-g,\dots ,\Bs_1+g\\
  &y^{j}_{p,k_2} = y_p +  k_2 \Delta y^j &k_2=-g,\dots ,\Bs_2+g\\
  &\ucoef^{j}[p,k_1,k_2]   \defeq u(x^{j}_{p,k_1},y^{j}_{p,k_2} ) \label{eq:blocksamples}
\end{align}

\begin{figure}[h]
  \centering
  \input{refinement}
  \caption{Dyadic grid refinement and coarsening of a single block. Refinement: First the block $\Block_p^j$ is refined by midpoint insertion and then split into four new blocks $(\Block_{p0}^j,\Block_{p1}^j,\Block_{p2}^j,\Block_{p3}^j)$. Coarsening: After a low pass filter is applied all midpoints are removed and merged into one block.}
  \label{fig:dyadicRefinement}
\end{figure}
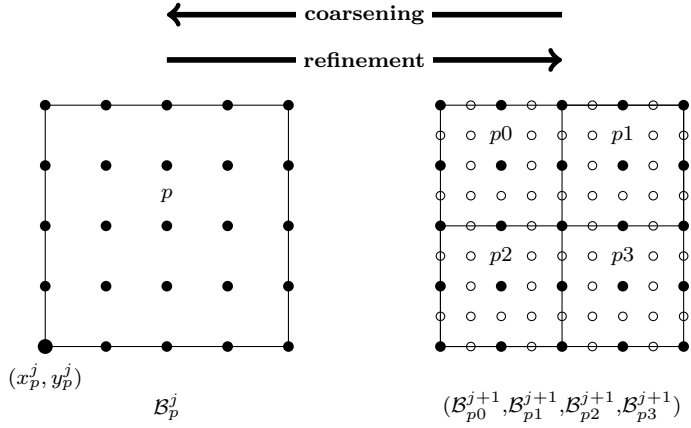

For a block \textit{refinement} $j\to j+1$ the lattice spacings are halved and dyadic
points are added to the block, as shown in \cref{fig:dyadicRefinement}.
The values at the refined blocks can be obtained by the refinement relation:
\begin{align}
 \widehat{\ucoef}^{j+1} [p,l_1,l_2] = \sum_{k_1=-g}^{\Bs_1+g}\sum_{k_2=-g}^{\Bs_2+g} \,
  h_{l_1-2k_1} h_{l_2-2k_2}\, \ucoef^{j} [p,k_1,k_2]\label{eq:int2D}
\end{align}
where $h_k$ denotes the weights of the one dimensional interpolation scheme:
\begin{equation}
  \label{eq:refrel}
  \varphi(x) = \sum_{k=-N}^N h_{k} \varphi(2x-k).
\end{equation}
In the following, we will refer to \cref{eq:int2D} as the \textit{prediction operation} $P_j^{j+1}:\ucoef^j\mapsto \widehat{\ucoef}^{j+1}$ as it was introduced for point value multiresolution in \cite{Harten1993}.
Using \cref{eq:refrel} one can show that \cref{eq:int2D}
is equivalent to the continuous refinement relation:
\begin{align}
  \label{eq:lagrange2D}
   \widehat{u}^{(p)}(x,y)& =\sum_{k_1=-g}^{\Bs_1+g}\sum_{k_2=-g}^{\Bs_2+g}  \ucoef^{j}[p,k_1,k_2]\varphi^j_{p,k_1,k_2}(x,y)\,,\\ \text{where}&\quad
  \varphi^j_{p,k_1,k_2}(x,y) = \varphi\left(\frac{x-x_{p,k_1}^{j}}{\Delta x^j}\right)\varphi\left(\frac{y-y_{p,k_2}^{j}}{\Delta y^j}\right)
\end{align}
is the two dimensional tensor product one dimensional \textit{interpolating scaling functions} $\varphi(x)$.

When a block is \textit{coarsened}, the tree level is decimated by one: $j+1\to j$ and
every second grid point is removed. The values at the coarser level are obtained by the coarsening relation:
\begin{align}
 \ucoef^{j-1} [p,l_1,l_2] = \sum_{k_1=-g}^{\Bs_1+g}\sum_{k_2=-g}^{\Bs_2+g} \,
  \tilde{h}_{2l_1-k_1} \tilde{h}_{2l_2-k_2}\, \ucoef^{j} [p,k_1,k_2]\label{eq:coarse}
\end{align}
In the notation of Harten \cite{Harten1993} coarsening is called \textit{decimation} and \cref{eq:coarse} is denoted by $D_{j}^{j-1}\colon \ucoef^{j}\to \ucoef^{j-1}$ in the following. After decimation the block will be merged with its neighboring blocks, as shown in \cref{fig:dyadicRefinement}. Similar to the continuous refinement relation \cref{eq:lagrange2D} there is a continuous counterpart for coarsening: the \textit{dual scaling function} $\tilde{\varphi}$,
which satisfies
\begin{equation}
  \label{eq:dualscalingf}
  \tilde{\varphi}(x) = \sum_{k=-N}^N \tilde{h}_{k} \tilde{\varphi}(2x-k).
\end{equation}

Instead of using \textit{Deslauriers-Dubuc} (DD) wavelets, as in the framework of Harten \cite{Harten1993},  we use lifted Deslauriers-Dubuc wavelets, i.e. \textit{biorthogonal Cohen--Daubechies–-Feauveau wavelets} of fourth order (CDF 4,4) with filter coefficients $h_{k}$ and dual filter coefficients $\tilde{h}_k$ listed in \cref{tabl:dd24}. The lifted DD wavelets allow a better scale separation and can be easily implemented replacing the loose downsampling filter by a low pass filter before coarsening the grid.

\begin{table}[htpb]
  \centering
  \setlength\tabcolsep{2pt}
  \begin{tabular}{ccccccccccccccc}
    \toprule
    continuous  & $k$     &    -6&-5&-4 & -3  &-2   & -1    & 0& 1&   2   & 3 & 4 & 5 & 6\\
    \midrule
    $\tilde{\varphi}$ &$\tilde{h}_k$ &$-\frac{1}{256}$  &0 &$\frac{9}{128}$ & $-\frac{1}{16}$  & $-\frac{63}{256}$ & $\frac{9}{16}$& $\frac{87}{64}$& $\frac{9}{16}$ &$-\frac{63}{256}$ & $-\frac{1}{16}$ & $\frac{9}{128}$ &0 &$-\frac{1}{256}$\\\addlinespace
    $\varphi$ &$h_k$  &&& & $-\frac{1}{16}$ &  0     & $\frac{9}{16}$ & 1& $\frac{9}{16}$& 0 & $-\frac{1}{16}$ &&&\\
    \bottomrule
  \end{tabular}
  \caption{Filter coefficients $h_k$ and dual filter coefficients $\tilde{h}_k$ of the Cohen-Daubechies–Feauveau (CDF 4,4) wavelets applied in the prediction/restriction operation. }
  \label{tabl:dd24}
\end{table}

\subsection{Computing Wavelet Coefficients and Adaptation Criterion}

With the definition in \cref{eq:int2D,eq:coarse} we have introduced a biorthogonal multiresolution basis $\{\tilde{\varphi}^j_{p,k_1,k_2},\allowbreak\varphi^j_{p,k_1,k_2}\}$, which can approximate any continuous function $u\in L^2(\Domain)$
arbitrary close. This aspect is the main property of
a multiresolution analysis and can be used to relate and compare samples at
different resolutions, i.e. different scales.

The difference between two consecutive approximations can be represented by a wavelet with its corresponding coefficients
\begin{align}
\label{eq-def:detail}
      d^j[p,k_1,k_2]
      =  \left[u^j -  P_{j-1}^{j}D_{j}^{j-1} u^j\right]_{p,k_1,k_2}
      =  \ucoef^j[p,k_1,k_2] -  \widehat \ucoef^j[p,k_1,k_2]
\end{align}
known as \textit{wavelet details}.

As shown in one space dimension by Unser in \cite{Unser1996} for biorthogonal wavelets the difference $d^j$ between two consecutive approximations is bounded for sufficiently smooth $L^2$ functions $u$. The bound depends on the local regularity of $u$ and the order $N$ (here $N=4$) of the scaling function:
\begin{equation}
  \norm{d^j} =  \norm{u - \widehat u} \le C_{\varphi,\tilde{\varphi}} (\Delta x^j)^N \norm{u^{(N)}}
  \label{eq:detail}
\end{equation}
where $\Delta x^j\sim 2^j$ is the step size, $C_{\varphi,\tilde{\varphi}}$ is a constant independent of $u$. This result can be understood as an  interpolation error, since the CDF4,4 filter can be viewed as an interpolation of the averaged data $D_j^{j-1}\ucoef^j$.
From \cref{eq:detail} we can thus {conclude}:
since the approximation error of the interpolation scheme depends on the smoothness
of the sampled function and the lattice spacing on the block, we can
increase the local lattice spacing of the block for blocks where the function is smooth
and keep the fine scales only for blocks where $ u_p$ is not smooth. This is achieved by coarsening
the block, as decreasing the tree level $j$ increases the lattice spacing. For an intended approximation error we therefore define the \textit{wavelet threshold} $\epsilon$
together with the \textit{coarsening indicator} $i_\epsilon(\Block_p^j)$:
\begin{equation}
  i_\epsilon(\Block_p^j) \defeq
  \begin{cases}
    1 &\text{ if} \max\limits_{
    \substack{k_1=0,\dots,\Bs_1\\k_2=0,\dots,\Bs_2}}
    \mid d^j[p,k_1,k_2]\mid<\epsilon\\
    0 & \text{otherwise}
  \end{cases}
\end{equation}
which marks the block for coarsening.
For vector valued quantities $\u=(u_1,u_2,\dots,u_K)$,
a block will be coarsened only if all components indicate coarsening.
The pseudocode in \cref{algo:w-adapt}
outlines the wavelet adaptation algorithm for vector-valued quantities.
For the sake of completeness we will put this algorithm in relation to the underlying
wavelet representation.
\begin{algorithm}[htp!]
 	\caption{Wavelet adaptation}
 	\label{algo:w-adapt}
  \begin{algorithmic}[1]
    \Require $ \u\colon\Grid\to \mathbb{R}^K$ where $\Grid\subset \mathbb{R}^d$ defined in \cref{eq-def:MultiresolutionMesh}
    and $d\in\{2,3\},\, K\in\mathbb{N}$
    \Require threshold $\epsilon \ge 0$ and minimal and maximum tree level $\Jmin,\Jmax$
    \vspace{0.2cm}
    \Function{\texttt{adapt}}{$ \u, \epsilon,\Jmin,\Jmax$}
      \State set $N_\text{blocks} = 0$
      \While{$\abs{\bigcup_j\Lambda^j}\ne N_\text{blocks}$} \Comment {\textit{coarsening is stopped if number of blocks has not changed}}
            \State Synchronize all ghost layers of the blocks
            \State Count the number of active blocks $N_\text{blocks}$
		        \ForAll{$p\in \bigcup_j \Lambda^j$}   \Comment \textit{loop over all active treecodes}
               \State {Compute the refinement indicator of the block:}

                  \State1) normalize every state vector component:
                  \State  $ \,$ $(\u^{(p)})_i\leftarrow  (\u^{(p)})_i/\norm{ (\u^{(p)})_i}$, $i=1,\dots,K$
                  \State2) compute $ \widehat \u^j_p$ using \cref{eq:int2D,eq:coarse} in a restriction and prediction step
                  \State3) compute the detail coefficients of all state vector components with \cref{eq:detail}
		              \vspace{0.2cm}

            \If{$i_\epsilon(\Block_p) = 1$ and $\Jmin \le j-1$}
		            \State tag the block $\Block_p$ for coarsening
		        \EndIf
		     \EndFor
         \State  Remove coarsening tag from blocks,
                if coarsening would result
                in a non graded
                \State grid.
         \State Coarse all blocks taked for coarsening
	   \EndWhile
     \State Balance the load between processors
     \State \textbf{return} $ \u^\epsilon$ \Comment \textit{ wavelet filtered field}
    \EndFunction
 	\end{algorithmic}
 \end{algorithm}

\subsection{Wavelet Representation in the Continuous Setting} 
For completeness of this manuscript we give a detailed description of the underlying wavelet representation in a concise way.
For the interested reader we recommend Ref. \cite{SweldensSchroeder1997} for a succinct introduction to biorthogonal wavelets and to Refs. \cite{Mallat1999,Donoho1992,DeslauriersDubuc1987,DeslauriersDubuc1989} for a more detailed description.

Assuming that we have sampled a continuous function $u\in L^2(\Domain)$ on an equidistant grid corresponding to refinement level $\Jmax$, we can block-decompose it in terms of \cref{eq:blocksamples}.
By choosing $j=\Jmax$ in \cref{eq:lagrange2D} and summing over all blocks $p$, we can thus represent $u$ in $L^2$ using a basis of dilated and translated scaling functions $\{\varphi_{p,k_1,k_2}^j\}$:
\begin{equation}
  \label{eq:finestscale}
  u=\sum_{p\in\Lambda^\Jmax} u^{(p)}=\sum_{\lambda\in\overline{\Lambda}^{\Jmax}} c_{\lambda}
 ^{\Jmax} \varphi_{\lambda}^{\Jmax}\,.
\end{equation}
Here we have introduced a multi-index $\lambda=(p,k_1,k_2)\in\overline{\Lambda}^j\defeq\Lambda^j\times\{0,\dots,\Bs_1\}\times\{0,\dots,\Bs_2\}$ for ease of notation.
With this notation we can rewrite \cref{eq:finestscale} in a wavelet series
\begin{equation}
  \label{eq:scalarfield}
   u =\sum_{\lambda\in\overline{\Lambda}^{\Jmin}} c_{\lambda}
  ^{\Jmin} \varphi_{\lambda}^{\Jmin} + \sum_{j=\Jmin}^{\Jmax-1} \sum_{\lambda\in\overline{\Lambda}^{j}}\sum_{\mu=1}^3
  d^j_{\mu\lambda}\psi_{\mu,\lambda}^j,
\end{equation}
where
the interpolating scaling basis $\{\varphi_{\lambda}^{\Jmin}\}_{\lambda\in\overline{\Lambda}^{\Jmin}}$ approximates $u$ at
the coarsest scale $\Jmin$ and the wavelet basis $\{\psi_{\mu,\lambda}^j\}_{\mu=1,2,3,\lambda\in\overline{\Lambda}^{j\ge\Jmin}}$ contains all the additional information necessary to construct $u$.

In the CDF4,4 setting we have biorthogonal scaling functions:
 \begin{align*}
   &\varphi_{\lambda}^j(x,y)=\varphi\left(\frac{x - x^j_{\lambda}}{\Delta x_{p}^j}\right)\varphi\left(\frac{y - y^j_{\lambda}}{\Delta y_{p}^j}\right),\, 
    \tilde{\varphi}_{\lambda}^j(x,y)=\tilde\varphi\left(\frac{x - x^j_{\lambda}}{\Delta x_{p}^j}\right) \tilde\varphi\left(\frac{y - y^j_{\lambda}}{\Delta y_{p}^j}\right)\quad
\\ &  \quad\text{with}\quad \sprod{\varphi^{j}_{\lambda_1}}{\tilde{\varphi}^{j}_{\lambda_2}}=\delta_{\lambda_1,\lambda_2}
 \end{align*} 
 and the associated three biorthogonal wavelets (in the $d$ -dimensional case we have $2^d-1$ wavelets see \citep[p.42]{van2010})
  \begin{align*}
   &\psi_{1,\lambda}^j(x,y)=\psi\left(\frac{x- x^j_{\lambda}}{\Delta x_{p}^j}\right)\varphi\left(\frac{y- y^j_{\lambda}}{\Delta y_{p}^j}\right),\,
   \tilde\psi_{1,\lambda}^j(x,y)=\tilde\psi\left(\frac{x- x^j_{\lambda}}{\Delta x_{p}^j}\right)\tilde\varphi\left(\frac{y- y^j_{\lambda}}{\Delta y_{p}^j}\right)\\
   &\psi_{2,\lambda}^j(x,y)=\varphi\left(\frac{x- x^j_{\lambda}}{\Delta x_{p}^j}\right)\psi\left(\frac{y- y^j_{\lambda}}{\Delta y_{p}^j}\right),\,
   \tilde\psi_{1,\lambda}^j(x,y)=\tilde\varphi\left(\frac{x- x^j_{\lambda}}{\Delta x_{p}^j}\right)\tilde\psi\left(\frac{y- y^j_{\lambda}}{\Delta y_{p}^j}\right)\\
   &\psi_{3,\lambda}^j(x,y)=\psi\left(\frac{x- x^j_{\lambda}}{\Delta x_{p}^j}\right)\psi\left(\frac{y- y^j_{\lambda}}{\Delta y_{p}^j}\right),\,
   \tilde\psi_{1,\lambda}^j(x,y)=\tilde\psi\left(\frac{x- x^j_{\lambda}}{\Delta x_{p}^j}\right)\tilde\psi\left(\frac{y- y^j_{\lambda}}{\Delta y_{p}^j}\right)
\\ &  \quad\text{with}\quad \sprod{\psi^{j_1}_{\mu_1,\lambda_1}}{\tilde{\psi}^{j_2}_{\mu_2,\lambda_2}}= \delta_{\mu_1,\mu_2}\delta_{\lambda_1, \lambda_2}\delta_{j_1, j_2}
 \end{align*}
for the horizontal ($\mu=1$), vertical ($\mu=2$) and diagonal ($\mu=3$) direction.
The wavelet and its dual are defined by the same scaling relations as $\varphi$ and $\tilde\varphi$:
\begin{equation}
  \label{eq:dualwavelet}
  {\psi}(x) = \sum_{k=-N}^N {g}_{k} {\psi}(2x-k) \quad \text{and}\quad
  \tilde{\psi}(x) = \sum_{k=-N}^N \tilde{g}_{k} \tilde{\psi}(2x-k).
\end{equation}
The filter coefficients are given by
$g_k=(-1)^{1-n}\tilde{h}_k$ and $\tilde g_k=(-1)^{1-n}{h}_k$ with $h_k,\tilde{h}_k$ listed in \cref{tabl:dd24}.
 The components of the scaling and wavelet coefficients  ($c_\lambda^j$ and $d_\lambda^j$) are determined via component-wise
 projection of \cref{eq:scalarfield} onto the dual basis $\{\tilde\varphi^j_\lambda, \tilde\psi^j_{\mu\lambda}\}$:
 \begin{align}
   \label{eq:coef}
   c_{\lambda}^{j} = \sprod{u}{\tilde{\varphi}_{\lambda}^j }, \qquad
   d_{\mu\lambda}^j = \sprod{u}{\tilde{\psi}_{\mu\lambda}^j }\qquad \text{where} \quad \lambda = (p,k_1,k_2)
 \end{align}
 assuming that $\sprod{\varphi^{j}_{\lambda_1}}{\tilde\psi^j_{\lambda_2}} =\sprod{\psi^{j}_{\lambda_1}}{\tilde\varphi^{j}_{\lambda_2}}=0 $
 are orthogonal.
 Note that $\sprod{a}{b}=\int_\Domain a(\vec{x})b(\vec{x})\d{\vec{x}}$ denotes the $L^2$-inner product.

In most wavelet adaptation schemes, one truncates \cref{eq:scalarfield}
such that only detail coefficients are kept which carry significant information. According to
\cite{SchneiderVasilyev2010} \textit{``this can be expressed as a non-linear filter"}, which
acts as a cut-off wavelet coefficients with small magnitude. The cut-off is given by
the threshold parameter $\epsilon >0$. However, in contrast to these schemes, our block-based adaptation groups the detail coefficients in blocks, i.~e. all details on the block are kept if at least one detail carries important information. This seems to be inefficient at first sight, because unnecessary information is kept, but grouping details in blocks is reasonable, since the block-based adaptation is computationally efficient for MPI distributed architectures. Moreover, groups of
significant details are often nearest neighbors, rather than a single significant detail in a block.
Therefore we define the set of blocks
 \begin{equation}
   I_\epsilon^j\defeq\left \{p \in \Lambda^j\mid \max\limits_{
   \substack{k_1=0,\dots,\Bs_1\\k_2=0,\dots,\Bs_2}}\abs{ d^j[p,k_1,k_2]}> \epsilon \right\}
 \end{equation}
 with significant details in the predefined tree level range $\Jmin\le j\le\Jmax$.
 In the spirit of our previous notation we thus define: $\overline{I}_\epsilon^j\defeq I_\epsilon^j\times\{0,\dots,\Bs_1\}\times\{0,\dots,\Bs_2\}$ for the set of all significant detail indices.
 The filtered block-based wavelet field in \cref{eq:scalarfield} can now be written as follows:
\begin{equation}
  \label{eq:scalarfield-threshold}
   u^\epsilon =\sum_{\lambda\in\overline{\Lambda}^\Jmin} c_{\lambda}
  ^{\Jmin} \varphi_{\lambda}^{\Jmin}+ \sum_{j=\Jmin}^{\Jmax-1} \sum_{\lambda\in \overline{I}_\epsilon^{j}}
  \sum_{\mu=0}^{3} d^j_{\mu,\lambda}\psi_{\mu,\lambda}^j
\end{equation}
In the following we will denote all fields with an upper index $\epsilon$, which
have been filtered with \cref{algo:w-adapt} and can be thus expressed as \cref{eq:scalarfield-threshold}.

For illustration we have computed the vorticity $\omega^\epsilon=\partial_x v_y^\epsilon-\partial_y v_x^\epsilon$ of a thresholded vector field $\vec{u}^\epsilon=(v_x^\epsilon,v_y^\epsilon,p^\epsilon)$ in \cref{fig:threshold-vorx} for various $\epsilon$ (more details can be found in \cref{sec:vortex_street}).
Here, $\epsilon>0$ and $\epsilon=0$ corresponds to a filtered and unfiltered field, respectively. For increasing $\epsilon$, less detail coefficients will be above the threshold and therefore
the number of blocks decreases. 
Taking the difference between the thresholded  \cref{eq:scalarfield-threshold} and the original field
\cref{eq:scalarfield} only details below the threshold are left. Hence the total error can be estimated and yields:
\begin{equation}
  \label{eq:scalarfield-err}
  \norm{u-u^\epsilon} \le
  \sum_{j=\Jmin}^{\Jmax-1} \sum_{\lambda\in {\overline{I}_\epsilon^{j}}^c}
  \sum_{\mu=0}^{3} \abs{d^j_{\mu,\lambda}}\norm{\psi_{\mu,\lambda}^j}\,.
\end{equation}
Because $\norm{\psi_{\mu,\lambda}^j}_\infty=1$ we finally get $\norm{u-u^\epsilon}_\infty\le C\epsilon$ for the total error in the $L^\infty$-norm. Similarly one can normalize $\psi_{\mu,\lambda}^j$
in the $L^2$-norm, which corresponds to re-weighting the thresholding criterion  $\abs{d^j_{\mu,\lambda}}<\epsilon$ with a level ($j$) and dimension ($d$) dependent threshold: $\abs{d^j_{\mu,\lambda}}<\epsilon_0 2^{-d(j-\Jmax)/2}\epsilon $ \cite{RousselSchneider2010}.
The additional constant $\epsilon_0 = 1$ for $d=3$ and $\epsilon_0=0.1$ for $d=2$
tunes the offset of the compression error. It is chosen such that the relative compression error $\errWavelet$ is close to, but still below $\epsilon$, i.e. it fulfills \cref{eq:errWavelet} in the $L^2$-norm.

\subsection{$L^2$ Inner Products Expressed in the Wavelet Basis} 
\label{sec:weightedInnerProduct-appendix}

\begin{figure}[htp!]
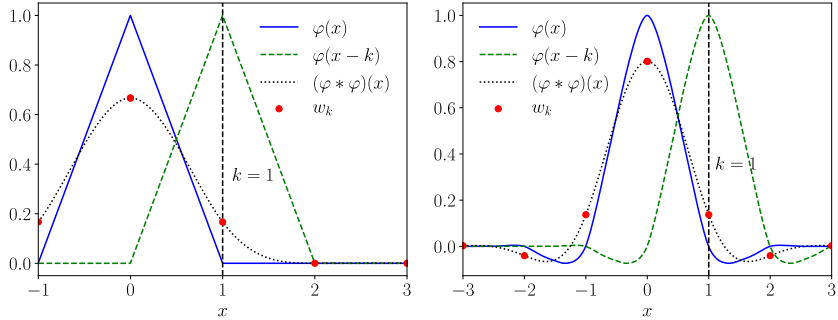

  \centering
    \includesvg[width=0.45\textwidth]{weightsDD2}
    \includesvg[width=0.45\textwidth]{weightsDD4}
  \caption{Autocorrelation functions of Deslauriers interpolating scaling functions $\varphi$ of order two (left) and order four (right).}
  \label{fig:autocorrelation}
\end{figure}

The  $L^2$ inner product is computed as a weighted
sum of two fields $\u$ and $\boldsymbol{v}$.
For this we first refine both fields onto a unified grid  with identical treecodes $\Lambda^j$,
as explained in \cref{sec:weightedInnerProduct}.
Then we are able to compute \cref{eq:weighted-inner-product} as a weighted sum over all blocks:
\begin{align}
	\label{eq:weighted-inner-product2}
		\sprod{\u}{\boldsymbol{v}}
		&=\sum_{j=1}^{\Jmax}\sum_{p\in \Lambda^j}\sprod{\u^{(p)}}{\boldsymbol{v}^{(p)}}
		=\sum_{j=1}^{\Jmax}\sum_{p\in \Lambda^j}\sprod{\boldsymbol{\ucoef}^{(p)}}{\vcoef^{(p)}}_{\boldsymbol{I}_K \otimes \boldsymbol{W}\otimes\boldsymbol{W}} \Delta x_p\Delta y_p\\
	 &\text{with weights: }(\boldsymbol{W})_{lm} =\sprod{\varphi_l^j}{\varphi_m^j}\,,\label{eq:pre-computed-weights}
\end{align}
Note that this quadrature rule is exact for $\epsilon=0$.
We denote by $\boldsymbol{I}_K \otimes \boldsymbol{W}\otimes\boldsymbol{W}$ the Kronecker product between the weight matrix $\boldsymbol{W}$ and the identity matrix $\boldsymbol{I}_K\in\mathbb{R}^{K,K}$. The weight matrix is pre-computed by \cref{eq:pre-computed-weights} and its non vanishing values $(\boldsymbol{W})_{ik}=w_{i-k}$ are shown in \cref{tabl:dd24weights}.
The listed matrix elements are the discrete values of the autocorrelation function between two compactly supported scaling functions $\varphi$, see \cref{fig:autocorrelation}. Therefore $\boldsymbol{W}$ is sparse, symmetric and circulant. Since $\boldsymbol{W}$ is also strictly diagonal dominant and all diagonal entries are positive, $\boldsymbol{W}$ and the Kronecker product of such matrices is also positive definite.
\begin{table}[htp!]
  \centering
  \begin{tabular}{cccccccc}
    \toprule
    $\abs{k}$ (Order)   & 0    & 1   &   2   & 3   & 4   & 5   \\
    \midrule
    $w_k$ ($N=2$)  & 2/3  & 1/6 &   \\\addlinespace
    $w_k$ ($N=4$)  & 0.8001, &
                        0.1370, &
                        -0.0402, &
                        0.0028, &
                        $\num{-7.5925e-05}$, &
                        $\num{-1.4829e-07}$\\
    \bottomrule
  \end{tabular}
  \caption{Values of the autocorrelation function $w_k=\int\varphi(x-k)\varphi(x)\d{x}$ of Deslauriers Dubuc interpolating functions
  of order $N=2$ and $N=4$ in the interior of the block. The values for DD4  are rounded to the 4th digit. }
  \label{tabl:dd24weights}
\end{table}

\section{Derivation of the Error Estimation given in eq.~(23)}
\label{sec:deriv-eq23}
Using \cref{eq:errorsum} the total error in \cref{eq-def:errWPOD} becomes
\begin{align}
    \label{eq:errwPOD-appendix1}
    \errWPOD\le
    \frac{
            \sum_{i=1}^{\NSnapshots}
            \norm{\u_i -\u_i^\epsilon}^2
        }{
            \sum_{i=1}^{\NSnapshots}
            \norm{\u_i}^2
        }
        +
        \frac{
        \sum_{i=1}^{\NSnapshots}
        \norm{\u_i^\epsilon -\tilde\u_i^\epsilon}^2
        }{
        \sum_{i=1}^{\NSnapshots}\norm{\u_i}^2}.
\end{align}
Furthermore we can simplify the first term with \cref{eq-def:waveleterror} inserting $\norm{\u_i - \u_i^\epsilon}\le \epsilon \norm{\u_i}$ into the nominator:
\begin{align*}
\frac{
    \sum_{i=1}^{\NSnapshots}
    \norm{\u_i - \u_i^\epsilon}^2
    }
    {\sum_{i=1}^{\NSnapshots}\norm{\u_i}^2}
=
    \frac{
        \sum_{i=1}^{\NSnapshots}(\epsilon \norm{\u_i})^2
    }{
        \sum_{i=1}^{\NSnapshots}\norm{\u_i}^2
    }
= \epsilon^2\,.
\end{align*}
The second term in \cref{eq:errwPOD-appendix1} can be expressed with the help of the eigenvalues of the correlation matrix. We use the identities: $\sum_{i=1}^{\NSnapshots}\norm{\u_i^\epsilon - \tilde\u_i^\epsilon}^2=\sum_{k=r+1}^{\NSnapshots} \lambda_k^\epsilon$ for perturbed eigenvalues $\lambda_k^\epsilon=\lambda_k+l_k\epsilon$ and
$\sum_{i=1}^{\NSnapshots}\norm{\u_i}^2=\sum_{k=1}^{\NSnapshots} \lambda_k$, yielding
\begin{align}
    \label{eq:errwPOD-appendix2}
    \frac{\sum_{i=1}^{\NSnapshots}\norm{\u_i^\epsilon(\vec{x}) - \tilde\u_i^\epsilon(\vec{x})}^2}{\sum_{i=1}^{\NSnapshots}\norm{\u_i(\vec{x})}^2}
    =
    \frac{\sum_{k=r+1}^{\NSnapshots}(\lambda_k+\epsilon l_k)}{\sum_{k=1}^{\NSnapshots}\lambda_k}
    = \errPOD(r,0) + \mathcal{M}_r \epsilon
    .
\end{align}
Here $\mathcal{M}_r={\sum_{k=r+1}^{\NSnapshots}l_k}/{\sum_{k=1}^{\NSnapshots}\lambda_k}$ is the \textit{perturbation coefficient} of the total error. Note that the perturbations $l_k$ are caused by the non vanishing mixed terms $\sprod{\varphi_\lambda^j}{\psi_{\mu,\lambda}^j}$ (see also \cite{CastrillonAmaratunga2002}), when computing the correlation matrix from thresholded snapshots $u_i^\epsilon$. Note that for orthogonal wavelets the first order perturbations would vanish.
For slowly decaying eigenvalues $\lambda_k$ the perturbation coefficient $\mathcal{M}_r$ is typically very small, since the sum of perturbations $l_k$ is small compared to the total energy. In this case it is reasonable to neglect the second term in \cref{eq:errwPOD-appendix2}:
\begin{equation}
    \errWPOD\lesssim \errPOD(r,0) + \epsilon^2\,.
\end{equation}
However, in general $\mathcal{M}_r$ does not vanish and we only have linear convergence in $\epsilon$:
\begin{equation}
    \errWPOD\le \errPOD(r,0) + \mathcal{M}_r\epsilon+\mathcal{O}(\epsilon^2)\,.
\end{equation}
Note that
$\mathcal{M}_r$ does not depend on $\epsilon$, as all epsilon dependence has been removed. So it is asymptotically a first order scheme in $\epsilon$ only. For a certain range we can observe second order, if $\mathcal{M}_r$ is sufficiently small. Eventually for $\epsilon$ sufficiently small the first order term will dominate.

\section{Technical Details and Supplementary Material}
\label{appxC}
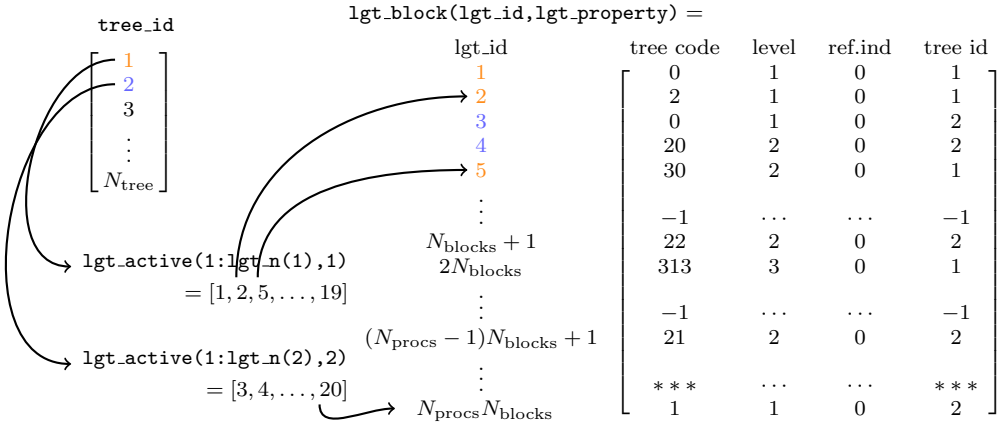
\begin{figure}[htp!]
	\centering
	\input{lgtdata}
	\caption{Example of the light data structure in \texttt{WABBIT}, to be compared with \cref{fig:treestructure}.
	For each tree \texttt{lgt\_active} stores a light-ID list of all active blocks.
	With the blocks light-ID (\texttt{lgt\_id}) all parameters in the forest (tree code,
	tree-ID, tree level, refinement status) can be accessed, from the corresponding
	row in \texttt{lgt\_block}. Note that the order of the light-ID depends on the
	process rank.
	}
	\label{fig:lgtstructure}
\end{figure}

\begin{figure}[htp!]
  \centering
  \includegraphics[width=0.48\textwidth]{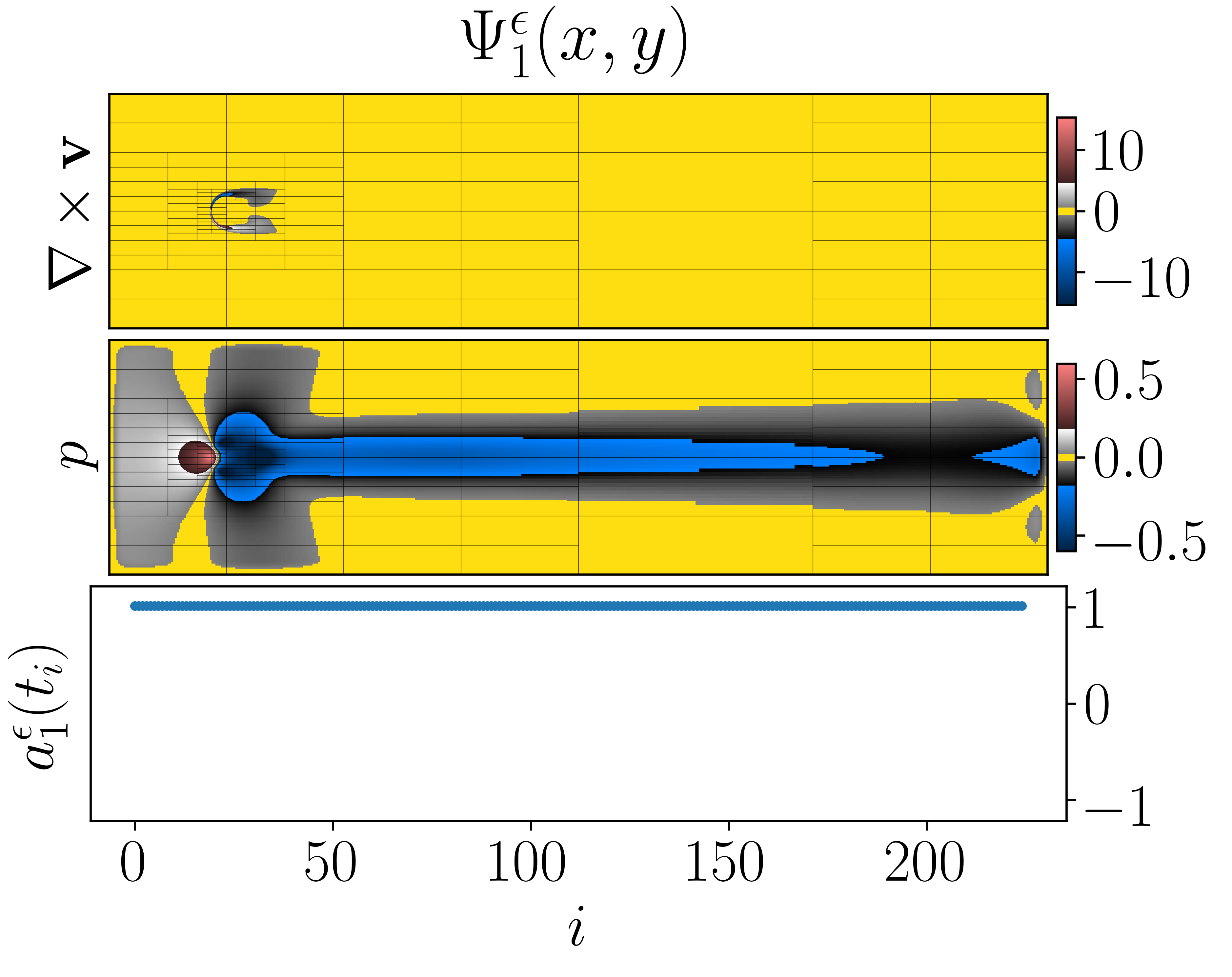}
  \includegraphics[width=0.48\textwidth]{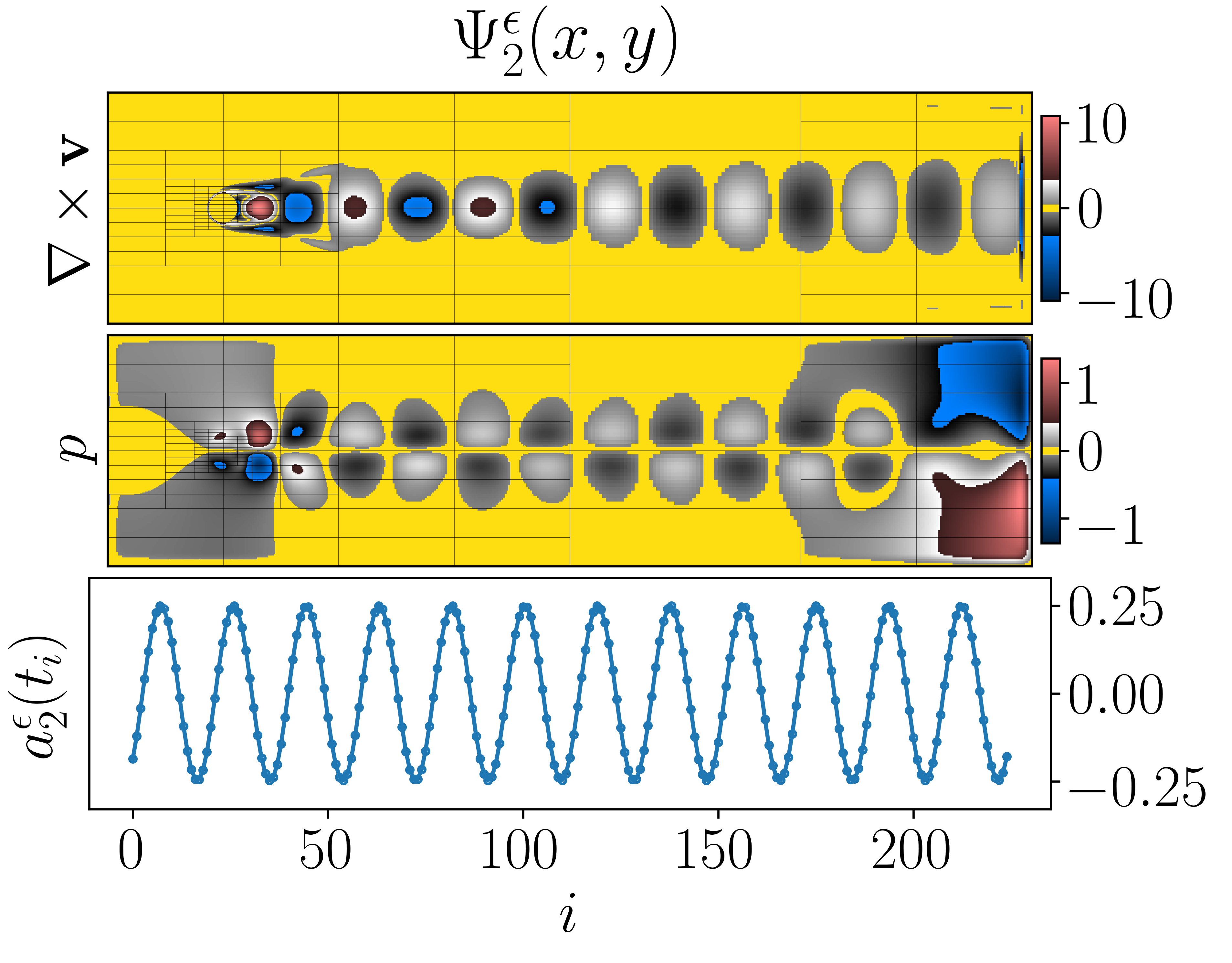}
  \includegraphics[width=0.48\textwidth]{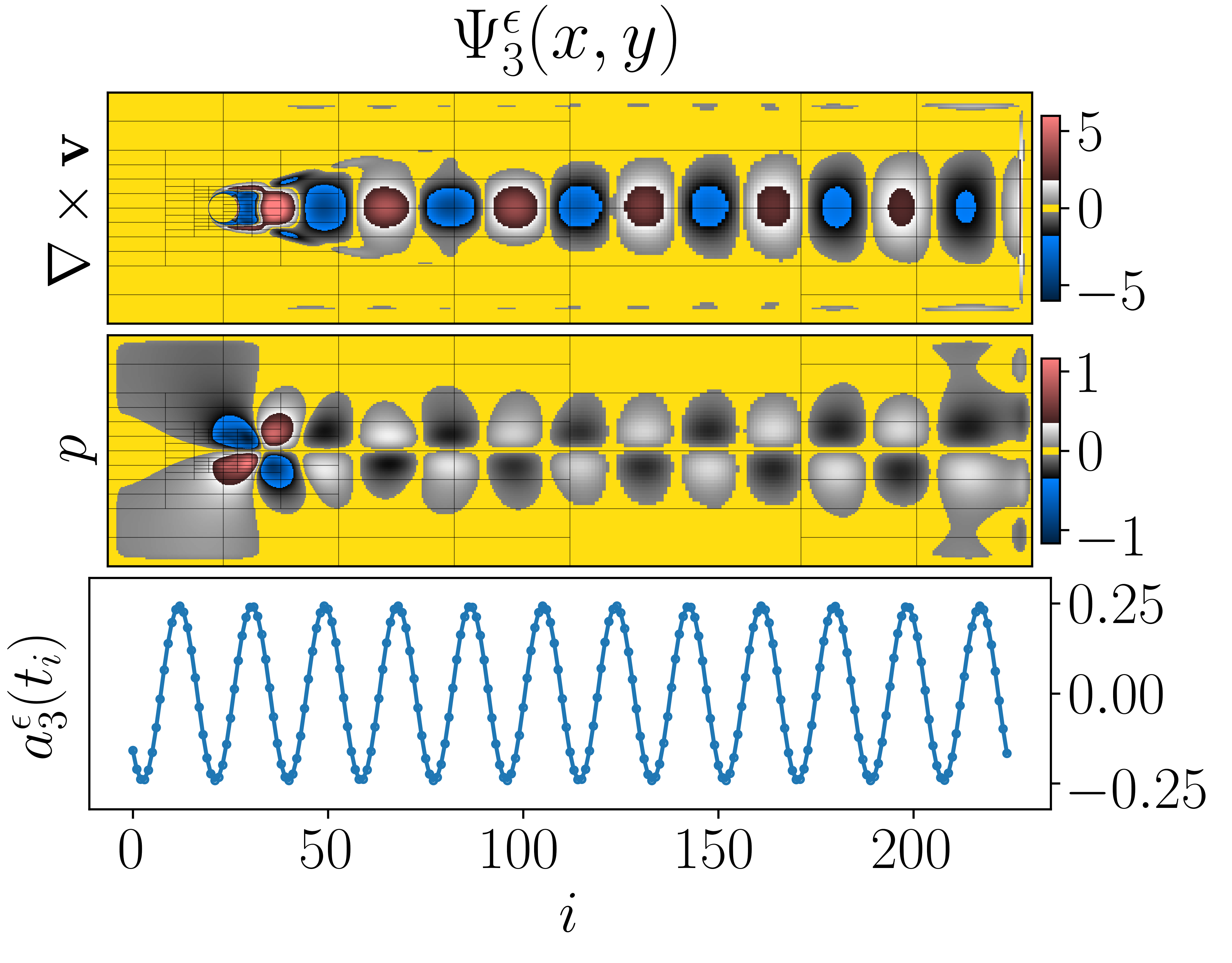}
  \includegraphics[width=0.48\textwidth]{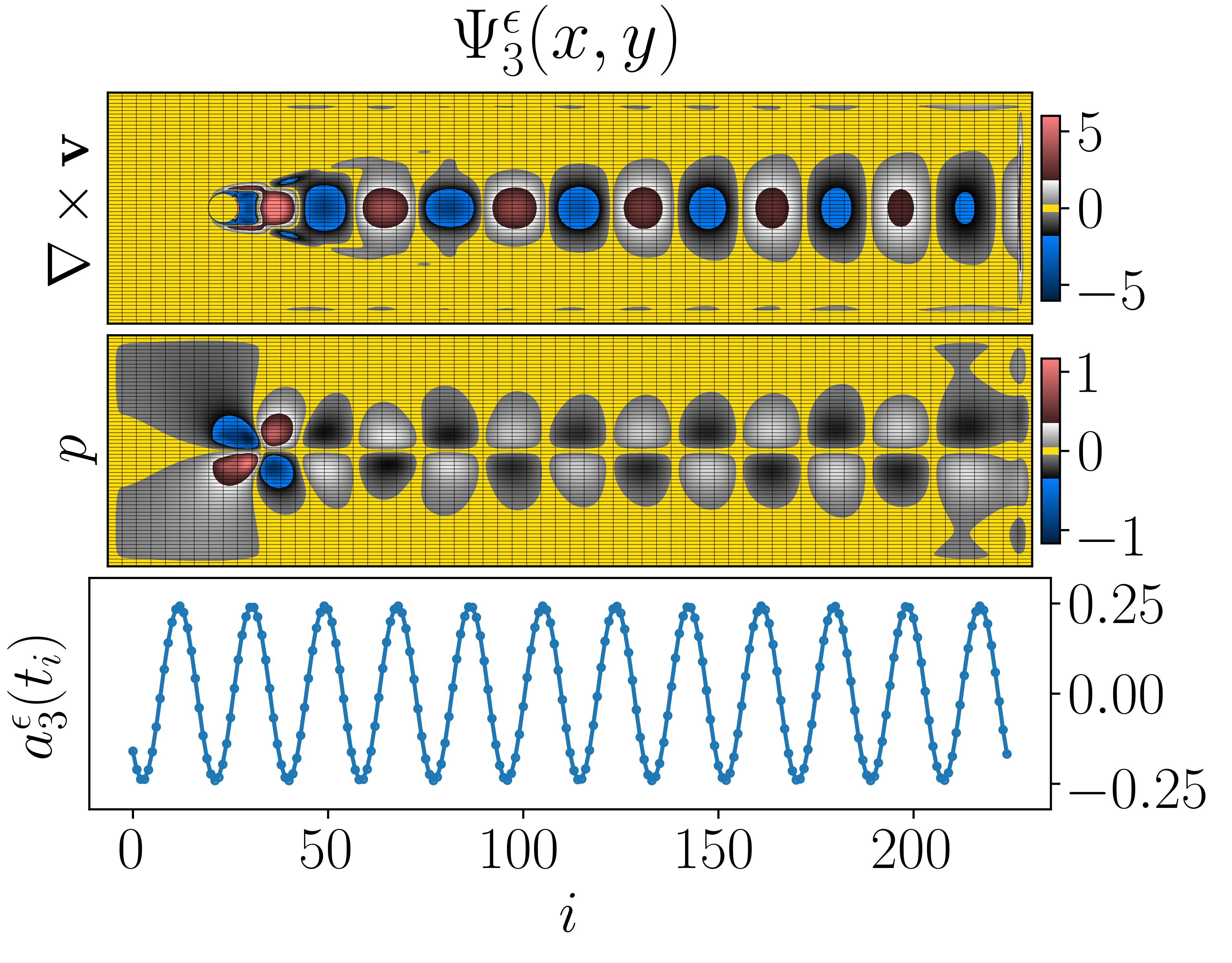}
  \caption{First three sparse modes $\vec{\Mode}^\epsilon_k$ ($k=1$ upper left, $k=2$ upper right, $k=3$ bottom left) with their corresponding amplitudes $a_k^\epsilon(t_i)$ computed with $\epsilon=\num{1.0e-2}$ and one dense ($\epsilon=0$) mode $\Mode^\epsilon_3$ for comparison (bottom, right). Each figure shows the modes{ }``vorticity" (labeled by $\nabla \times \vec{v}$), computed from the two velocity components of the modes, and the pressure component (labeled by $p$).
  Note that the first mode represents the base flow, which is non oscillating,
whereas the other modes always have an oscillating structure, with a frequency
increasing with the mode number.
When comparing the dense mode $\Mode^\epsilon_3$ of the non-adaptive case in the lower right of \cref{fig:vort_modes} with the adapted modes on the lower left, no qualitative differences can be observed, except the local changes in the resolution.}
  \label{fig:vort_modes}
\end{figure}

%% file: refinement.tex
\colorlet{colP1}{gray!60}
\colorlet{colP2}{gray!20}
\colorlet{colG}{blue!40}        
\colorlet{colredundant}{red!50} 
\usetikzlibrary{shapes.misc}

\tikzset{cross/.style={cross out, draw=black,minimum size=4pt, inner sep=0pt, outer sep=0pt}}
\begin{tikzpicture}[scale=0.4]

	\coordinate (A) at (3.5,2.5);
	\coordinate (B) at (10.5,8.5);

	\coordinate (P1a) at (5,7);
	\coordinate (P1b) at (9,4);
	\coordinate (P2a) at (P1b);
	\coordinate (P2b) at (17,1);
	\coordinate (P3a) at (P1b);
	\coordinate (P3b) at (17,9);
	\coordinate (P4a) at (3,3);
	\coordinate (P4b) at (17,11);

	\coordinate (P5a) at (3,3);
	\coordinate (P5b) at (1,1);
	\coordinate (P6a) at (P1a);
	\coordinate (P6b) at (1,11);
	\coordinate (P4a) at (P3b);
	\coordinate (P4b) at (15,11);

   \draw (16,11) rectangle (24,19);
   \draw (16,11) rectangle (20,15);
   \draw (20,15) rectangle (24,19);
   \draw (3,11) rectangle (11,19);



	\foreach \x in {3,...,11}
	{
		\foreach \y in {11,...,19}
		{
		\ifthenelse{\intcalcMod{\y-1}{2}=0 \AND \intcalcMod{\x-1}{2}=0}
						{
						\fill (\x,\y)  circle[radius=5pt];
						}{

						}
		}
	}

	\foreach \x in {16,...,24}
	{
		\foreach \y in {11,...,19}
		{
		\ifthenelse{\intcalcMod{\y-1}{2}=0 \AND \intcalcMod{\x}{2}=0}
						{
						\fill (\x,\y)  circle[radius=5pt];
						}{
						\draw (\x,\y) circle [radius =4pt] ;
						}
		}
	}
	\node [fill=white] at (18,18) {\footnotesize $p0$};
	\node [fill=white] at (22,18) {\footnotesize $p1$};
	\node [fill=white] at (18,14) {\footnotesize $p2$};
	\node [fill=white] at (22,14) {\footnotesize $p3$};
	\node [fill=white] at (7,16) {$p$};
        \node at (7,9) {$\mathcal{B}_p^{j}$};
				\node at (20,9) {($\mathcal{B}_{p0}^{j+1}$,$\mathcal{B}_{p1}^{j+1}$,$\mathcal{B}_{p2}^{j+1}$,$\mathcal{B}_{p3}^{j+1}$)};
				\coordinate (xblock) at (3,11);
				\fill (xblock) circle (7pt);
				\node at (3,10) {$(x_p^j,y_p^j)$};

\path[->] (7,20.5) edge[line width=0.742mm] node[ fill=white, anchor=center, pos=0.5,font=\bfseries] {refinement} (20,20.5);
\path[<-] (7,22) edge[line width=0.742mm] node[ fill=white, anchor=center, pos=0.5,font=\bfseries] {coarsening} (20,22);

\end{tikzpicture}

%% file: lgtdata.tex
\colorlet{colT1}{orange!90!white}
\colorlet{colT2}{blue!60!white}

\newcommand{\tikzmark}[3][]{\tikz[overlay,remember picture] \node [anchor=base,#1](#2) {#3};}
\begin{minipage}[c]{0.1\textwidth}
\begin{align*}
  \texttt{tree\_id}\\
  \left [
  \begin{array}{c}
    \tikzmark[colT1]{treeid1}{1} \\
    \tikzmark[colT2]{treeid2}{2} \\
    3\\
    \vdots \\
    N_\mathrm{tree}
  \end{array}
\right ]
\end{align*}

\begin{align*}
  \tikzmark[colT1]{lgtid1}{\;}
  \texttt{lgt\_active(1:lgt\_n(1),1)}\\ =
  \left [\tikzmark[colT2]{T1id1}{}1,\tikzmark[colT2]{T1id2}{\phantom{2}} 2,
  \tikzmark[colT2]{T1id3}{\phantom{1}}5, \dots , 19 \right ]\\
  &\\
  \tikzmark[colT2]{lgtid2}{\;}
  \texttt{lgt\_active(1:lgt\_n(2),2)}\\ =
  \left [ \tikzmark[colT2]{T2id1}{}3, \tikzmark[colT2]{T2id2}{\phantom{1}}4,  \dots , \tikzmark[colT2]{T2id20}{\phantom{1}}20 \right ]\\
\end{align*}
\end{minipage}%
\begin{minipage}[c]{0.6\textwidth}
\begin{align*}
     &    \texttt{lgt\_block(lgt\_id,lgt\_property)}=\\
     & \begin{blockarray}{*{5}{c} l}
    \begin{block}{*{5}{>{$\footnotesize}c<{$}} l}
      lgt\_id &tree code  & level & ref.ind & tree id \\
    \end{block}
    \begin{block}{>{$\footnotesize}c<{$}[*{5}{c}]}
    \tikzmark[colT1]{T1L1}{1}&0       &1      &0      &1\\
    \tikzmark[colT1]{T1L2}{2}&2       &1      &0      &1\\
      \tikzmark[colT2]{T2L1}{3}&0       &1      &0      &2\\
      \tikzmark[colT2]{T2L2}{4}&20      &2      &0      &2\\
      \tikzmark[colT1]{T1L3}{5}&30      &2      &0      &1\\
      \vdots &-1        &  \cdots& \cdots &  -1 \\
      $N_\mathrm{blocks}+1$ &22      &2 & 0 & 2\\
      $2N_\mathrm{blocks}$ &313       &3 & 0 & 1\\
      \vdots &-1        &  \cdots& \cdots &  -1 \\
      $(N_\mathrm{procs}-1)N_\mathrm{blocks}+1$ &21   &2 &0 & 2\\
      \vdots & ***       &  \cdots& \cdots &  *** \\
 \tikzmark[colT2]{T2L20}{\phantom{1}}     $N_\mathrm{procs}N_\mathrm{blocks}$ &1   &1 &0 & 2\\
    \end{block}
  \end{blockarray}
\end{align*}
\end{minipage}
\begin{tikzpicture}[overlay, remember picture,scale=0.7]
    \draw[,->,thick] (treeid1) to [in=-180,out=180] (lgtid1);
    \draw[,->,thick] (treeid2) to [in=-180,out=180](lgtid2);
    \draw[,->,thick] (T1id2.north) to  [in=-180,out=90](T1L2);
    \draw[,->,thick] (T1id3.north) to  [in=-180,out=90](T1L3);
    \draw[,->,thick] (T2id20.south) to [in=-180,out=270] (T2L20);
\end{tikzpicture}